\documentclass[12pt,a4paper]{article}
\usepackage{amsmath,amssymb}
\usepackage{latexsym}
\usepackage{graphicx}
\usepackage{color}
\usepackage{multirow}
\usepackage{indentfirst}
\usepackage{subfigure}
\usepackage{mathrsfs}
\usepackage[pdftex]{hyperref}
 \usepackage[numbers,sort&compress]{natbib}

\newtheorem{exam}{\hspace{6mm}Example}[section]

\begin{document}
\baselineskip=2pc

\begin{center}
{\large \bf  A quasi-conservative discontinuous Galerkin method for multi-component flows using the non-oscillatory kinetic flux}
\end{center}

\centerline{
Dongmi Luo%
\footnote{Institute of Applied Physics and Computational Mathematics, Beijing 100088, China. E-mail: dongmiluo@stu.xmu.edu.cn.},
Jianxian Qiu%
\footnote{School of Mathematical Sciences and Fujian Provincial
Key Laboratory of Mathematical Modeling and High-Performance
Scientific Computing, Xiamen University, Xiamen, Fujian 361005, China. E-mail: jxqiu@xmu.edu.cn.},
Jun Zhu%
\footnote{College of Sciences, Nanjing University of Aeronautics and Astronautics, Nanjing, Jiangsu 210016, China. E-mail: zhujun@nuaa.edu.cn.},
Yibing Chen%
\footnote{Institute of Applied Physics and Computational Mathematics, Beijing 100088, China, 
E-mail: chen\_yibing@iapcm.ac.cn.}
}

\vspace{20pt}

\begin{abstract}
In this paper, a high order quasi-conservative discontinuous Galerkin (DG) method using the non-oscillatory kinetic flux is proposed for  the 5-equation model of compressible multi-component flows with Mie-Gr\"uneisen equation of state. 
The method mainly consists of three steps: firstly, the DG method with the non-oscillatory kinetic flux is used to solve the conservative equations of the model; secondly, inspired by Abgrall's idea, we derive a DG scheme for the volume fraction equation which can avoid the unphysical oscillations near the material interfaces; 
finally, a multi-resolution WENO limiter and a maximum-principle-satisfying limiter are employed to ensure oscillation-free near the discontinuities, and preserve the physical bounds for the volume fraction, respectively.
Numerical tests show that the method can achieve high order for smooth solutions and keep non-oscillatory at discontinuities. Moreover, the velocity and pressure are oscillation-free at the interface and the volume fraction can stay in the interval [0,1].
\end{abstract}


\pagenumbering{arabic}

\newpage

\section{Introduction}
\label{sec1}
\setcounter{equation}{0}
\setcounter{figure}{0}
\setcounter{table}{0}

Numerical simulation of compressible multi-component flows with immiscible interfaces has been an active research topic in the computational fluid dynamics because of their application to a wide range of field, such as inertial confinement fusion, underwater bubble dynamic and so on. The major difficulty of the simulations of multi-component flows is how to track the material interfaces clearly. 

The numerical approaches can be split into two major groups with respect to the treatment of  the material interface in the Eulerian framework. One is the sharp interface method (SIM) and the other is the diffuse interface method (DIM). Sharp interface methods \cite{wang2010, wang2006, qiu2007, qiu2008, zhu2011, fedkiw1999, liu2003, liu2005, lu2016, osher1988,chen1986}
view the multi-material interfaces as genuine discontinuities. Thus the sharp interfaces are strictly maintained. 
However, none of these methods is able to dynamically create interfaces and to solve interfaces separating pure medium and mixtures as stated in \cite{H11}.

In contrast, in the diffuse interface approach the interfaces are viewed as a numerically diffused zone, 
and an artificial mixture zone is created.
A number of different models \cite{H01,allaire2002, kreeft2010, shyue1999, shyue2001, shyue1998, H18, abgrall2001} have been developed so far based on this idea, including 4-equation model, 5-equation model, 7-equation model and so on. 
However, these models are usually non-conservative, which leads both theoretical and computational challenging problems \cite{abgrall2001}. The special strategies are required to handle these non-conservative terms in order to keep the pressure and velocity non-oscillatory at the interfaces. The quasi-conservative approach developed by Abgrall in \cite{H01} is an effective means to deal with this problem. Based on Godunov method for solution evolution, 
Shyue extended the idea to the different equations of state (EOSs), such as stiffened gas equation \cite{shyue1998}, Van der Waals \cite{shyue1999}, Mie-Gr\"uneisen \cite{shyue2001}. Besides the traditional Godunov method, an alternative is the gas kinetic scheme (GKS) \cite{xu2001}, which provides more physical information of the flow and is free from constructing Riemann solver. 
In the past decades, the GKS has been well developed to solve for multi-component flows \cite{ni2011,lian2000,pan2017,xu1997,li2019,li2020,liu2017}. {
A second-order gas-kinetic scheme for multicomponent flow was presented in \cite{xu1997,lian2000} based on the BGK equation for each component with its own equilibrium state. Chen and Jiang proposed a non-oscillatory kinetic (NOK) scheme for the ideal gas \cite{chen2009} and stiffened gas \cite{chen2011}. Ni and Sun \cite{ni2011} proposed a $\gamma$-DGBGK scheme for compressible multiconponent flows simulation.  Recently, an improved GKS for multicomponent flows \cite{li2020} is proposed to increase the computational efficiency. 
In these papers mentioned above, most of them are the second-order schemes at most. The works in \cite{ni2011,liu2017,pan2017} can achieve high order, but only are applied to the ideal gas or stiffened gas.}

For the diffuse interface method, the numerical diffusion may lead to a very bad representation of the interfaces, especially when long time computations are needed.
A way to circumvent the numerical diffusion is to adopt a high order method, such as spectral volume method \cite{liu2017}, weighted essentially non-oscillatory (WENO) method \cite{rehman2018,nonomura2017,coralic2014,johnsen2006}, discontinuous Galerkin (DG) method which we are interested in. 
DG method has been applied to solve a variety of different models \cite{henrydefrahan2015, saleem2018, gryngarten2013, franquet2012, cheng2020}.  
There exist a few research works in the 5-equation model. Saleem, Ali and Qamar \cite{saleem2018} adopted the second-order Runge-Kutta DG (RKDG) method 
for solving the reduced 5-equation model \cite{kreeft2010}. In their work, the Lax-Friedrichs (LF) flux and the local LF flux were used to compute the numerical flux and a WENO limiter was utilized to eliminate oscillations at discontinuities. However, one can observe that the velocity and pressure produce the oscillations at the interface from the results of the interface only problem since the limiter was applied to the conservative variables.
Gryngarten and Menon \cite{gryngarten2013} applied the local DG method \cite{yan2011} to the 5-equation model \cite{allaire2002} with Peng-Robinson EOS, where the non-conservative equations were rewritten into conservative formula with source terms. A moment limiter \cite{krivodonova2007} was applied to the conserved and primitive variables. The numerical flux used in their study is the HLLC approximate Riemann solver for the conservation of mass, momentum and energy. But the additional equations must be solved which increases the computations.

In this paper, a high order quasi-conservative DG method for compressible multi-component flows with Mie-Gr\"uneisen EOS based on the 5-equation model \cite{allaire2002} is developed. 
The method can obtain the high order in smooth regions, keep oscillation-free at discontinuities, including the material interface, which is different from the work in \cite{saleem2018}, and guarantee the volume fraction in [0,1]. In addition, we do not need the extra equations to solve and reduce the computations compared to the work in \cite{gryngarten2013}.
Following the idea of the quasi-conservative method introduced by Abgrall \cite{H01}, the quasi-conservative DG method with NOK flux has three steps. Firstly, we adopt DG method to discretize the conservative equations in space. In order to treat  Mie-Gr\"uneisen EOS, the NOK flux \cite{chen2011,liu2017} is utilized to compute the numerical flux in our work instead of the traditional numerical flux. Secondly, according to the discretizations of the conservative equations, the necessary condition that avoids the unphysical oscillation near the material interfaces is derived, which is also the discretization method for the volume fraction equation in \eqref{eq1}. 
At last, the new multi-resolution WENO limiter \cite{zhu2020} is employed to prevent the oscillations at discontinuities. In order to keep the pressure and velocity oscillation-free at the interfaces, we applied the limiter to the primitive variables as in \cite{gryngarten2013}  and the maximum-principle-satisfying limiter developed by Zhang and Shu \cite{H16,H19} is applied to ensure that the volume fraction does not go out of the range. 
Thus, a high order quasi-conservative discontinuous Galerkin method for multi-component flows with Mie-Gr\"uneisen EOS using the NOK flux is developed. 

The organization of the paper is as follows. The governing equations and EOSs are described in Section \ref{goveq}. In Section \ref{sec2}, the DG method for multi-component flows, identification of troubled cells, and limiters are described in detail. One- and two-dimensional numerical examples are presented to demonstrate the accuracy and the oscillation-free of the method in Section \ref{sec3}. In Section \ref{sec4}, the conclusions are given.

\section{Govorning equation}
\label{goveq}
In one dimension, the 5-equation model for an immiscible two-material compressible flow \cite{allaire2002} is considered, which is in the form of:

\begin{equation}
\label{eq1}
\begin{cases}
W_t+F(W)_x=0,\\
\frac{\partial Y}{\partial t}+v\frac{\partial Y}{\partial x}=0,
\end{cases}
\end{equation}
where $W=(\rho_1Y_1, \rho_2Y_2, \rho v, E)^T$ and $F(W)=(\rho_1Y_1v, \rho_2Y_2v, \rho v^2+P, v(E+P))^T$; $\rho_1$ and $\rho_2$ are the partial density of the fluids 1 and 2, respectively; $P$ is the pressure, $v$ is the velocity and $E=\rho e+\frac{1}{2}\rho v^2$ is the total energy with $\rho e$ being the internal energy; $Y_1=Y$  is the volume fraction of fluid 1, lies in the interval $[0,1]$, and $Y_1+Y_2=1$. The total density, momentum and energy of the mixture are defined as
\begin{align}
\label{mixrho}
 \rho&=\rho_1Y_1+\rho_2Y_2,\quad \rho v=Y_1\rho_1v_1+Y_2\rho_2v_2,\\
 \label{mixe}
  E&=Y_1\rho_1e_1+\frac{1}{2}\rho_1Y_1(v_1)^2+Y_2\rho_2e_2+\frac{1}{2}\rho_2Y_2(v_2)^2.
 \end{align}

In order to close the equation \eqref{eq1}, a mixture EOS is needed. In this work, each of the fluids is modeled by Mie-Gr\"uneisen EOS, i.e.
\[
P(\rho,e)=\Gamma(\rho)[\rho e-\rho e_{ref}(\rho)]+P_{ref}(\rho),
\]
where $\Gamma$ is the Gr\"uneisen coefficient, $P_{ref}(\rho) $ and $e_{ref}(\rho)$ are the reference pressure and internal energy, respectively.
This is a general EOS since it can produce the different types of EOSs:\\
(1) Ideal gas EOS
\begin{equation}
\label{ig}
\begin{cases}
\Gamma(\rho)=\gamma-1,\\
P_{ref}(\rho)=0,\\
e_{ref}(\rho)=0;
\end{cases}
\end{equation}
(2) Stiffened gas EOS
\begin{equation}
\label{sg}
\begin{cases}
\Gamma(\rho)=\gamma-1,\\
P_{ref}(\rho)=-\gamma B,\\
e_{ref}(\rho)=0;
\end{cases}
\end{equation}
(3) Jones-Wilkins-Lee EOS (JWL EOS)
\begin{equation}
\label{jwl}
\begin{cases}
\Gamma(\rho)=\Gamma_0,\\
P_{ref}(\rho)=\frac{\mathcal{A}}{\mathcal{R}_1\rho_0}\exp(-\frac{\mathcal{R}_1\rho_0}{\rho})+\frac{\mathcal{B}}{\mathcal{R}_2\rho_0}\exp(-\frac{\mathcal{R}_2\rho_0}{\rho})-e_0,\\
e_{ref}(\rho)=\mathcal{A}\exp(-\frac{\mathcal{R}_1\rho_0}{\rho})+\mathcal{B}\exp(-\frac{\mathcal{R}_2\rho_0}{\rho});
\end{cases}
\end{equation}
where $\mathcal{A}, \mathcal{R}_1, \rho_0, \mathcal{B}, \mathcal{R}_2 $ and $e_0$ are the material-dependent parameters.\\
(4) Cochran-Chan EOS (CC EOS)
\begin{equation}
\label{cc}
\begin{cases}
\Gamma(\rho)=\Gamma_0,\\
P_{ref}(\rho)=-\frac{\mathcal{A}}{\rho_0(1-\epsilon_1)}[(\frac{\rho_0}{\rho})^{1-\epsilon_1}-1]+\frac{\mathcal{B}}{\rho_0(1-\epsilon_2)}[(\frac{\rho_0}{\rho})^{1-\epsilon_2}-1]-e_0,\\
e_{ref}(\rho)=\mathcal{A}(\frac{\rho_0}{\rho})^{-\epsilon_1}-\mathcal{B}(\frac{\rho_0}{\rho})^{-\epsilon_2};
\end{cases}
\end{equation}
where $\mathcal{A}, \epsilon_1, \rho_0, \mathcal{B}, \epsilon_2 $ and $e_0$ are the material-dependent parameters.\\
(5) Shock-Wave EOS (Shock EOS)
\begin{equation}
\label{sw}
\begin{cases}
\Gamma(\rho)=\Gamma_0(\frac{V}{V_0})^{\alpha},\; V=\frac{1}{\rho},\; V_0=\frac{1}{V_0},\\
P_{ref}(\rho)=P_0+\frac{c_0^2(V_0-V)}{[V_0-s(V_0-V)]^2},\\
e_{ref}(\rho)=e_0+\frac{1}{2}[P_{ref}(\rho)+P_0](V_0-V);
\end{cases}
\end{equation}
where $s, c_0, \rho_0, \alpha, P_0 $ and $e_0$ are the material-dependent parameters.

A wide range of real materials can be modeled by these EOSs. A typical set of numerical values for some sample materials is listed in Table \ref{tab1}. Mie-Gr\"uneisen EOS can be also rewritten as
\[
P(\rho,\rho e)=(\gamma(\rho)-1)\rho e-\gamma(\rho)\pi(\rho),
\]
where $\gamma(\rho)=\Gamma(\rho)+1, \pi(\rho)=\frac{\Gamma(\rho)\rho e_{ref}(\rho)-P_{ref}(\rho)}{\Gamma(\rho)+1}.$

\begin{table}
\caption{Material-dependent quantities used in this paper. See \cite{shyue2001} for other material parameters.  }
\renewcommand{\multirowsetup}{\centering}
\begin{center}
\begin{tabular}{|c|c|c|c|c|c|c|c|c|c|c|c|c|}
\hline
JWL EOS & $\rho_0(kg/m^3)$ & $\mathcal{A}$(GPa)     & $\mathcal B$(GPa) & $\mathcal R_1$ & $\mathcal R_2$ &$\Gamma_0$ &$\alpha$\\
Water & 1004 & 1582 & -4.67 & 8.94 & 1.45 & 1.17 & 0\\
CC EOS & $\rho_0(kg/m^3)$ & $\mathcal A$(GPa) & $\mathcal B$(GPa) & $\epsilon_1$ & $\epsilon_2$ & $\Gamma_0$ & $\alpha$ \\
Copper & 8900 & 145.67 & 147.75 & 2.99 & 1.99 & 2 & 0\\
TNT & 1840 & 12.87 & 13.42 & 4.1 & 3.1 & 0.93 & 0\\
Shock EOS & $\rho_0(kg/m^3)$ & $c_0(m/s)$ & $s$ & $\Gamma_0$ & $\alpha$ & $P_0$ & $e_0$\\
Molybdenum & 9961 & 4770 & 1.43 & 2.56 & 1 & 0 & 0\\
MORB & 2660 &2100 & 1.68 & 1.18 & 1 & 0 &0\\
 \hline
\end{tabular}
\end{center}
\label{tab1}
\end{table}

To close system \eqref{eq1} for the mixing cells, the isobaric closure assumption  \cite{allaire2002} is supplemented here, which assumes that there is no pressure jump across a material interface, i.e.
\[
P_1=P_2=P.
\]
 Thus, according to \eqref{mixrho} and \eqref{mixe}, the internal energy of the mixture is given by
 \begin{equation}
 \rho e=\sum\limits_kY_k\rho_ke_k=\sum\limits_kY_k\frac{P_k+\gamma_k(\rho_k)\pi_k(\rho_k)}{\gamma_k(\rho_k)-1}.
 \end{equation}
 Using $P_1=P_2=P$, we can obtain 
 \[
 \frac{P+\gamma\pi}{\gamma-1}=\rho e=\sum\limits_kY_k\frac{P+\gamma_k(\rho_k)\pi_k(\rho_k)}{\gamma_k(\rho_k)-1}.
 \]
 Therefore, we have the two following equations:
 \begin{equation}
 \frac{1}{\gamma-1}=\sum\limits_k\frac{Y_k}{\gamma_k(\rho_k)-1},\; \frac{\gamma\pi}{\gamma-1}=\sum\limits_k\frac{Y_k\gamma_k(\rho_k)\pi_k(\rho_k)}{\gamma_k(\rho_k)-1}.
 \end{equation}
 
 Finally, the mixing sound speed \cite{allaire2002} can be written as follows:
 \[
 c^2=(\gamma-1)\sum\limits_{k}\frac{z_kc_k^2}{\gamma_k-1},
 \]
 where $c_k$ is the sound speed of the $k$th material and $z_k$ is mass fraction of fluid $k$ defined as $z_k=\frac{\rho_kY_k}{\sum\limits_{l}\rho_lY_l}$.

\section{The numerical scheme}
\label{sec2}
\setcounter{equation}{0}
\setcounter{figure}{0}
\setcounter{table}{0}

In this section, we describe a quasi-conservative RKDG method for the numerical
solution of compressible multi-components in the form of (\ref{eq1}) on a uniform mesh, which contains three steps. 

Step 1. Discretize the conservative equations in \eqref{eq1} using RKDG method \cite{cockburn1990, cockburn1989jcp, cockburn1989, cockburn1998} with NOK flux which is suitable for Mie-Gr\"uneisen EOS. 

Step 2. Following the idea of Abgrall's quasi-conservative method, define a numerical scheme for the equation of volume fraction which can prevent the oscillation of pressure and velocity near the material interfaces.

Step 3. Add the limiters, which include the limiters for oscillation-free near the discontinuities and the maximum-principle-satisfying limiters for the volume fraction.



\subsection{DG method for the conservative equations}

For simplicity, we take one dimension for example. For two dimensions, we can implement it similarly.
Assume the domain $\Omega$ is divided into $N$ nonoverlapping cells $\{I_j=(x_{j-\frac{1}{2}},x_{j+\frac{1}{2}}),j=1,\cdots,N\}$, and $\Delta x=x_{j+\frac{1}{2}}-x_{j-\frac{1}{2}}$. The DG finite element space is defined as
\[
V_h^k=\{p(x,t):p|_{I_j}\in P^k(I_j)\},
\]
where $P^k(I_j)$ is the space of polynomials of degree $\leq k$ defined on $I_j$.
Notice that $P^k(I_j)$ can be expressed as
\[
P^k(I_j) = \text{span}\{\varphi_1(x),\cdots,\varphi_L(x)\},
\]
where $L=k+1$ for one dimensional case, 
and $\{\varphi_1(x),\; \cdots,\; \varphi_L(x)\}$ is a basis of $P^k(I_j)$. The first three basis functions in one dimension we employ on the cell $I_j$ are
\begin{align}
\varphi_1(x)=1,\; \varphi_2(x)=\frac{x-x_j}{\Delta x}, \; \varphi_3(x)=(\frac{x-x_j}{\Delta x})^2-\frac{1}{12}, \; \forall x\in I_j.
\end{align}

The semi-discrete DG approximation for the conservative equations in (\ref{eq1}) is to
 find the numerical solution $u_h(\cdot,t)\in V_h^k$, $t \in (0, T]$ such that
 \begin{align}
  \label{v1}
  \int_{I_j} \frac{\partial W_h}{\partial t} \psi dx+(F\psi)|_{I_j}-\int_{I_j} F\psi_xdx=0,
\qquad  \forall \psi \in P^k(I_j),
\end {align}

Expressing $W_h$ as
 \begin{equation}
 \label{numer}
 W_h(x,t)=\sum_{l=1}^L W_j^{(l)}(t)\varphi_l(x),\quad \forall x\in I_j,
 \end{equation}
applying the Gauss quadrature rule to the third terms in (\ref{v1}) and replacing the flux $F$ by the numerical flux $\Hat F$, we obtain
\begin{equation}
\label{vv}
\begin{cases}
\int_{I_j} (\sum\limits_{l=1}^{L}\frac{dW^{(l)}_j(t)}{dt}\varphi_l(x)) \psi dx+\Hat F_{j+\frac{1}{2}}\psi_{j+\frac{1}{2}}^--\Hat F_{j-\frac{1}{2}}\psi_{j-\frac{1}{2}}^+\\
-\sum\limits_{G}F(W_h(x_{G}))w_G\psi_x(x_G)|I_j|=0,\quad \forall \psi\in V_h^k,\\
\int_{I_j} (W_h(x,0)-W(x,0))\psi dx=0,\quad \forall \psi\in V_h^k,
\end{cases}
\end{equation}
where $|I_j|$ is the volume of the element $I_j$, and $x_G$ and $w_G$ represent the Gaussian points and the weights on $I_j$, respectively. The summation $\sum\limits_G$ is taken over the Gauss points on $I_j$. The numerical flux has the form $\Hat F_{j+\frac{1}{2}}=\Hat F(u_{j+\frac{1}{2}}^{-},u_{j+\frac{1}{2}}^{+})$,
and $u_{j+\frac{1}{2}}^{-}=u(x_{j+\frac{1}{2}}^-)$ and $u_{j+\frac{1}{2}}^{+}=u(x_{j+\frac{1}{2}}^+)$ are defined as the values from the left
and right limit of $x_{j+\frac{1}{2}}$, respectively. 

In this work,  the NOK flux \cite{chen2011, liu2017} which can deal with the general EOS, is employed. It consists of two parts
\[
\Hat F_{j+\frac{1}{2}}=\eta F^K_{j+\frac{1}{2}}+(1-\eta)F^E_{j+\frac{1}{2}},
\]
where $\eta\in [0,1].$
The non-equilibrium part is
\[
F^K_{j+\frac{1}{2}}=F^+_{j+\frac{1}{2}}+F^-_{j+\frac{1}{2}},
\]
where
\begin{equation}
F^{\pm}_{j+\frac{1}{2}}=<u^1>_{{j+\frac{1}{2}},L/R}\begin{pmatrix}\rho_1Y_1 \\ \rho_2Y_2 \\ \rho v \\ E \end{pmatrix}_{j+\frac{1}{2}}^{\mp}+\begin{pmatrix} 0 \\ 0 \\ P_{j+\frac{1}{2}}^{\mp}<u^0>_{j+\frac{1}{2},L/R}  \\ \frac{1}{2}P_{j+\frac{1}{2}}^{\mp}<u^1>_ {j+\frac{1}{2},L/R}+\frac{1}{2}P_{j+\frac{1}{2}}^{\mp}v_{j+\frac{1}{2}}^{\mp}<u^0>_ {j+\frac{1}{2},L/R}\end{pmatrix},
\end{equation}
\[
<u^0>_{j+\frac{1}{2},L/R}=\frac{1}{2} \hbox{erfc} (\mp\sqrt{\lambda_{j+\frac{1}{2}}}v_{j+\frac{1}{2}}^{\mp}),
\]
\[
<u^1>_{j+\frac{1}{2},L/R}=v_{j+\frac{1}{2}}^{\mp}<u^0>_{j+\frac{1}{2},L/R}\pm\frac{1}{2}\frac{\hbox{e}^{-\lambda_{j+\frac{1}{2}}(v_{j+\frac{1}{2}}^{\mp})^2}}{\sqrt{\pi\lambda_{j+\frac{1}{2}}}},
\]

\[
\lambda_{j+\frac{1}{2}}=\min\{\frac{1}{(c^-_{j+\frac{1}{2}})^2},\frac{1}{(c^+_{j+\frac{1}{2}})^2} \},
\]
$c$ is the speed of sound.

In order to avoid oscillations of the pressure and velocity near a contact discontinuity, the equilibrium part should satisfy the consistent condition. The primitive variables are computed by
\begin{equation}
\begin{pmatrix}
\bar \rho_1\\
\bar \rho_2\\
\bar v\\
\bar P\\
\bar Y_1
\end{pmatrix}_{j+\frac{1}{2}}=
\begin{pmatrix}
(\rho_1)_{j+\frac{1}{2}}^-<u^0>_{j+\frac{1}{2},L}+(\rho_1)_{j+\frac{1}{2}}^+<u^0>_{j+\frac{1}{2},R}\\
(\rho_2)_{j+\frac{1}{2}}^-<u^0>_{j+\frac{1}{2},L}+(\rho_2)_{j+\frac{1}{2}}^+<u^0>_{j+\frac{1}{2},R}\\
<u^1>_{j+\frac{1}{2},L}+<u^1>_{j+\frac{1}{2},R}\\
P_{j+\frac{1}{2}}^-<u^0>_{j+\frac{1}{2},L}+P_{j+\frac{1}{2}}^+<u^0>_{j+\frac{1}{2},R}\\
(Y_1)_{j+\frac{1}{2}}^-<u^0>_{j+\frac{1}{2},L}+(Y_1)_{j+\frac{1}{2}}^+<u^0>_{j+\frac{1}{2},R}
\end{pmatrix}.
\end{equation}
Then we take
\begin{equation}
F^E_{j+\frac{1}{2}}=\begin{pmatrix}
\bar\rho_1 \bar Y_1\bar v\\
\bar\rho_2 \bar Y_2\bar v\\
\bar\rho \bar v^2+\bar P\\
\bar v(\bar E+\bar P)
\end{pmatrix}_{j+\frac{1}{2}},
\end{equation}
and $\bar E$ is determined by EOS.

\subsection{DG method for volume fraction equations}

Following the procedure of Abgrall \cite{H01} to avoid the oscillations of the pressure and velocity, 
we consider the interface only problem, and assume the velocity $v$ and the pressure $P$ are constants, i.e. $v=v_0,\;P=P_0$. Thus, $<u^0>_L+<u^0>_R=1$ and $<u^1>_L+<u^1>_R=v_0$ in the NOK flux. Firstly, we introduce
$$\tilde Z_{j+\frac{1}{2}}=Z_{j+\frac{1}{2}}^-<u^1>_{j+\frac{1}{2},L}+Z_{j+\frac{1}{2}}^+<u^1>_{j+\frac{1}{2},R},$$
$$a_l=\int_{I_j}(\varphi_l(x))^2dx,\; l=1,\cdots,L$$
for notation.
Then from the current spatial discretization, the semi-discretized scheme of the continuity equation can be written in the form as following:
\begin{align}
\label{rho1d1}
\frac{d(\rho_1Y_1)_j^{(l)}}{dt}=-&\frac{1}{a_l}[\eta(\widetilde {(\rho_1Y_1)}_{j+\frac{1}{2}}\varphi_l(x_{j+\frac{1}{2}}^-)-\widetilde {(\rho_1Y_1)}_{j-\frac{1}{2}}\varphi_l(x_{j-\frac{1}{2}}^+))+(1-\eta)v_0(\overline {(\rho_1Y_1)}_{j+\frac{1}{2}}\varphi_l(x_{j+\frac{1}{2}}^-)-\notag\\
&\overline{(\rho_1Y_1)}_{j-\frac{1}{2}}\varphi_l(x_{j-\frac{1}{2}}^+))-
v_0\int_{I_j}(\rho_1Y_1)(\varphi_l(x))_x dx], \; l=1,\cdots, L.
\end{align}

\begin{align}
\label{rho1d2}
\frac{d(\rho_2Y_2)_j^{(l)}}{dt}=-&\frac{1}{a_l}[\eta(\widetilde {(\rho_2Y_2)}_{j+\frac{1}{2}}\varphi_l(x_{j+\frac{1}{2}}^-)-\widetilde {(\rho_2Y_2)}_{j-\frac{1}{2}}\varphi_l(x_{j-\frac{1}{2}}^+))+(1-\eta)v_0(\overline {(\rho_2Y_2)}_{j+\frac{1}{2}}\varphi_l(x_{j+\frac{1}{2}}^-)-\notag\\
&\overline{(\rho_2Y_2)}_{j-\frac{1}{2}}\varphi_l(x_{j-\frac{1}{2}}^+))-
v_0\int_{I_j}(\rho_2Y_2)(\varphi_l(x))_x dx], \; l=1,\cdots, L.
\end{align}

Since $\rho=\rho_1Y_1+\rho_2Y_2$, we can get
\begin{align}
\label{rho1d3}
\frac{d\rho_j^{(l)}}{dt}=-&\frac{1}{a_l}[\eta(\tilde \rho_{j+\frac{1}{2}}\varphi_l(x_{j+\frac{1}{2}}^-)-\tilde \rho_{j-\frac{1}{2}}\varphi_l(x_{j-\frac{1}{2}}^+))+(1-\eta)v_0(\Bar\rho_{j+\frac{1}{2}}\varphi_l(x_{j+\frac{1}{2}}^-)-\Bar\rho_{j-\frac{1}{2}}\varphi_l(x_{j-\frac{1}{2}}^+))-\notag\\
&v_0\int_{I_j}\rho(\varphi_l(x))_x dx], \; l=1,\cdots, L.
\end{align}

Similarly, the discretization for the momentum equation can be written as
\begin{align}
\label{rhou1d1}
\frac{d(\rho v)_j^{(l)}}{dt}=-&\frac{1}{a_l}[\eta v_0(\tilde \rho_{j+\frac{1}{2}}\varphi_l(x_{j+\frac{1}{2}}^-)-\tilde \rho_{j-\frac{1}{2}}\varphi_l(x_{j-\frac{1}{2}}^+))+(1-\eta)v_0^2(\Bar\rho_{j+\frac{1}{2}}\varphi_l(x_{j+\frac{1}{2}}^-)-\Bar\rho_{j-\frac{1}{2}}\varphi_l(x_{j-\frac{1}{2}}^+))-\notag\\
&v_0^2\int_{I_j}\rho(\varphi_l(x))_x dx], \; l=1,\cdots, L.
\end{align}

Based on the equation (\ref{rho1d3}) and (\ref{rhou1d1}), we can derive $\frac{dv_j^{(l)}}{dt}=0$. 
The discretization for energy is
\begin{align}
\label{e1d1}
\frac{dE_j^{(l)}}{dt}=-&\frac{1}{a_l}[\eta(\tilde E_{j+\frac{1}{2}}\varphi_l(x_{j+\frac{1}{2}}^-)-\tilde E_{j-\frac{1}{2}}\varphi_l(x_{j-\frac{1}{2}}^+))+(1-\eta)v_0(\Bar E_{j+\frac{1}{2}}\varphi_l(x_{j+\frac{1}{2}}^-)-\Bar E_{j-\frac{1}{2}}\varphi_l(x_{j-\frac{1}{2}}^+))-\notag\\
&v_0\int_{I_j}E(\varphi_l(x))_x dx], \; l=1,\cdots, L.
\end{align}

Inserting the equation of state to (\ref{e1d1}), we can observe that the pressure keeps constant at the next time if the following condition is satisfied,
\begin{align}
\label{gm1d1}
\frac{d\kappa_j^{(l)}}{dt}=-&\frac{1}{a_l}[\eta(\tilde \kappa_{j+\frac{1}{2}}\varphi_l(x_{j+\frac{1}{2}}^-)-\tilde \kappa_{j-\frac{1}{2}}\varphi_l(x_{j-\frac{1}{2}}^+))+(1-\eta)v_0(\Bar\kappa_{j+\frac{1}{2}}\varphi_l(x_{j+\frac{1}{2}}^-)-\Bar\kappa_{j-\frac{1}{2}}\varphi_l(x_{j-\frac{1}{2}}^+))-\notag\\
&v_0\int_{I_j}\kappa(\varphi_l(x))_x dx], \; l=1,\cdots, L,\\
\label{gp1d1}
\frac{d\chi_j^{(l)}}{dt}=-&\frac{1}{a_l}[\eta(\tilde \chi_{j+\frac{1}{2}}\varphi_l(x_{j+\frac{1}{2}}^-)-\tilde \chi_{j-\frac{1}{2}}\varphi_l(x_{j-\frac{1}{2}}^+))+(1-\eta)v_0(\Bar\chi_{j+\frac{1}{2}}\varphi_l(x_{j+\frac{1}{2}}^-)-\Bar\chi_{j-\frac{1}{2}}\varphi_l(x_{j-\frac{1}{2}}^+))-\notag\\
&v_0\int_{I_j}\chi(\varphi_l(x))_x dx], \; l=1,\cdots, L,
\end{align}
where $\kappa=\frac{1}{\gamma-1}$ and $\chi=\frac{\gamma \pi}{\gamma-1}$. Following the work in \cite{shyue2001,chen2011,liu2017}, if the condition is replaced by
\begin{align}
\label{y1d1}
\frac{dY_j^{(l)}}{dt}=-&\frac{1}{a_l}[\eta(\tilde Y_{j+\frac{1}{2}}\varphi_l(x_{j+\frac{1}{2}}^-)-\tilde Y_{j-\frac{1}{2}}\varphi_l(x_{j-\frac{1}{2}}^+))+(1-\eta)v_0(\Bar Y_{j+\frac{1}{2}}\varphi_l(x_{j+\frac{1}{2}}^-)-\Bar Y_{j-\frac{1}{2}}\varphi_l(x_{j-\frac{1}{2}}^+))-\notag\\
&v_0\int_{I_j}Y(\varphi_l(x))_x dx], \; l=1,\cdots, L,
\end{align}
the conditions (\ref{gm1d1})-(\ref{gp1d1}) will be satisfied. It is easy to observe that the conditions (\ref{y1d1}) can be viewed as the discretization of the equation
\[Y_t+(vY)_x=0.
\]
However, the equation is different from the advection equation in (\ref{eq1}). If discretizing the species equation as (\ref{y1d1}), we will get the wrong solutions. Noting that the species equation can be rewritten as
\[
Y_t+vY_x=Y_t+(vY)_x-Yv_x=0,
\]
and discretized as follows,
\begin{align}
\label{y1d2}
\frac{dY_j^{(l)}}{dt}=-&\frac{1}{a_l}[\eta(\tilde Y_{j+\frac{1}{2}}\varphi_l(x_{j+\frac{1}{2}}^-)-\tilde Y_{j-\frac{1}{2}}\varphi_l(x_{j-\frac{1}{2}}^+))+(1-\eta)v_0(\Bar Y_{j+\frac{1}{2}}\varphi_l(x_{j+\frac{1}{2}}^-)-\Bar Y_{j-\frac{1}{2}}\varphi_l(x_{j-\frac{1}{2}}^+))-\notag\\
&v_0\int_{I_j}Y(\varphi_l(x))_x dx-Y(x_j)(\Hat v_{j+\frac{1}{2}}\varphi_l(x_{j+\frac{1}{2}}^-)-\Hat v_{j-\frac{1}{2}}\varphi_l(x_{j-\frac{1}{2}}^+)-\int_{I_j}v(\varphi_l(x))_xdx)], \notag\\&l=1,\cdots, L,
\end{align}
where $\Hat v_{j+\frac{1}{2}}=<u^1>_{j+\frac{1}{2},L}+<u^1>_{j+\frac{1}{2},R}$. Then (\ref{y1d2}) is degenerated to (\ref{y1d1}) near the contact discontinuities. Thus, the velocity and the pressure are oscillation-free.

Finally the semi-discrete schemes (\ref{vv}) and \eqref{y1d2} are discretized in time. Here, we use
an explicit, the third order TVD Runge-Kutta scheme \cite{EH21}. Casting (\ref{vv}) and \eqref{y1d2} in the form
\[
\frac{\partial u_h}{\partial t} =L_h(u_h,t),
\]
the scheme reads as
\begin{align}
 \label{tt}
&u_h^*=u_h^n+\Delta t_n L_h(u_h^n,t_n), \notag\\
&u_h^{**}=\frac{3}{4}u_h^n+\frac{1}{4}(u_h^*+\Delta t_n L_h(u_h^*,t_n+\Delta t_n)),  \\
&u_h^{n+1}=\frac{1}{3}u_h^n+\frac{2}{3}(u_h^{**}+\Delta t_n L_h(u_h^{**},t_n+\frac{1}{2}\Delta t_n )). \notag
\end{align}

\subsection{The limiting procedure}
\label{limitandtran}
In this subsection, two kinds of limiters are described briefly. One is the limiter to keep oscillation-free at discontinuities and the other one is the maximum-pricinple-satisfying limiter for the volume fraction, since it should satisfy $Y\in [0,1].$
\subsubsection{The limiter to control oscillations} 
As is well known, nonlinear limiters must be applied to control the spurious oscillations in the numerical solution for strong shocks, which has two steps following the work of Qiu and Shu \cite{EH13}. 

Step 1. We identify the ``troubled cells" using the minmod-type TVB limiter as in \cite{EH10, EH13, luo2019}. All the primitive variables are taken as the indicator variables. 

Step 2. We add the nonlinear limiters in the troubled cells. In this paper, the new type of multi-resolution WENO limiters developed in \cite{zhu2020,zhu2020a} is adopted. In order to keep the pressure non-oscillatory, we limit the primitive variables component-wisely here.

Next, we describe the method to detect ``troubled cells" and the new multi-resolution WENO limiter briefly.

For simplicity, we assume $u(x)\in V_h^k$ is the primitive variable such as $\rho_1, \rho_2,v,P,Y$ on $I_j$.
Denote 
\[
u^{-}_{j+\frac{1}{2}}=u_j^{(1)}+\tilde{u}_j,\; u^{+}_{j-\frac{1}{2}}=u_j^{(1)}-\tilde{\tilde{u}}_j, \; \Bar u_j=\frac{1}{\Delta x}\int_{I_j}udx.
\]
These values are modified by the standard minmod limiter
\[
\tilde{u}_j^{mod}=\tilde{m}(\tilde{u}_j,\Delta_+\Bar u_j,\Delta_-\Bar u_j),\; \tilde{\tilde{u}}_j^{mod}=\tilde{m}(\tilde{\tilde{u}}_j,\Delta_+\Bar u_j,\Delta_-\Bar u_j),
\]
where 
$\Delta_+\Bar u_j=\Bar u_{j+1}-\Bar u_j$, $\Delta_-\Bar u_j=\Bar u_{j}-\Bar u_{j-1}$, 
and $\tilde m$ is defined as
\begin{equation}
\widetilde{m}(a_1,a_2,a_3)=
\left
\{
\begin{array}{ll}
a_1,&\; \hbox{if}\quad |a_1|\leq  M\Delta x^2, \\
m(a_1,a_2,a_3),&\; \hbox{otherwise},
\end{array}
\right.
\end{equation}
\begin{equation}
m(a_1,a_2,a_3)=
\left
\{
\begin{array}{ll}
\hbox{sign}(a_1) \min(|a_1|,|a_2|,|a_3|),&\; \hbox{if}\;\; \hbox{sign}(a_1)=\hbox{sign}(a_2)=\hbox{sign}(a_3),\\
0,&\; \hbox{otherwise},
\end{array}
\right.
\end{equation}
and $M> 0$ is a constant. The choice of $M$ depends on the solution of the problem; see, e.g., \cite{cockburn1989} for detailed discussion. We use $M=1$ in our computation. Finally, $I_j$ is marked as a troubled cell for further reconstructions if one of the minmod functions does not return the first argument.

{
 In order to keep the velocity and pressure non-oscillatory, we limit the primitive variables component-wisely here. Then the limited primitive variables are used to compute the numerical fluxes and evolve the equation to obtain the new solutions at the next time in the conservative form. 
Assume $I_j$ is a troubled cell. 
The procedure of the limiting for the scalar case \cite{zhu2020} is given in the following.}

{\bf Step 1.} Define a series of polynomials of different degrees on the troubled cell $I_j$.

{\bf Step 1.1.} For a second-order spatial approximation, a zeroth degree polynomial $q_1(x)$ and a linear polynomial $q_2(x)$ are constructed, which satisfy 
\[
\int_{I_j}q_1(x)\varphi_1(x)dx=\int_{I_j}u(x)\varphi_1(x)dx,
\]
and 
\[
\int_{I_j}q_2(x)\varphi_l(x)dx=\int_{I_j}u(x)\varphi_l(x)dx,\; l=1,2.
\]

{\bf Step 1.2.} For a third-order spatial approximation, a quadratic polynomial $q_3(x)$ is constructed which satisfies
\[
\int_{I_j}q_3(x)\varphi_l(x)dx=\int_{I_j}u(x)\varphi_l(x)dx,\; l=1, 2, 3.
\]

{\bf Step 2.} Get equivalent expressions for these constructed polynomials of different degrees.

{\bf Step 2.1.} For the second-order approximation, we obtain a polynomial $p_{1,1}(x)$ by
\[
p_{1,1}(x)=\frac{1}{\gamma_{1,1}}q_2(x)-\frac{\gamma_{0,1}}{\gamma_{1,1}}p_{0,1}(x)
\]
with $\gamma_{0,1}+\gamma_{1,1}=1$ and $\gamma_{1,1}\neq 0,$ where $p_{0,1}(x)=q_1(x).$

{\bf Step 2.2.} For the third-order approximation, we define $p_{1,2}(x)=\omega_{1,1}p_{1,1}(x)+\omega_{0,1}p_{0,1}(x),$ and obtain a polynomial $p_{2,2}(x)$ through
\[
p_{2,2}(x)=\frac{1}{\gamma_{2,2}}q_3(x)-\frac{\gamma_{1,2}}{\gamma_{2,2}}p_{1,2}(x)
\]
with $\gamma_{1,2}+\gamma_{2,2}=1$ and $\gamma_{2,2}\neq 0.$

In these expressions, $\gamma_{l,l_2}$ and $\omega_{l,l_2}$ for $l=l_2-1,l_2; l_2=1,\cdots, k$ are the linear weights and nonlinear weights, respectively.

{\bf Step 3.} Compute the smoothness indicators $\beta_{l,l_2}$ by
\begin{equation}
\label{beta01}
\beta_{l,l_2}=\sum\limits_{s=1}^l\int_{I_j}\Delta x^{2s-1}(\frac{d^s}{dx^s}p_{l,l_2}(x))^2dx,\; l=l_2-1,l_2; l_2=1,2.
\end{equation}
However, $\beta_{0,1}$ can not be computed by \eqref{beta01}, which is defined below. We first define the linear polynomial $q_{j-1}(x)$ on $I_{j-1}$ by
\[
\int_{I_{j-1}}q_{j-1}(x)\varphi_l(x)dx=\int_{I_{j-1}}u(x)\varphi_l(x)dx,\; l=1,2.
\]
and similarly, the linear polynomial $q_{j+1}(x)$ on $I_{j+1}$ by
\[
\int_{I_{j+1}}q_{j+1}(x)\varphi_l(x)dx=\int_{I_{j+1}}u(x)\varphi_l(x)dx,\; l=1,2.
\]

Then, the smoothness indicators are computed by
\[
\zeta_{j-1}=\int_{I_{j}}\Delta x(\frac{d}{dx}q_{j-1}(x))^2dx,\; \zeta_{j+1}=\int_{I_{j}}\Delta x(\frac{d}{dx}q_{j+1}(x))^2dx.
\]
Thus, $\beta_{0,1}$ is defined as $\beta_{0,1}=\min(\zeta_{j-1},\zeta_{j+1}).$

{\bf Step 4.} Compute the nonlinear weights
\[
\omega_{l_1,l_2}=\frac{\bar \omega_{l_1,l_2}}{\sum\limits_{s=1}^{l_2}\bar \omega_{s,l_2}},\; \bar \omega_{l_1,l_2}=\gamma_{l_1,l_2}(1+\frac{\tau_{l_2}}{\upsilon+\beta_{l_1,l_2}}),\; l_1=l_2-1,l_2; l_2=1,2,
\]
where $\tau_{l_2}=(\beta_{l_2,l_2}-\beta_{l_2-1,l_2})^2,\; l_2=1,2,$ and $\upsilon$ is set to be $10^{-10}$ in all the computations.

{\bf Step 5.} Finally the new constructed polynomial $u^{new}(x)$ on the cell $I_j$ is given by
\[
u^{new}(x)=\sum\limits_{l=l_2-1}^{l_2}\omega_{l,l_2}p_{l,l_2}(x),\; l_2=1,2,
\]
for the second-order, third-order, respectively.

\subsubsection{The maximum-pricinple-satisfying limiter}

After limiting described above, the volume fraction $Y$ may still have a non-valid value, such as $Y<0$ or $Y>1$ in some cells.
Therefore, a genuinely high order accurate maximum-principle-satisfying scheme \cite{H19,H16} is employed in this paper. The procedure is described briefly in the following. 

Assume the volume fraction $Y(x)$ is the polynomial defined on $I_j$ and $\Bar Y$ is the cell average on $I_j$. Then we modify $Y(x)$ such that $Y(x)\in [\epsilon,1-\epsilon]$ for all $x\in S$ where $S$ is the set of the Legendre Gauss-Lobatto quadrature points for $I_j$. For all $j$, assume $\Bar Y\in [\epsilon,1-\epsilon]$, we use the modified polynomial $\tilde Y(x)$ instead of $Y(x)$, i.e.,
\begin{align}
\tilde Y(x)=\theta(Y(x)-\Bar Y)+\Bar Y,\; \theta=\min\Big\{\Big|\frac{1-\epsilon-\Bar Y}{Y_{max}-\Bar Y}\Big|, \Big|\frac{\epsilon-\Bar Y}{Y_{min}-\Bar Y}\Big|,1 \Big\},
\end{align}
where $Y_{max}=\max\limits_{x\in S}Y(x),\; Y_{min}=\min\limits_{x\in S}Y(x).$
 It is clear that the volume fraction $\tilde Y(x)$ should be in $[\epsilon,1-\epsilon]$ after this limiting. The parameter $\epsilon$ is set to be $10^{-8}$ in this work.
 
 For two-component flows, it is easy to see that the volume fraction of the fluid 2 also stays in $[\epsilon, 1-\epsilon]$ due to $Y_2=1-Y_1$
 if the volume fraction $Y_1=Y(x)$ of the fluid 1 is limiting using the method described above. However, for more than two fluids, it is a little different. We take three-component flows for example. Assume $Y_1(x)$, $Y_2(x)$ and $Y_3(x)$ are the volume fraction of the fluid 1, 2 and 3, respectively. The limiting procedure is given as follows.\\
 { \bf Step 1.}  
 Let $Y_{12}(x)=Y_1(x)+Y_2(x)$ and use the new volume fraction $Y_{12}(x)$ to define the parameter $\theta_1$
 \[
 \theta_1=\min\Big\{\Big|\frac{1-\epsilon-\Bar Y_{12}}{Y_{12,max}-\Bar Y_{12}}\Big|, \Big|\frac{\epsilon-\Bar Y_{12}}{Y_{12,min}-\Bar Y_{12}}\Big|,1 \Big\}.
 \]
 { \bf Step 2.} Similarly, define the parameters $\theta_i$
 \[
 \theta_i=\min\Big\{\Big|\frac{1-\epsilon-\Bar Y_{i-1}}{Y_{i-1,max}-\Bar Y_{i-1}}\Big|, \Big|\frac{\epsilon-\Bar Y_{i-1}}{Y_{i-1,min}-\Bar Y_{i-1}}\Big|,1 \Big\},\; i=2,3.
 \]
 { \bf Step 3.} Finally, use the modified polynomials $\tilde Y_1(x)$ and $\tilde Y_2(x)$ instead of $ Y_1(x)$ and $Y_2(x)$, i.e.
 \begin{align}
\tilde Y_i(x)=\theta(Y_i(x)-\Bar Y_i)+\Bar Y_i,\; \theta=\min\Big\{\theta_1, \theta_2, \theta_3, 1 \Big\}, i=1,2.
\end{align}
 
 From the procedure described above, we need to compute a common coefficient $\theta$, then $\theta$ is applied to modify the volume fraction polynomials. Moreover, the procedure can be extended to more than three fluids easily.

\section{Numerical examples}
\label{sec3}
\setcounter{equation}{0}
\setcounter{figure}{0}
\setcounter{table}{0}

In this section we present numerical results obtained with the quasi-conservative DG method described
in the previous sections for a selection of one- and two-dimensional examples.
Recall that the method has been described in one dimension. Its implementation in two dimensions
is similar. The CFL number in time step selection is set to be 0.3 for $P^1$ elements, 0.15 for
$P^2$ elements. For the examples with Mie-Gr\"uneisen EOS, the material-dependent parameters are given in Table \ref{tab1}.

\subsection{One-dimensional examples}

\begin{exam}{\em
\label{exam4.1}
To assess the accuracy of the new method, we first consider a one-dimensional convection of change in volume fraction with the equation of state \eqref{sg}, which is also studied in \cite{gryngarten2013}. And the parameters are set to be
$\gamma_1=1.4,\; \gamma_2=1.9,\; B_1=1,\; B_2=0.$
The initial condition is given by
\begin{align*}
\rho(x,0)=1,\; v(x,0)=1,\; P(x,0)=1,Y(x,0)=0.5+0.499\sin(\pi x)
\end{align*}
 with a periodic boundary condition. The computational domain is on (0,2). We compute the solution up to $T = 1$. 
 The error of the volume fraction is listed in Table \ref{ex4.1}, which shows the convergence of the second order for $P^1$ elements, the third order for $P^2$ elements for the quasi-conservative DG method.

\begin{table}
\caption{Example~\ref{exam4.1}: Solution error with periodic boundary conditions and $T=1$.}
\renewcommand{\multirowsetup}{\centering}
\begin{center}
\begin{tabular}{|c|c|c|c|c|c|c|c|c|c|c|c|c|}
\hline
$k$ & $N$  &10      & 20 & 40 & 80 &160 &320\\
\hline
\multirow{6}{1cm}{1}
 & $L^1$ &1.965e-2      & 5.453e-3 & 1.184e-3 & 2.923e-4  & 7.375e-5 & 1.852e-5\\
 &  Order      & \quad  & 1.849       & 2.203      &  2.018        & 1.987 &1.994 \\
 &$L^2$   & 2.682e-2      & 6.777e-3 & 1.424e-3 &  3.247e-4  & 8.196e-5 & 2.059e-5\\
 &Order   & \quad   & 1.985     & 2.251      &   2.133      & 1.986 & 1.993 \\
 &$L_{\infty}$ & 5.202e-2 & 1.355e-2 & 3.152e-3  &  4.859e-4 & 1.182e-4 & 2.975e-5\\
 &Order    & \quad  & 1.941      & 2.104     &   2.700      & 2.039 & 1.990   \\
 \hline
 \multirow{6}{1cm}{2}

 & $L^1$  & 1.471e-3      &1.949e-4 & 2.365e-5 &  3.000e-6 & 3.599e-7 & 4.477e-8 \\
 &  Order      & \quad & 2.916    & 3.043     & 2.979    & 3.059 & 3.007  \\
 &$L^2$  & 1.750e-3       & 2.117e-4 & 2.595e-5 &  3.268e-6 & 3.987e-7 & 4.969e-8\\
 &Order       & \quad &3.047     & 3.028       & 2.989    & 3.035 & 3.004 \\
 &$L_{\infty}$ & 2.944e-3 & 2.843e-4 & 3.590e-5  & 4.858e-6 & 5.620e-7 & 7.023e-8 \\
 &Order    & \quad       &3.372    & 2.985     & 2.886   & 3.112 &3.000   \\
 \hline 
\end{tabular}
\end{center}
\label{ex4.1}
\end{table}

}\end{exam}

\begin{exam}{\em
\label{exam4.2}
To verify the non-oscillation property for the pressure and velocity fields, in this example we consider the interface only problem with the initial condition given by
\begin{equation*}
(\rho,v,P,\gamma,B)=
\begin{cases}
(1,1,1,1.4,1),  \;&x\leqslant 0,\\
(0.125,1,1,1.9,0), \;&x>0.
\end{cases}
\end{equation*}
The stiffened gas equation of state \eqref{sg} is used and the numerical results with $100$ points at $T=2$ are plotted in Fig \ref{figinterface}.

 From the figure, one can observe that the new DG method can preserve the oscillation-free property of the pressure and velocity at the material interface. Moreover, from the close-up of the density at the interface in Fig \ref{figinterface}, it is clear that the high order method have better resolution than lower order method.

 \begin{figure}[hbtp]
 \begin{center}
 \mbox{
 {\includegraphics[width=8cm]{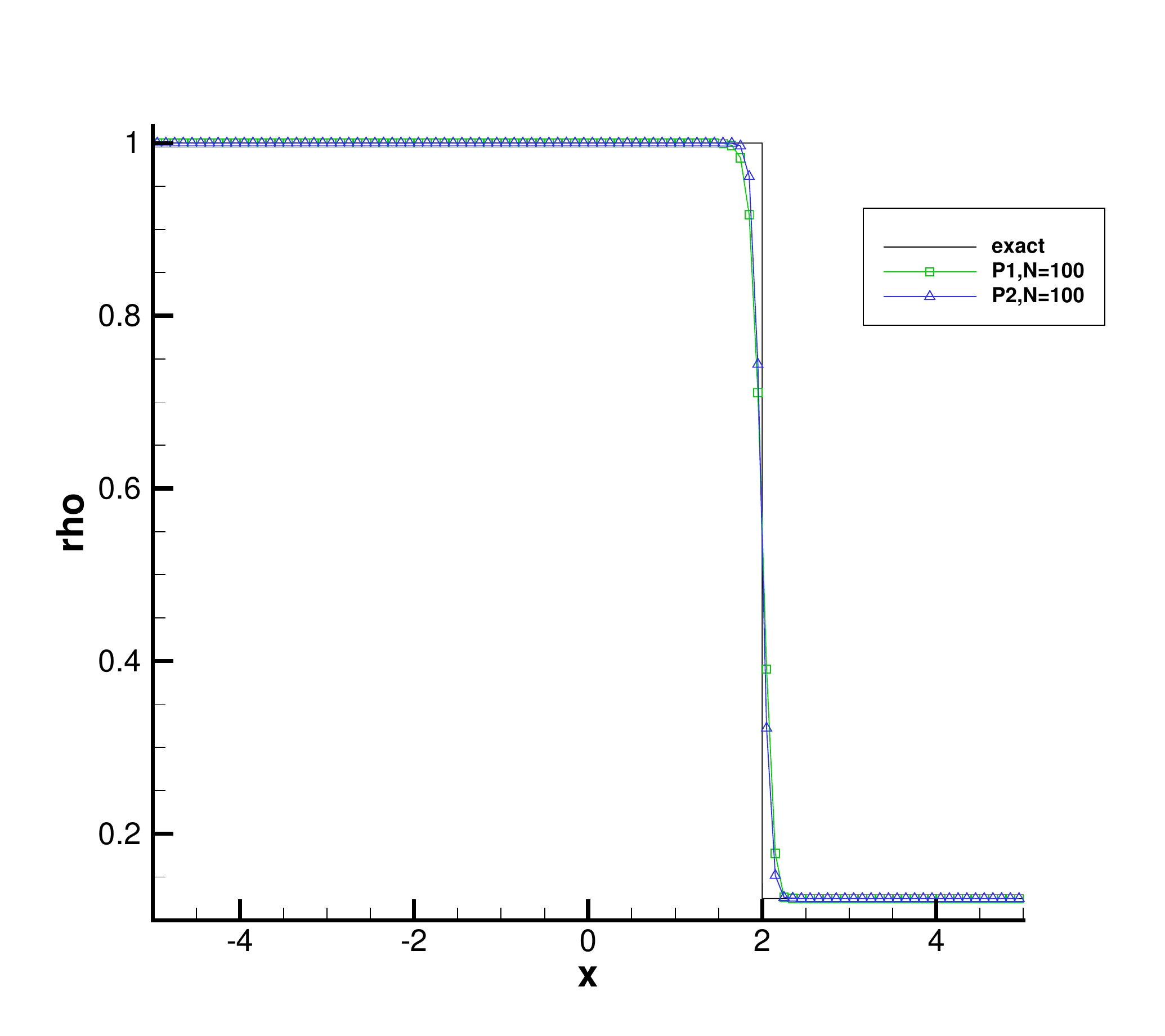}}\quad
{\includegraphics[width=8cm]{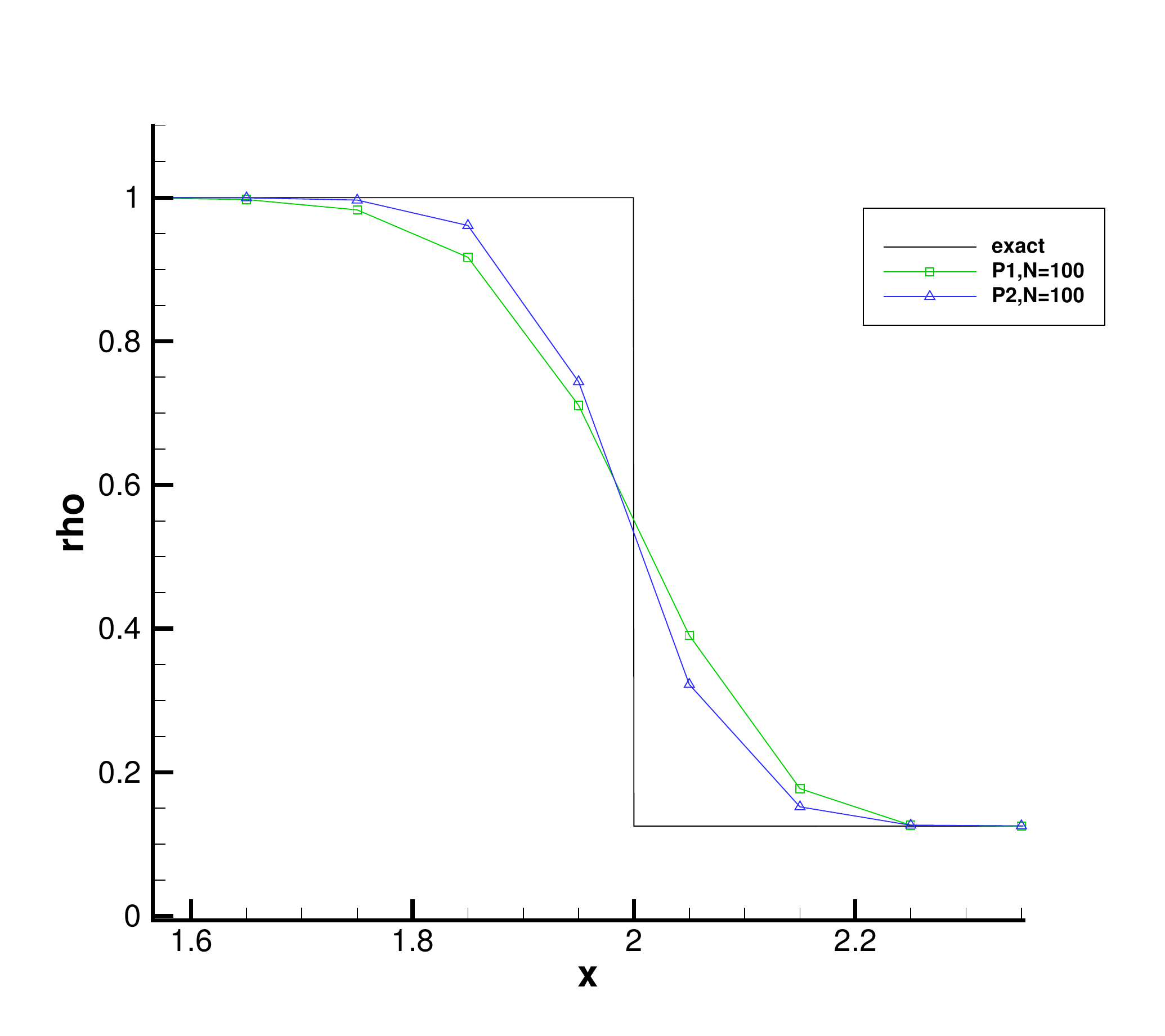}}

   }

   \mbox{
   {\includegraphics[width=8cm]{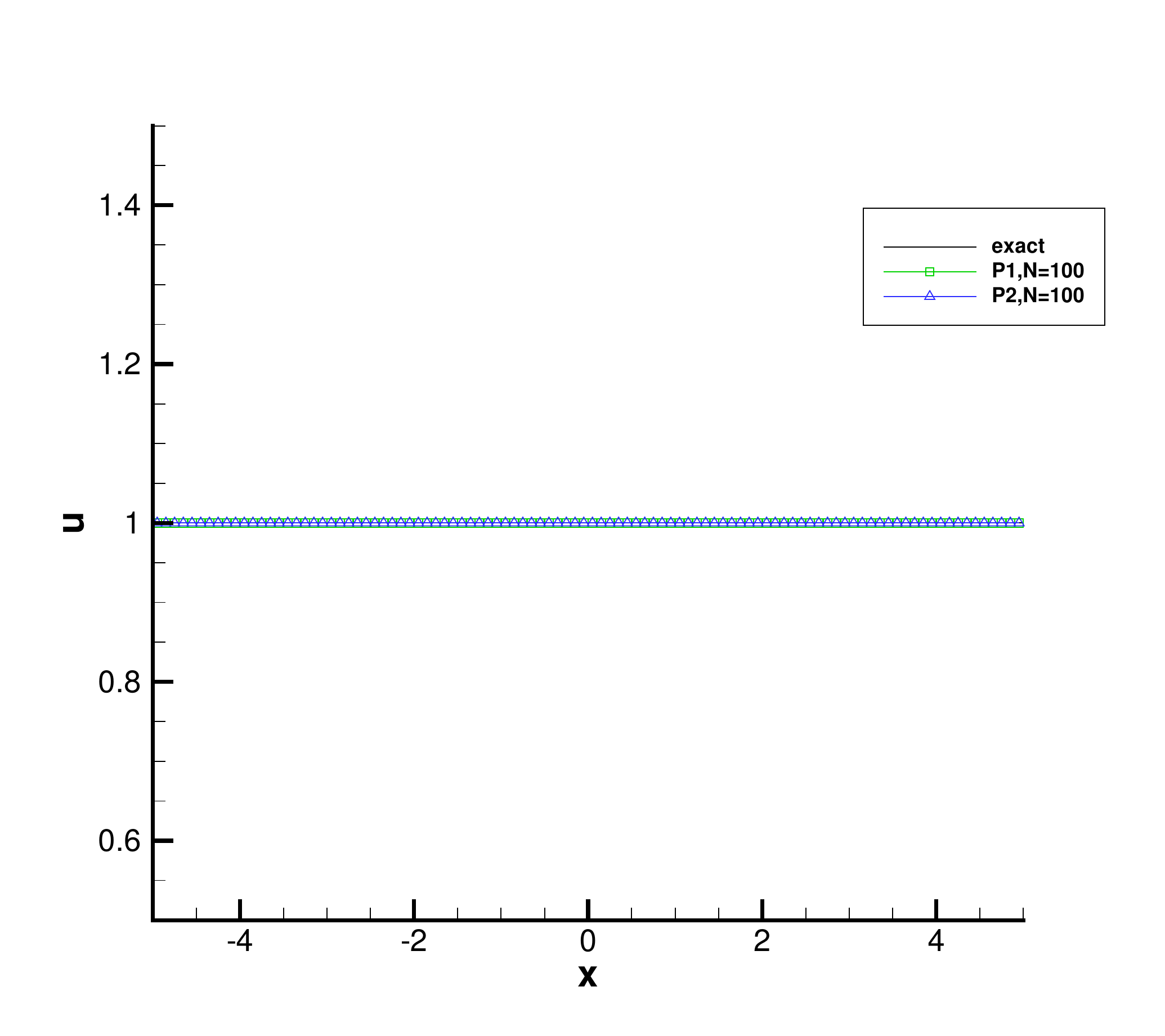}}\quad
   {\includegraphics[width=8cm]{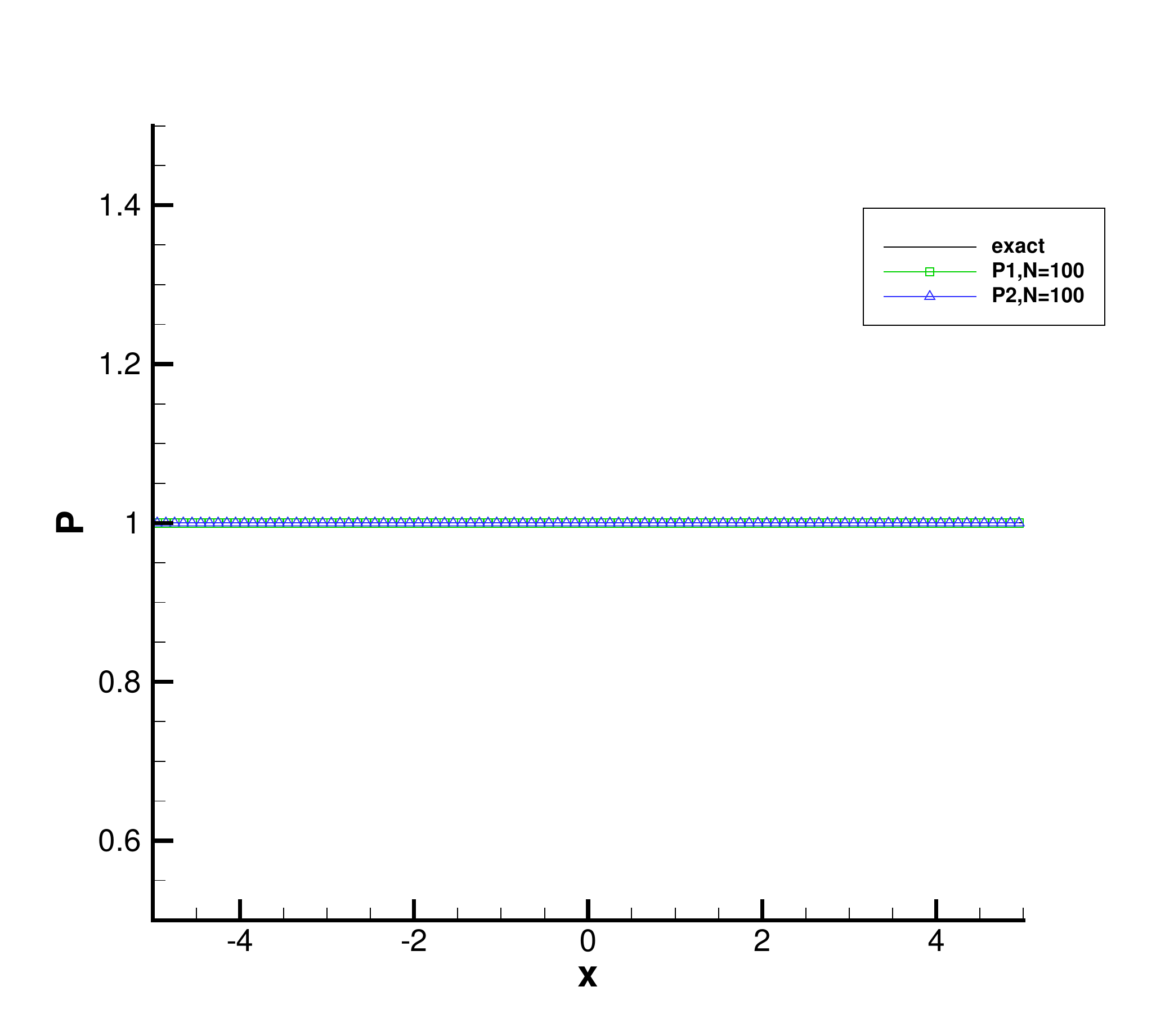}}
   }

   \caption{Example~\ref{exam4.2} interface only problem. $N=100$}
   \label{figinterface}
   \end{center}
   \end{figure}

}\end{exam}

\begin{exam}{\em
\label{exam4.6}
In this example the gas-liquid shock tube test with a strong shock wave is considered and the stiffened gas EOS \eqref{sg} is employed. This is a very challenging test case with a strong shock wave since the shock and the material interface are close and the pressure ratio is excessively high.  The initial condition is
\begin{equation*}
(\rho,v,P,\gamma,B)=
\begin{cases}
(10^3,0,10^9,4.4,6\times 10^8),  \; & x\leqslant 0.5,\\
(50,0,10^5,1.4,0), \;  &x>0.5,
\end{cases}
\end{equation*}
and the computational domain is (-0.2,1). The final time is $T=0.0002$. 

Fig. \ref{figex4.6} is computed with $5000$ points. From the close-up density profile near the shock and material interface, we can observe that the solutions with $P^2$ elements have better resolution than ones with $P^1$ elements.

\begin{figure}[hbtp]
 \begin{center}
 \mbox{
 {\includegraphics[width=8cm]{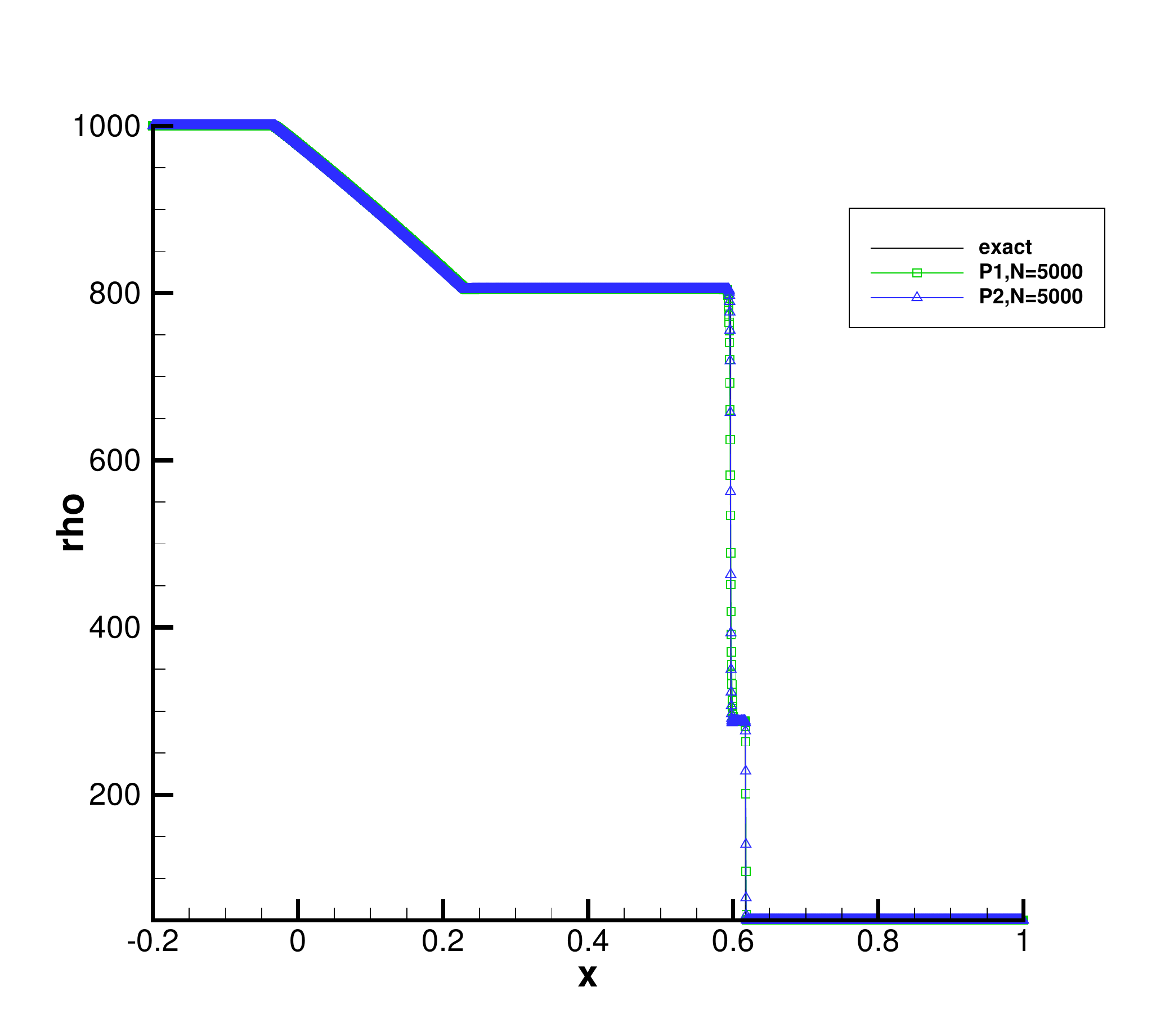}}\quad
{\includegraphics[width=8cm]{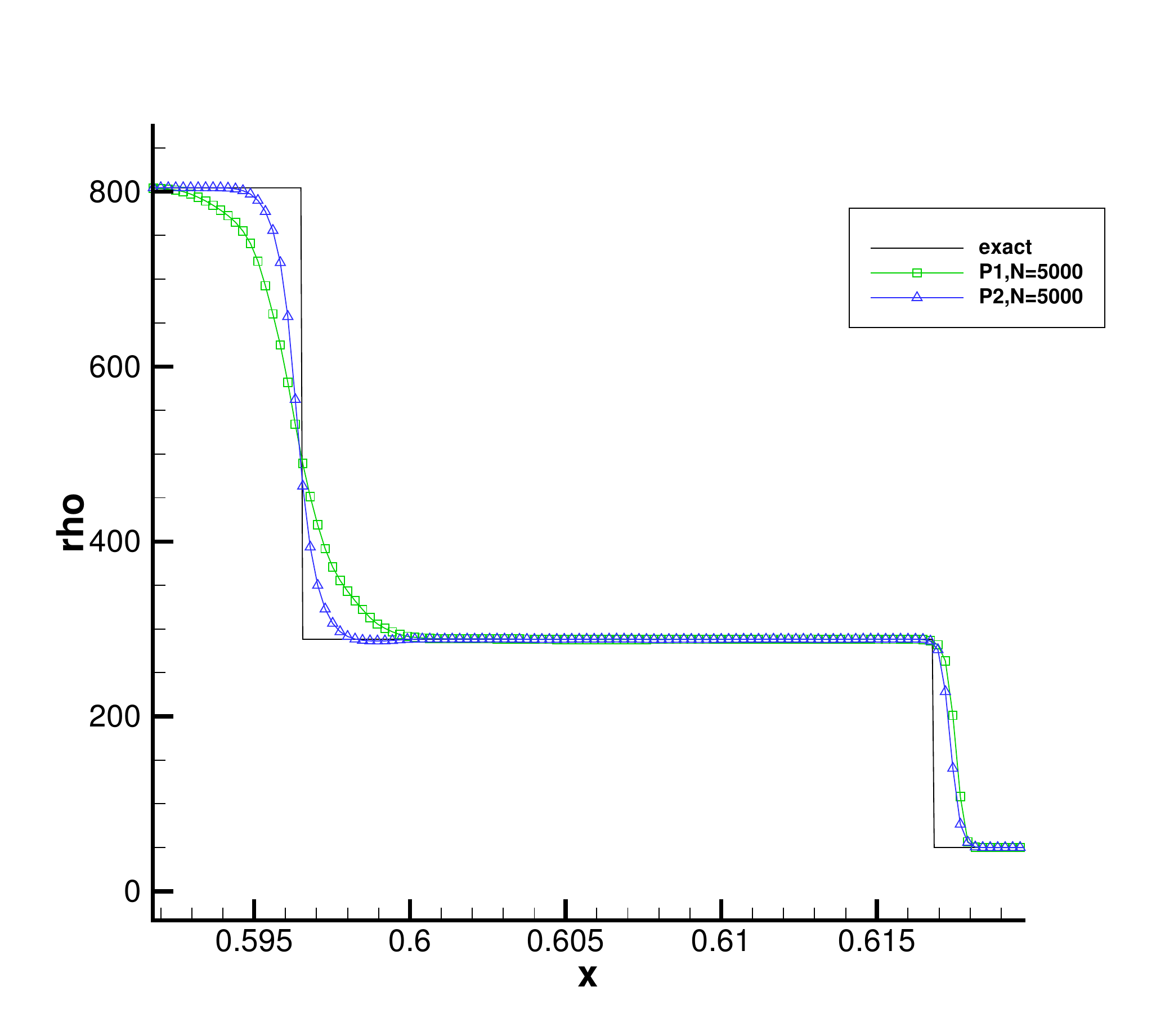}}
}

 \mbox{
 {\includegraphics[width=8cm]{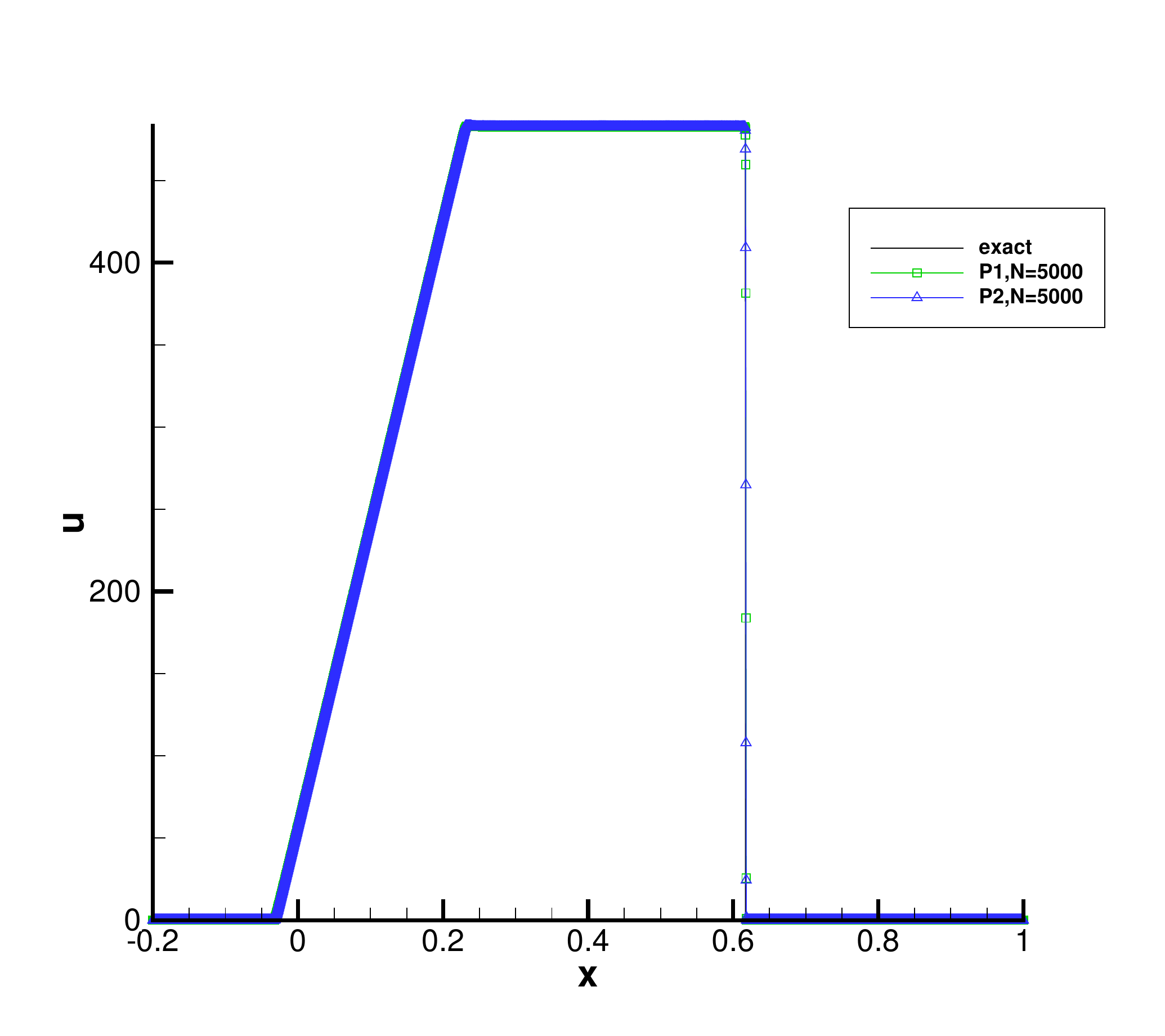}}\quad
 {\includegraphics[width=8cm]{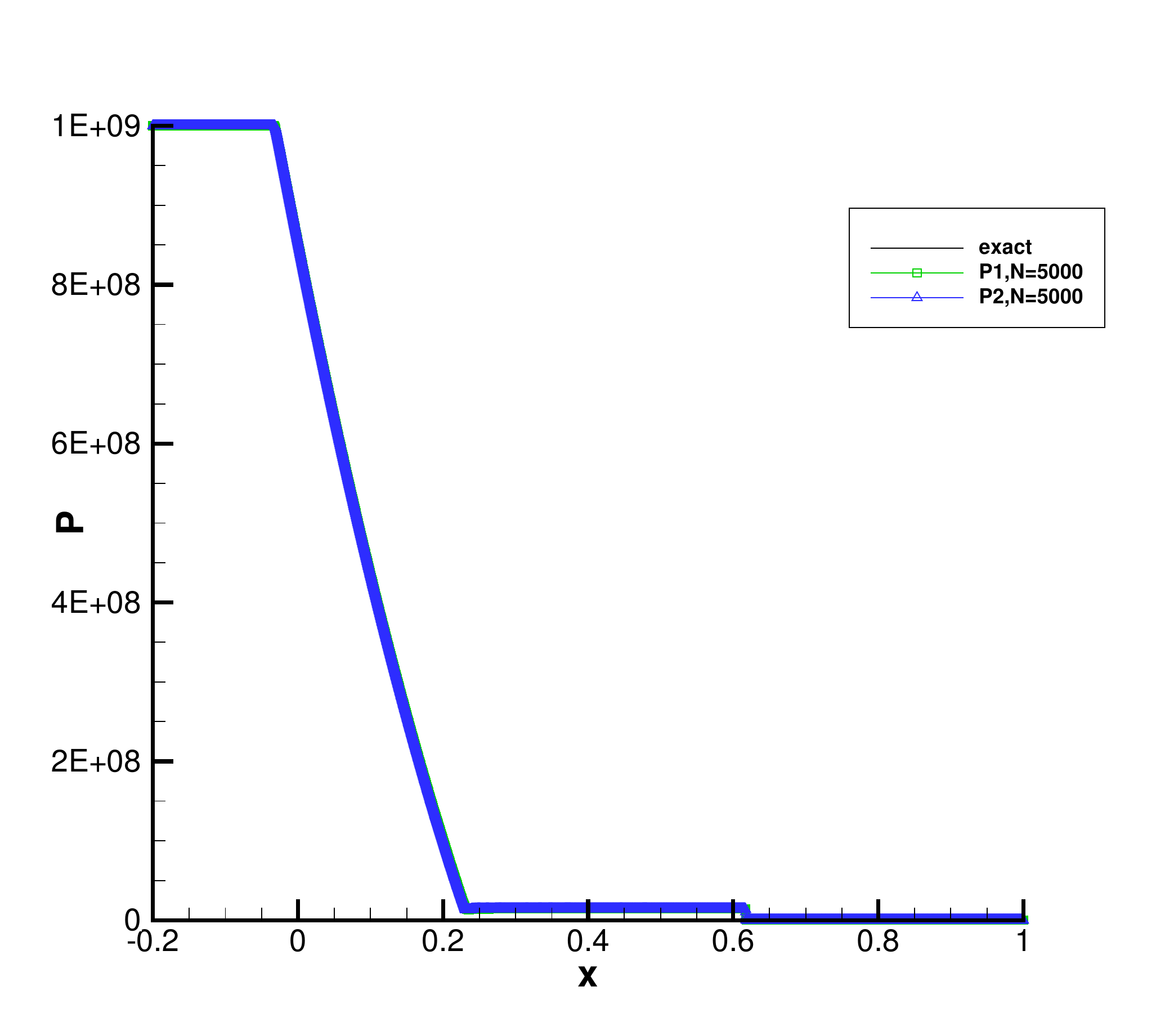}}
}

   \caption{Example~\ref{exam4.6} $N=5000$}
   \label{figex4.6}
   \end{center}
   \end{figure}

}
\end{exam}

\begin{exam}{\em
\label{copexp}
In order to show that the quasi-conservative DG method works with Mie-Gr\"uneisen EOS, we test 
a two-component impact problem, which is also studied in \cite{shyue2001,H18}. In this problem, to model the material properties of the copper and solid explosive, the same CC EOSs \eqref{cc} are used, but with a different set of material-dependent quantities for each of them. At the beginning, the copper has an initial velocity of $1500$ m/s, while the explosive is at rest.
The computational domain is $(0,1)$ and the initial condition is given by
\begin{equation*}
(\rho,v,P)=
\begin{cases}
(8900,1500,10^{5}),  \;&x\leqslant 0.5,\\
(1840,0,10^5), \;&x>0.5.
\end{cases}
\end{equation*}
The boundary conditions are constant states on both the left and right sides of the domain. The integration is stopped at $T=85 \;\mu s$. 

The exact solution for this problem consists of a rightward-moving shock, a leftward-moving shock and a material interface in between. The results with 200 uniform points are demonstrated in Fig \ref{figcopexp}, where the solid line is the fine grid solution computed by $\Delta x=\frac{1}{2000}$ with $P^1$ elements. From that one can observe these nonlinear structures are all resolved well.

\begin{figure}[hbtp]
 \begin{center}
 \mbox{
{\includegraphics[width=8cm]{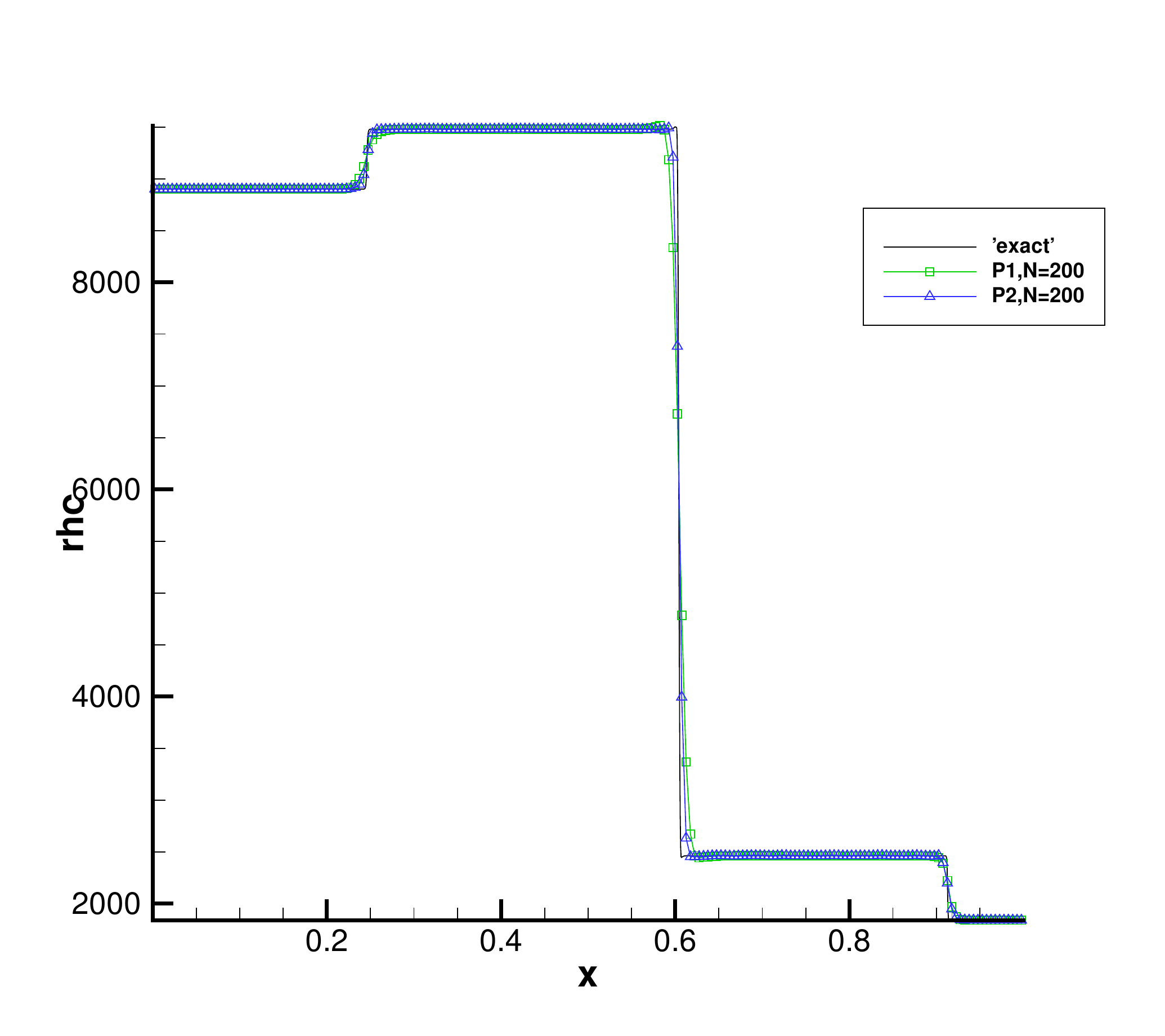}}\quad
{\includegraphics[width=8cm]{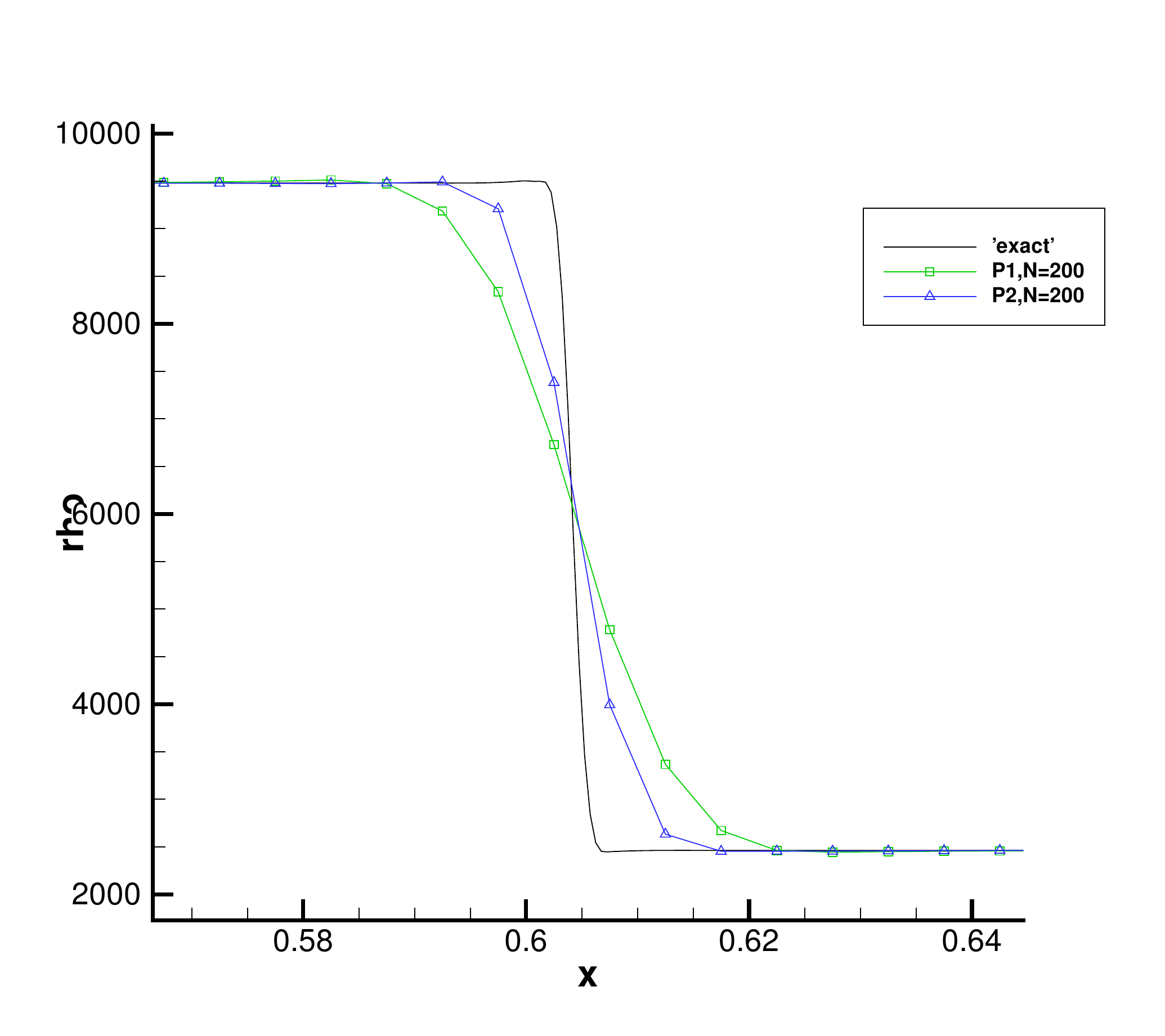}}

   }

   \mbox{
   {\includegraphics[width=8cm]{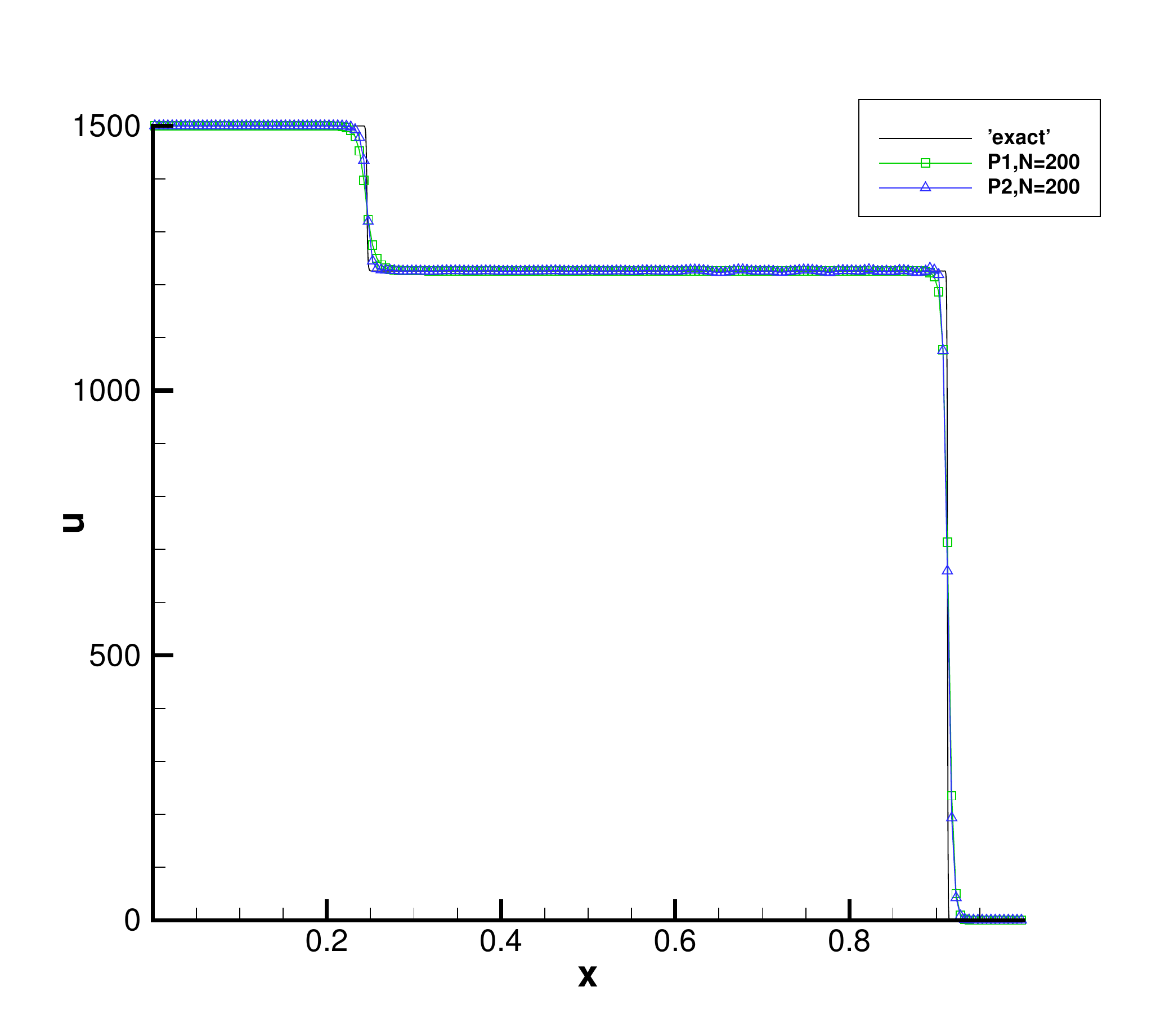}}\quad
      {\includegraphics[width=8cm]{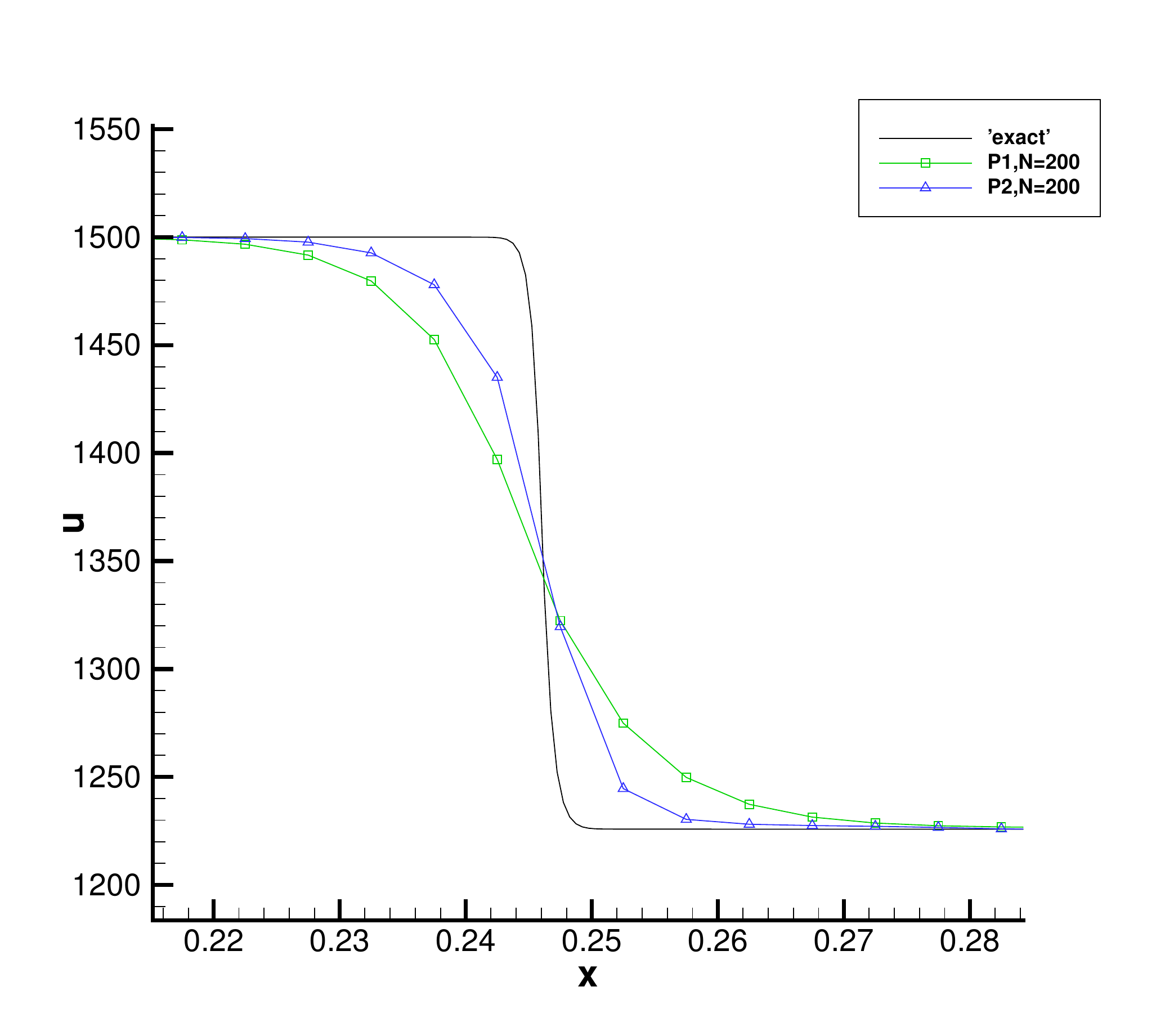}}

   }

 \mbox{
   {\includegraphics[width=8cm]{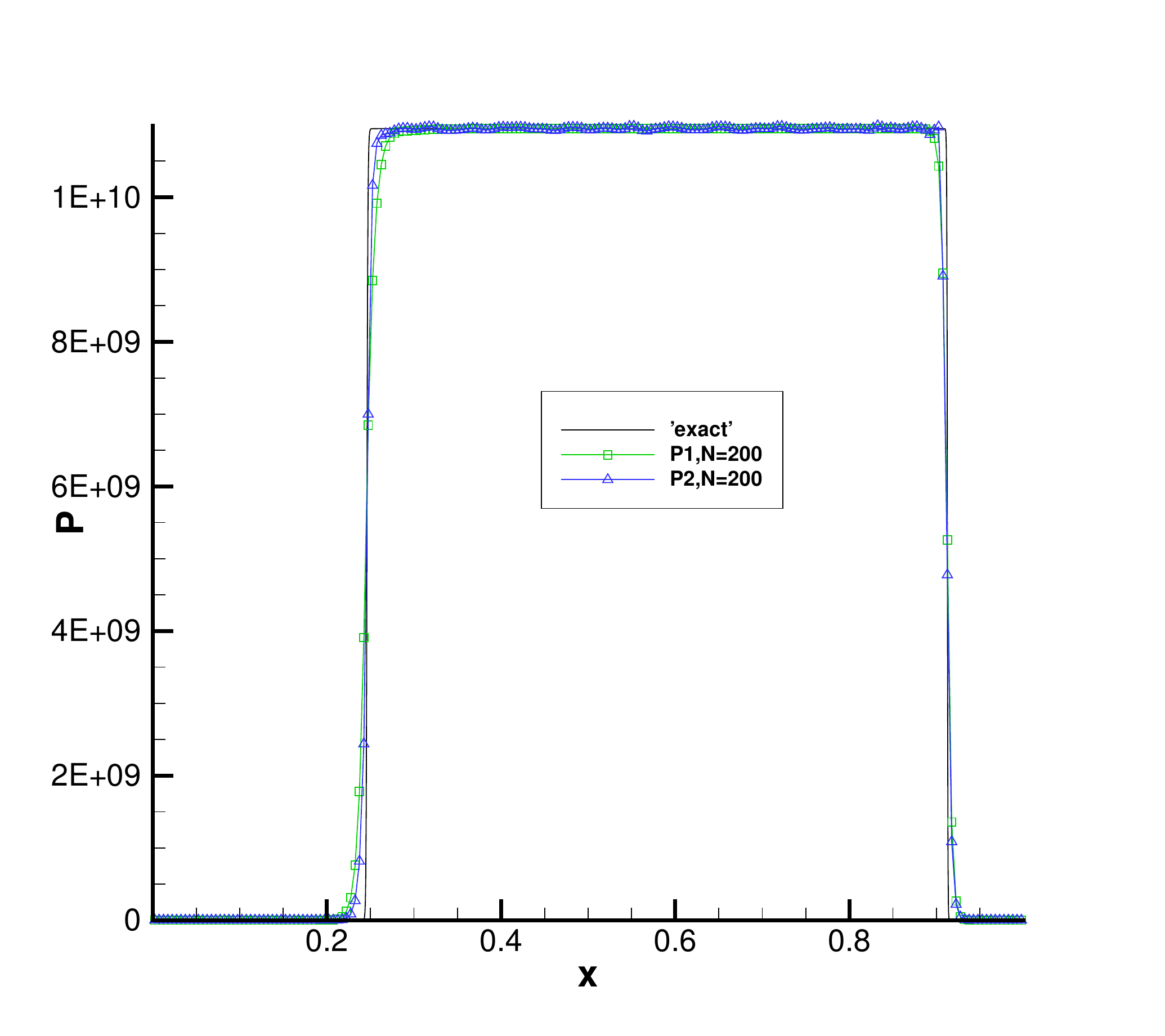}}\quad
      {\includegraphics[width=8cm]{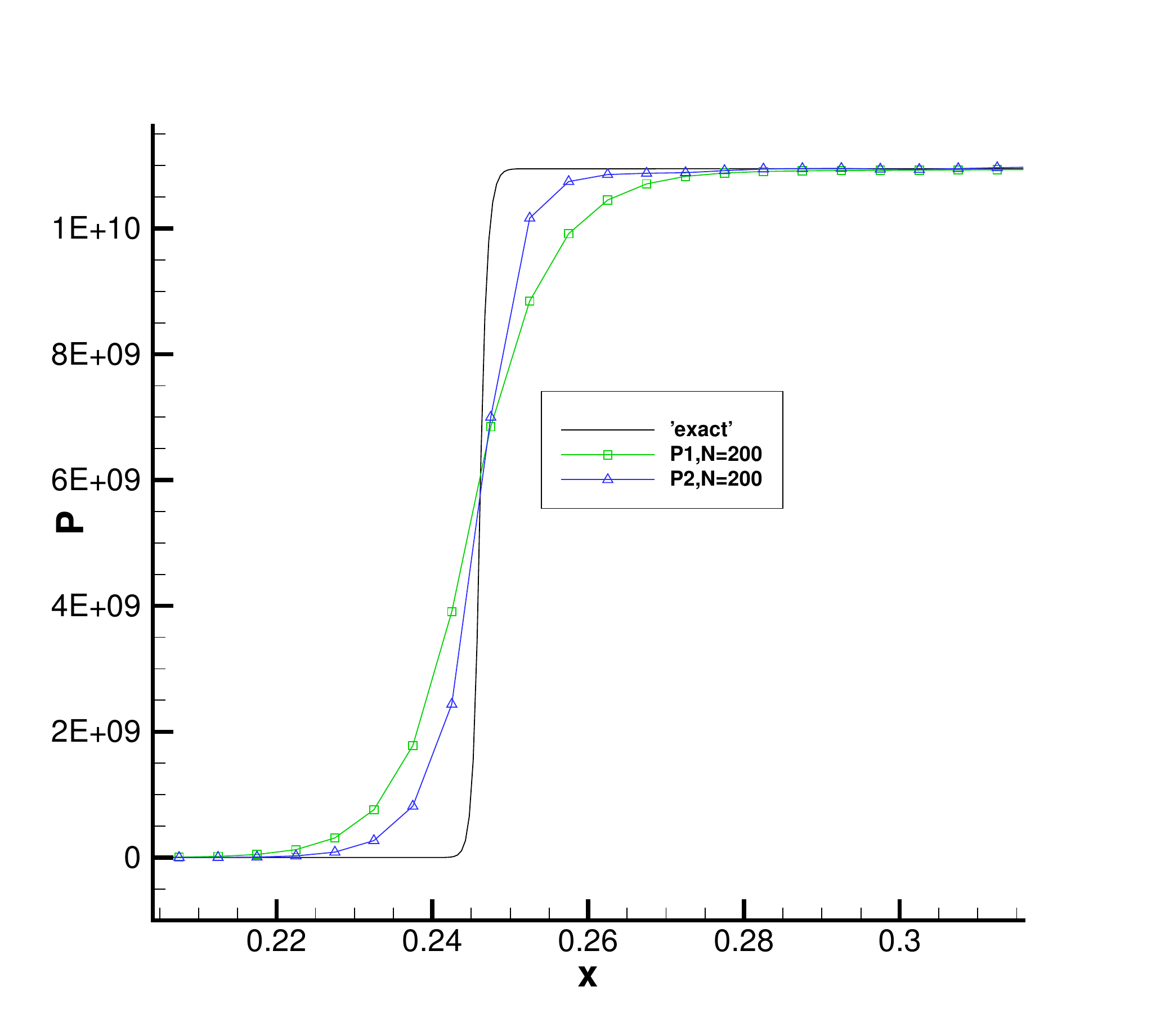}}

   }

   \caption{Example~\ref{copexp} $N=200$, Left: $P^1$ elements;  Right: $P^2$ elements}
   \label{figcopexp}
   \end{center}
   \end{figure}

}
\end{exam}

\subsection{Two-dimensional examples}

\begin{exam}{\em
\label{exam4.12d}
In order to test the accuracy in two-dimensional case, similar to Example \ref{exam4.1}, we first consider a two-dimensional convection of change in volume fraction with the equation of state \eqref{sg} and the parameters are 
$\gamma_1=1.4, \gamma_2=1.9, B_1=1, B_2=0.
$
The initial condition is given by
\begin{align*}
\rho(x,0)=1,\; \mu(x,0)=1,\nu(x,0)=1,\; P(x,0)=1,Y(x,0)=0.5+0.499\sin(\pi (x+y))
\end{align*}
 with a periodic boundary condition, where $\mu$ and $\nu$ are the velocities in the $x$-direction and $y$-direction, respectively. The computational domain is taken as $(x,y)\in(0,2)\times(0,2)$. We compute the solution up to $T = 1$. 
 The error of the volume fraction is listed in Table \ref{ex4.12d}, which shows the convergence of the second order for $P^1$ elements, the third order for $P^2$ elements for the quasi-conservative DG method in two dimensions.

\begin{table}
\caption{Example~\ref{exam4.12d}: Solution error with periodic boundary conditions and $T=1$.}
\renewcommand{\multirowsetup}{\centering}
\begin{center}
\begin{tabular}{|c|c|c|c|c|c|c|c|c|c|c|c|c|}
\hline
$k$ & $N\times M$  &$10\times 10$      & $20\times20$ & $40\times40$ & $80\times80$ &$160\times160$&$320\times320$ \\
\hline
\multirow{6}{1cm}{1}
 & $L^1$ &1.013e-1      & 1.902e-2 & 4.018e-3 & 1.047e-3  & 2.669e-4& 6.730e-5 \\
 &  Order      & \quad  & 2.413      & 2.243      &  1.940        & 1.972 & 1.988\\
 &$L^2$   & 1.011e-1      & 2.014e-2 & 4.676e-3 &  1.161e-3  & 2.949e-4 & 7.456e-5\\
 &Order   & \quad   & 2.328     & 2.107      &   2.010      & 1.977& 1.984 \\
 &$L_{\infty}$ & 1.350e-1 & 3.297e-2 & 7.824e-3  &  2.016e-3 & 4.202e-4 & 1.066e-4\\
 &Order    & \quad  & 2.034      & 2.075     &   1.956      & 2.262 & 1.979  \\
 \hline
 \multirow{6}{1cm}{2}

 & $L^1$  & 2.195e-2      &1.732e-3 & 1.977e-4 &  3.436e-5 & 3.027e-6 & 3.738e-7 \\
 &  Order      & \quad & 3.664    & 3.131     & 2.525    & 3.505  &3.018 \\
 &$L^2$  & 2.520e-2       & 1.862e-3 & 2.166e-4 &  6.261e-5 & 3.324e-6& 4.138e-7 \\
 &Order       & \quad &3.758     & 3.104       & 1.791    & 4.235& 3.006\\
 &$L_{\infty}$ & 4.294e-2 & 2.447e-3 & 2.984e-4  & 4.221e-5 & 4.666e-6  &5.832e-7\\
 &Order    & \quad       &4.133    & 3.036     & 2.822   & 3.177   &3.000 \\
 \hline 
\end{tabular}
\end{center}
\label{ex4.12d}
\end{table}

}\end{exam}

\begin{exam}{\em
\label{exam4.10}
To show the performance of our method with high pressure ratio in two dimensions, we consider the simulation of a model underwater explosion problem \cite{lee2009, shyue2006}. In this test, the computation domain is taken as $(x,y)\in (-2,2)\times (-1.5,1)$. Initially, the horizontal air-water interface is located at the $y=0$ and the center of a circular gas bubble with the radius 0.12 in the water is located at $(0,-0.3)$. Above the air-water interface, the fluid is a perfect gas at the standard atmospheric condition and below the air-water interface in region outside the gas bubble the fluid is water. Thus the initial condition is
\begin{equation*}
(\rho,\mu,\nu,P,\gamma,B)=
\begin{cases}
(1.225,0,0,101325,1.4,0),  \; & y > 0,\\
(1250,0,0,10^9,1.4,0), \;  &x^2+(y+0.3)^2\leqslant 0.12^2,\\
(1000,0,0,101325,4.4,6\times 10^8), \; &else.
\end{cases}
\end{equation*}
And the reflecting boundary conditions are employed on the bottom of the domain, while non-reflecting boundary conditions are used on the remaining sides \cite{lee2009}.

 From the initial condition, it is obvious that both the gas and water are in a stationary position at the beginning, but due to the pressure difference between the fluids, breaking of the bubble results in a circularly outward-going shock wave in water, an inward-going rarefaction wave in gas, and an interface lying in between that separates the gas and the water. Soon after, this shock wave is diffracted through the nearby air-water surface, causing the subsequent deform of the interface topology from a circle to oval-like shape.
 
The contours of the density and pressure are plotted in Figs. \ref{figbwstiff1} and \ref{figbwstiff2} at four different times $T=0.2, 0.4, 0.8$ and $1.2$ ms obtained by our method with a uniform $640\times 400$ mesh. From the density and pressure plots, one can clearly see that the improvement on the use of the high order method to the sharpness near the interfaces. The cross-sections of the density and pressure for the same run along line $x=0$ are shown in Figs. \ref{figbwstiff3} and \ref{figbwstiff4}, which give some information about the differences between $P^1$ and $P^2$ elements at the selected times.

\begin{figure}[hbtp]
 \begin{center}
 \mbox{

 {\includegraphics[width=8cm]{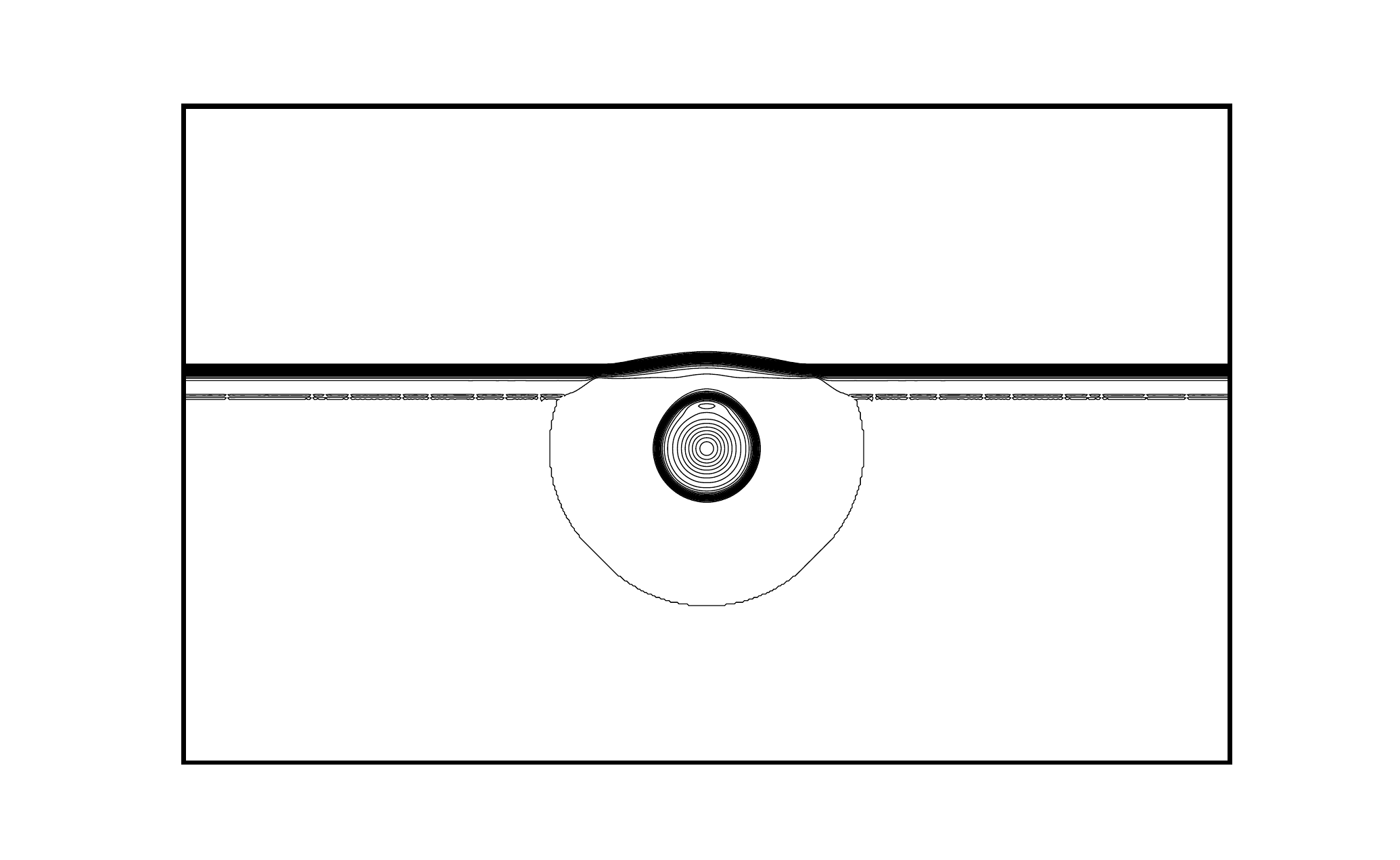}}\quad
   {\includegraphics[width=8cm]{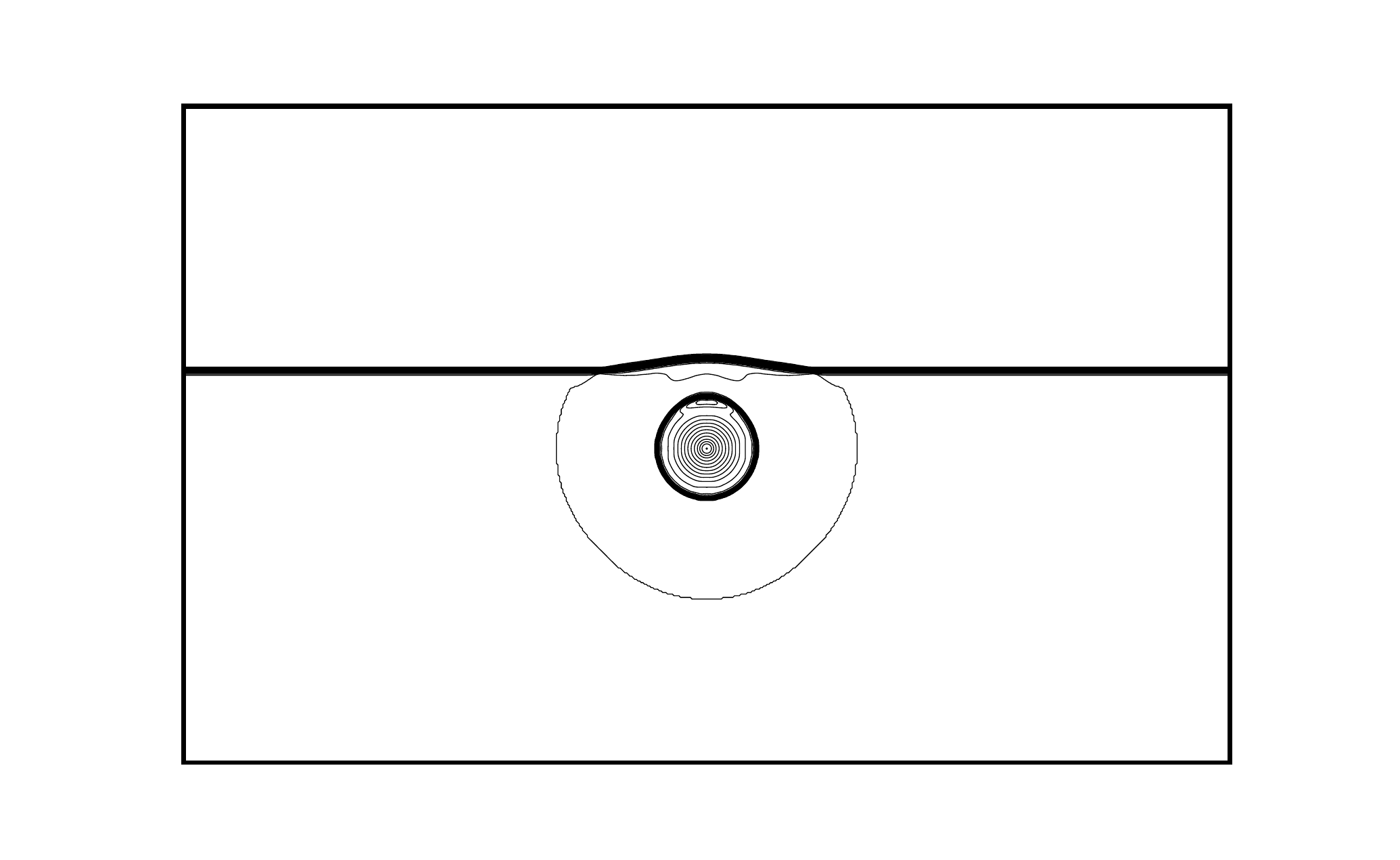}}
   }

   \mbox{

 {\includegraphics[width=8cm]{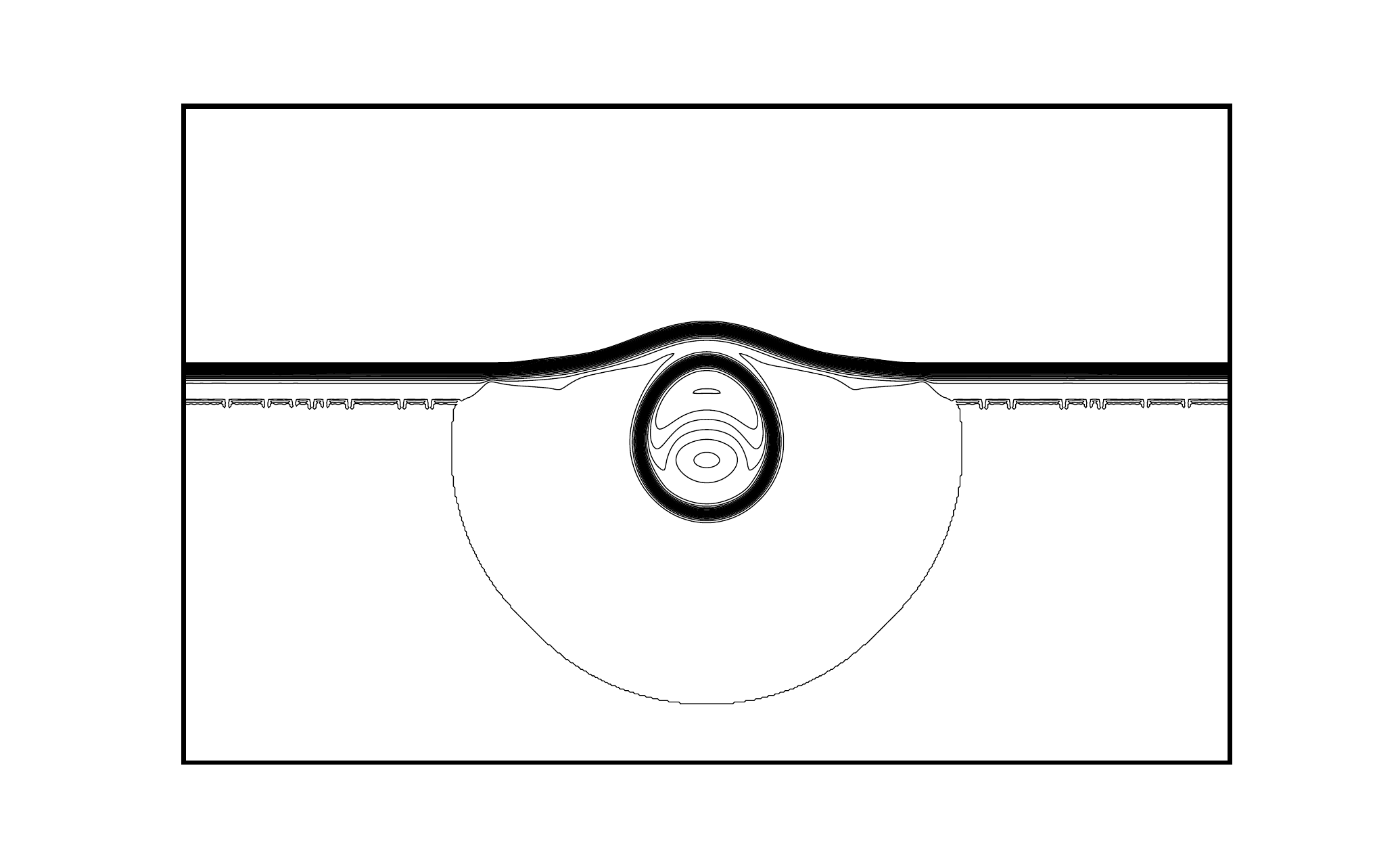}}\quad
   {\includegraphics[width=8cm]{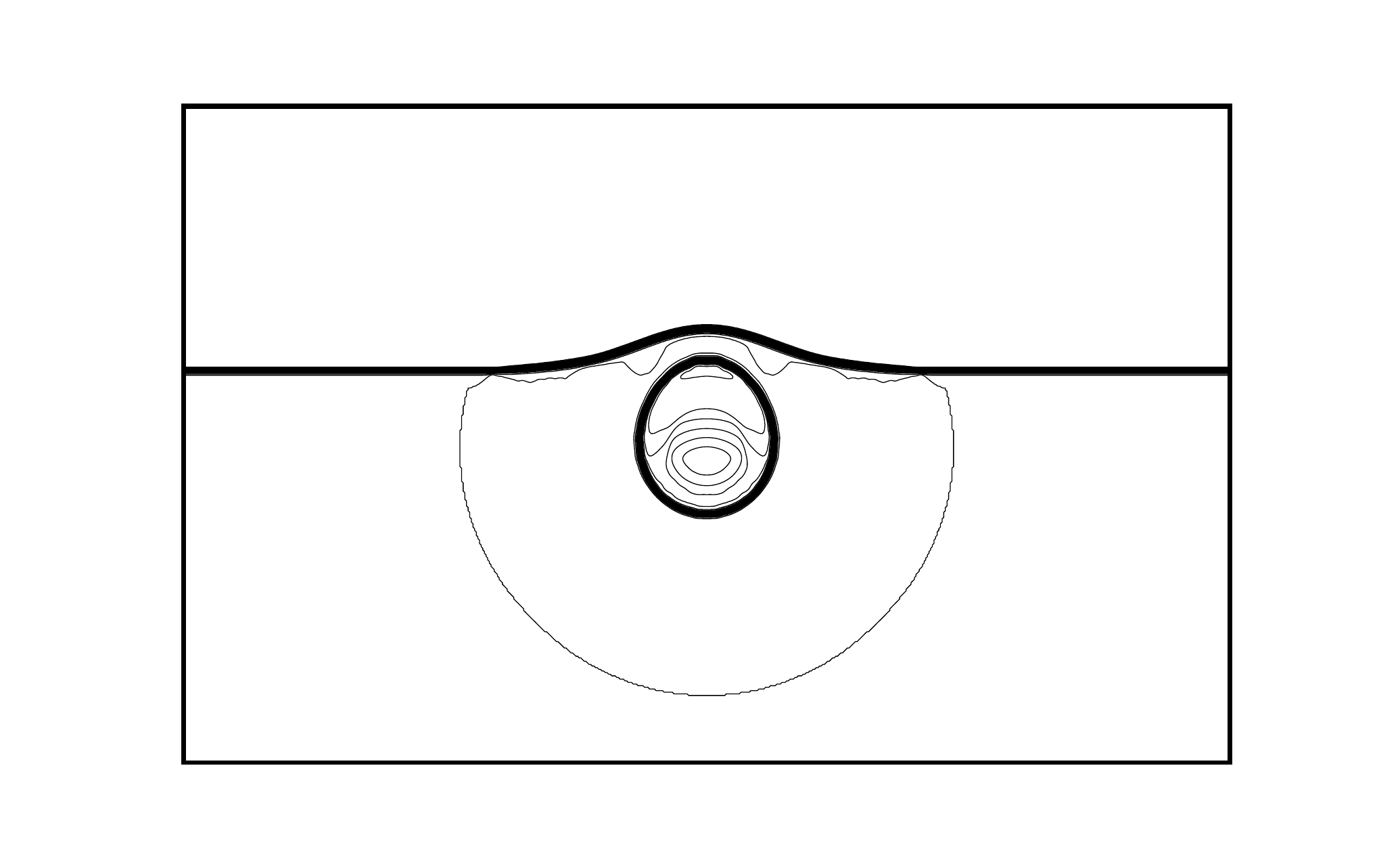}}
   }

   \mbox{

 {\includegraphics[width=8cm]{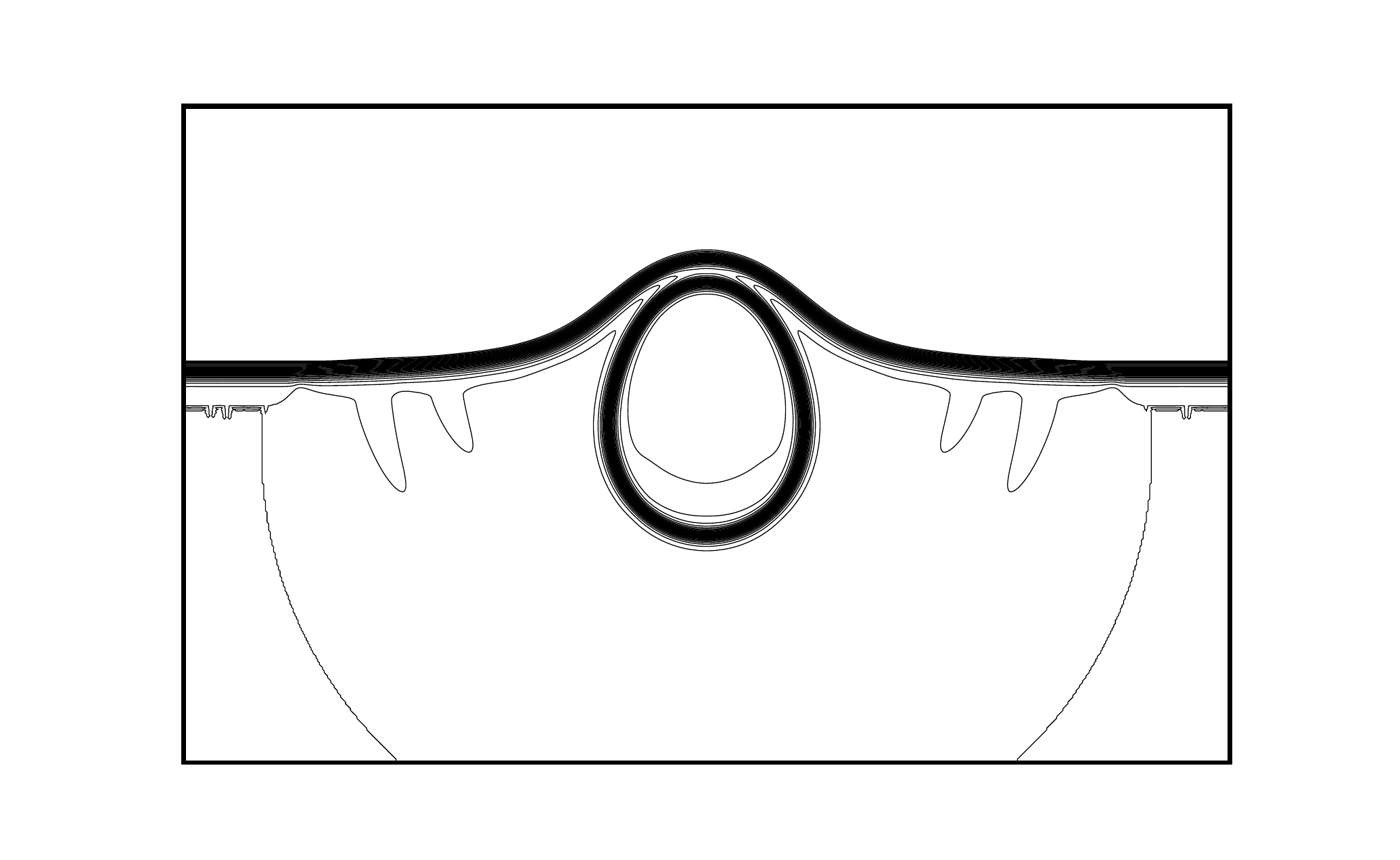}}\quad
   {\includegraphics[width=8cm]{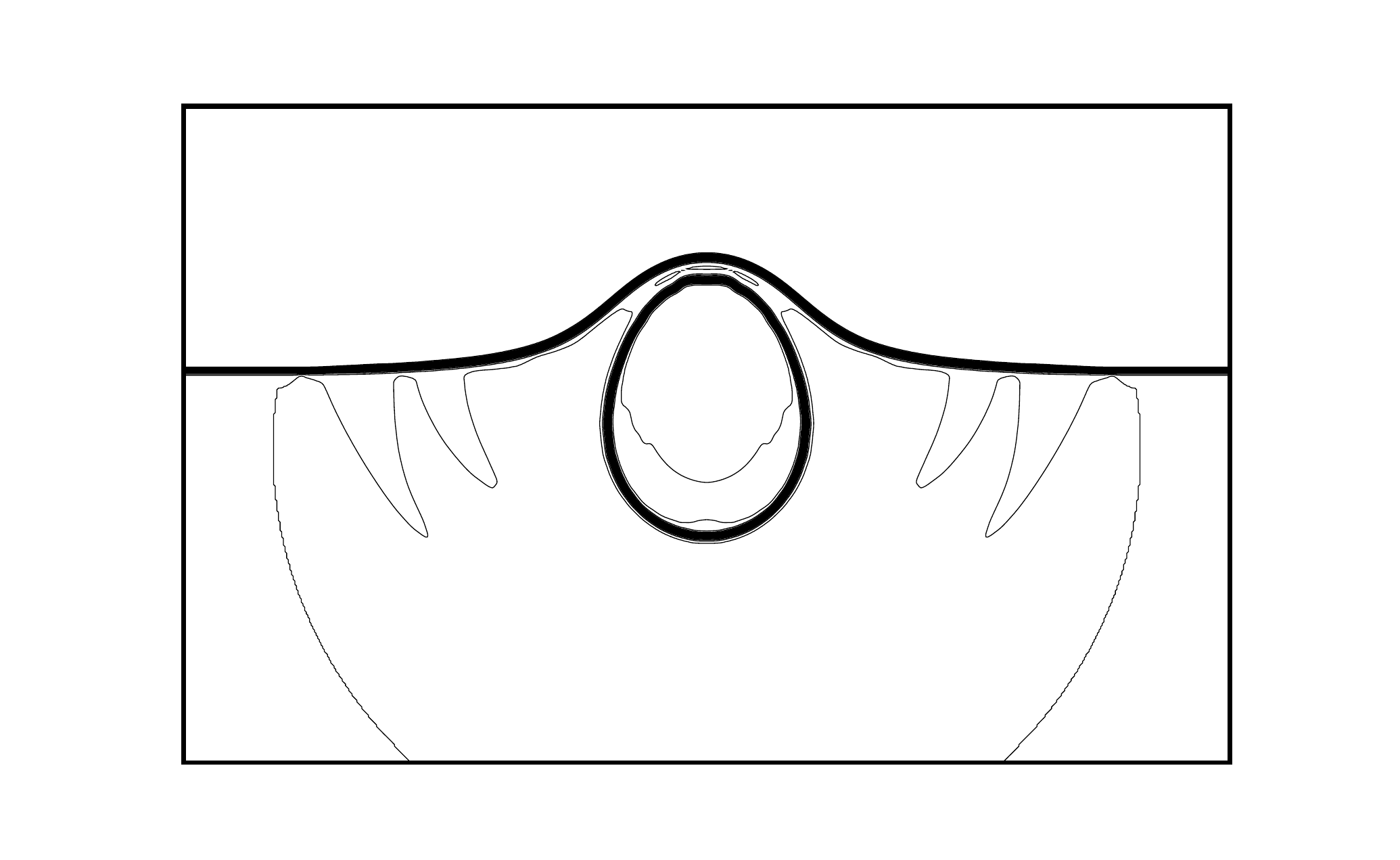}}
   }

   \mbox{

 {\includegraphics[width=8cm]{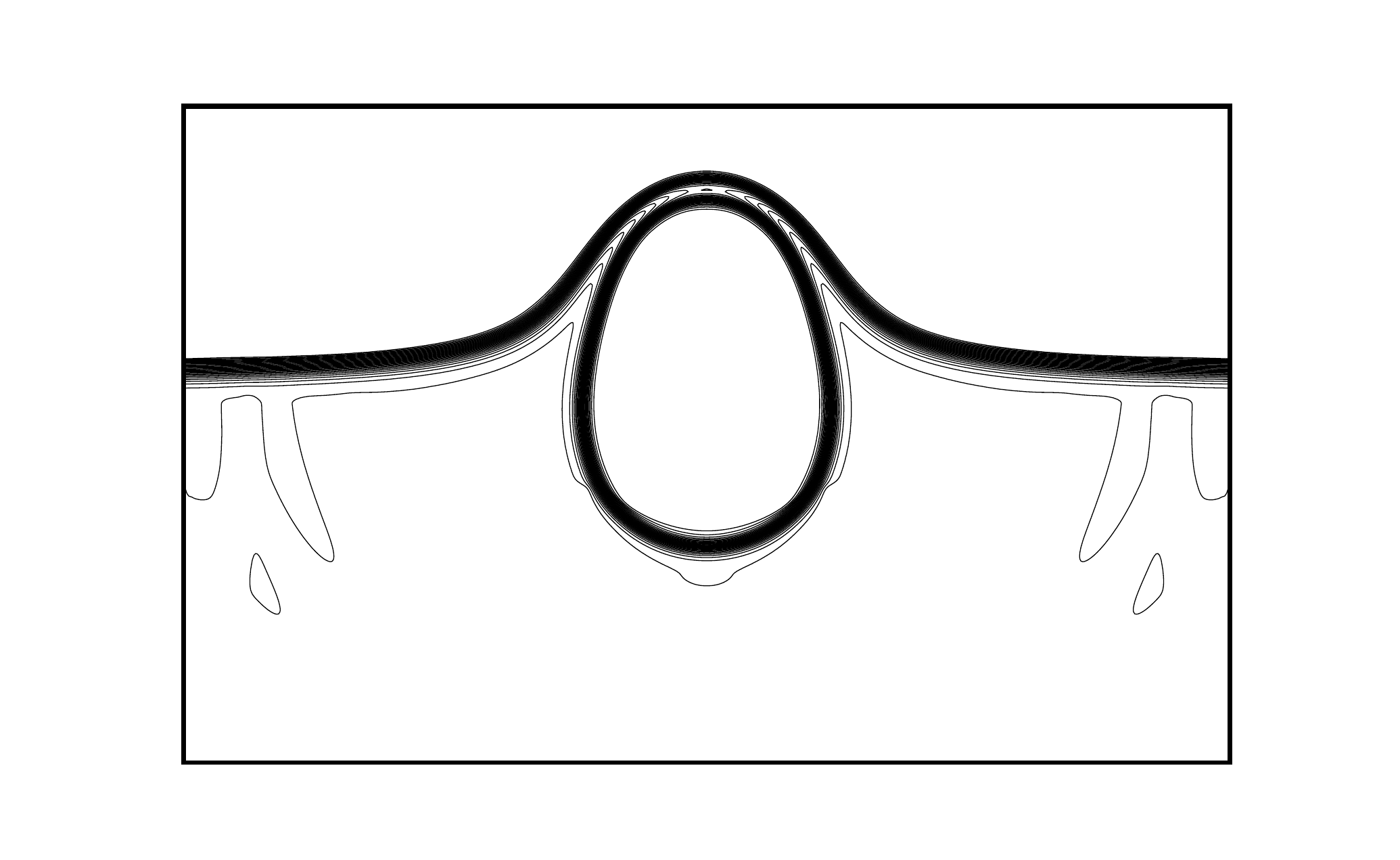}}\quad
   {\includegraphics[width=8cm]{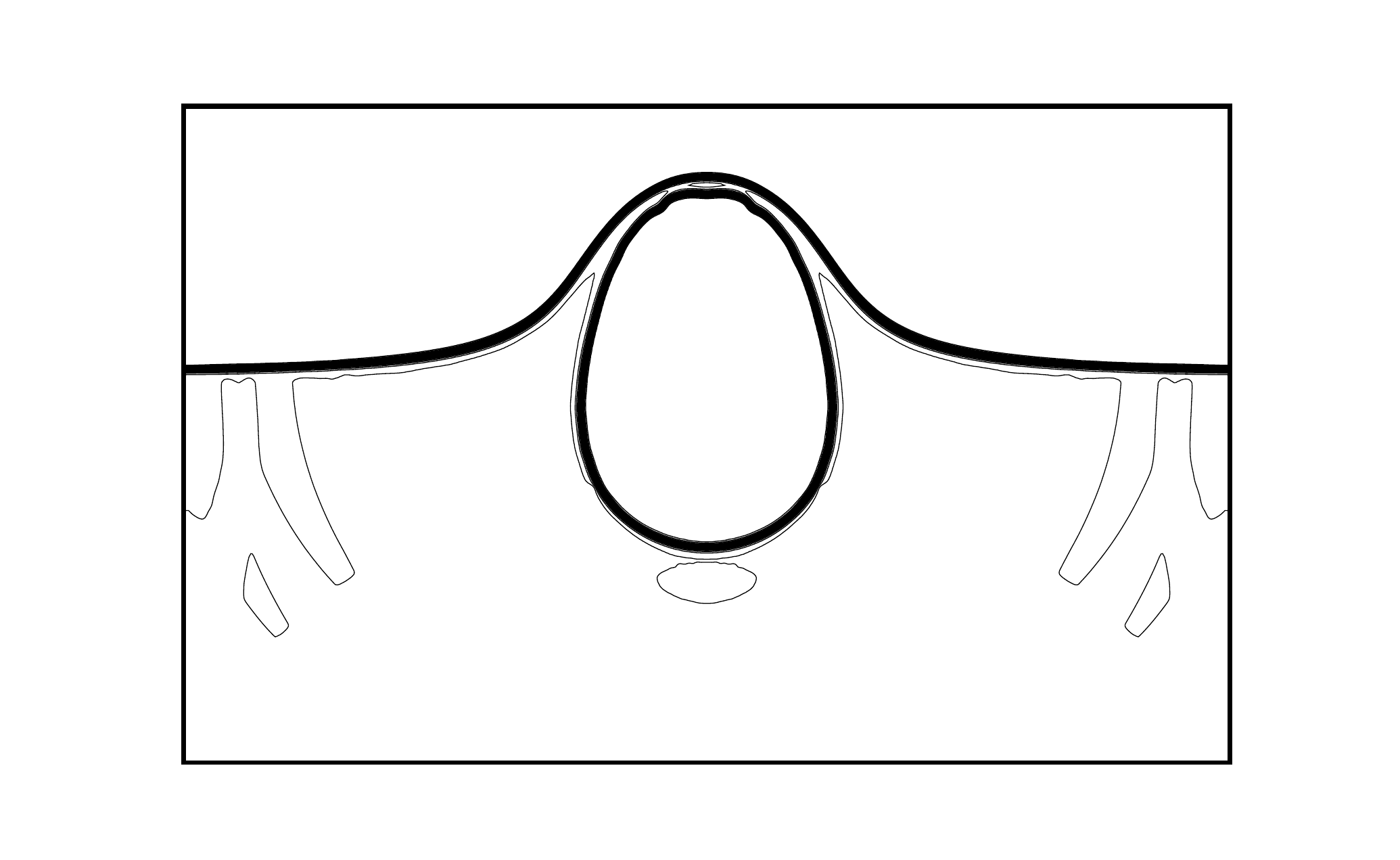}}
   }

   \caption{Example~\ref{exam4.10} The density contours. From top to bottom: $T=$ 0.2, 0.4, 0.8, 1.2 ms. Left: $P^1$ elements; Right: $P^2$ elements}
   \label{figbwstiff1}
   \end{center}
   \end{figure}

   \begin{figure}[hbtp]
 \begin{center}
 \mbox{
 {\includegraphics[width=6cm]{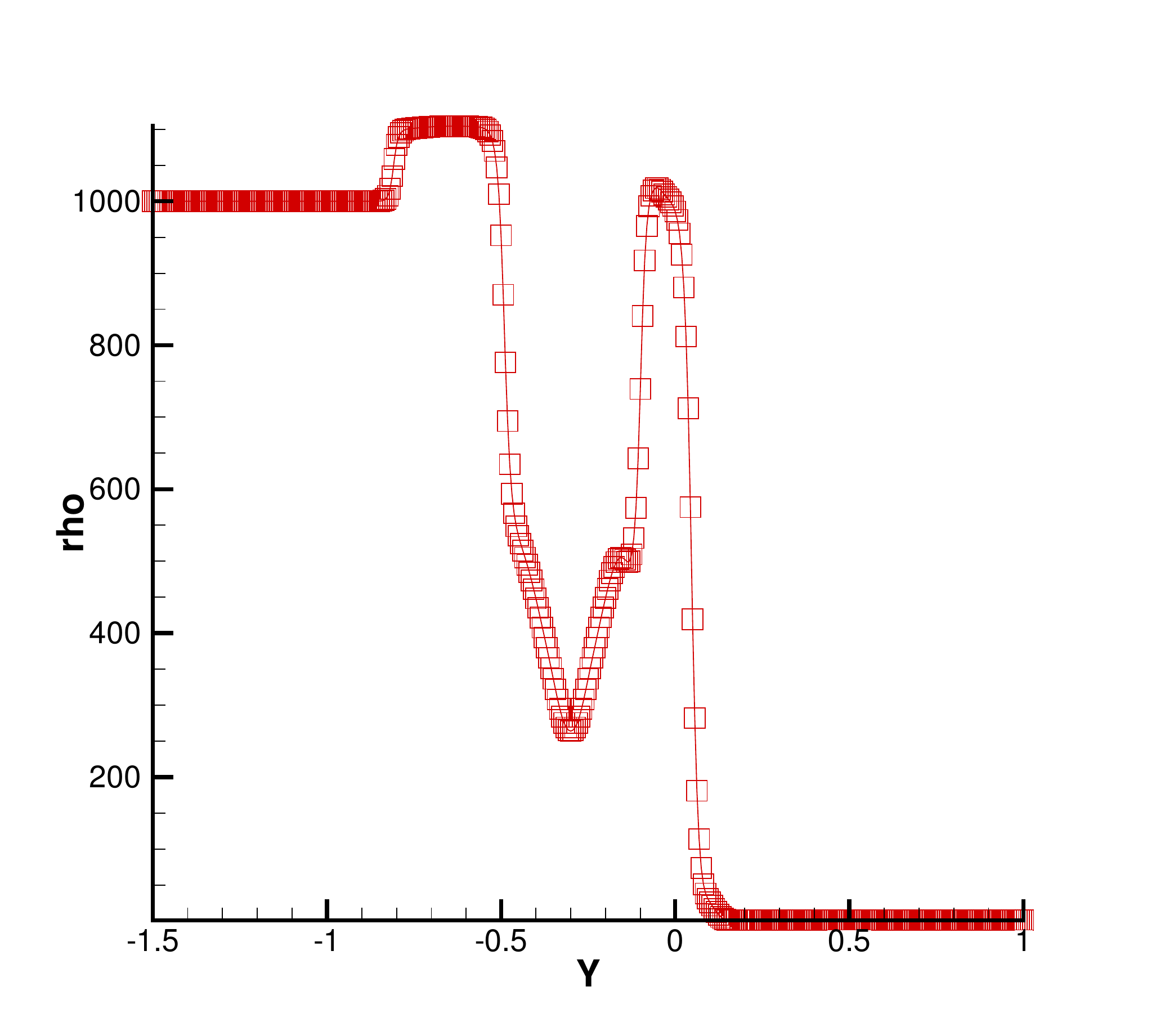}}\quad
   {\includegraphics[width=6cm]{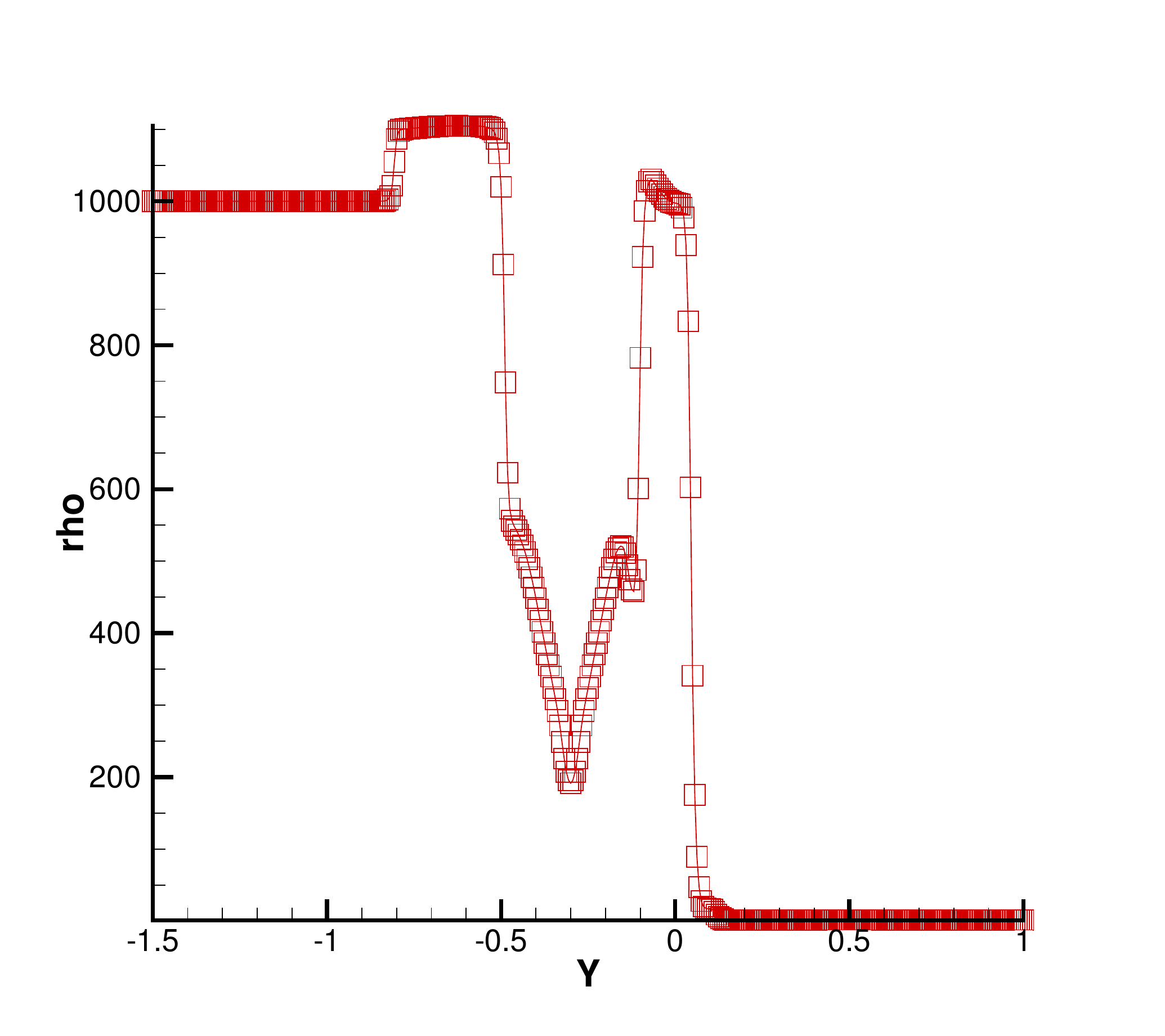}}
   }

   \mbox{

 {\includegraphics[width=6cm]{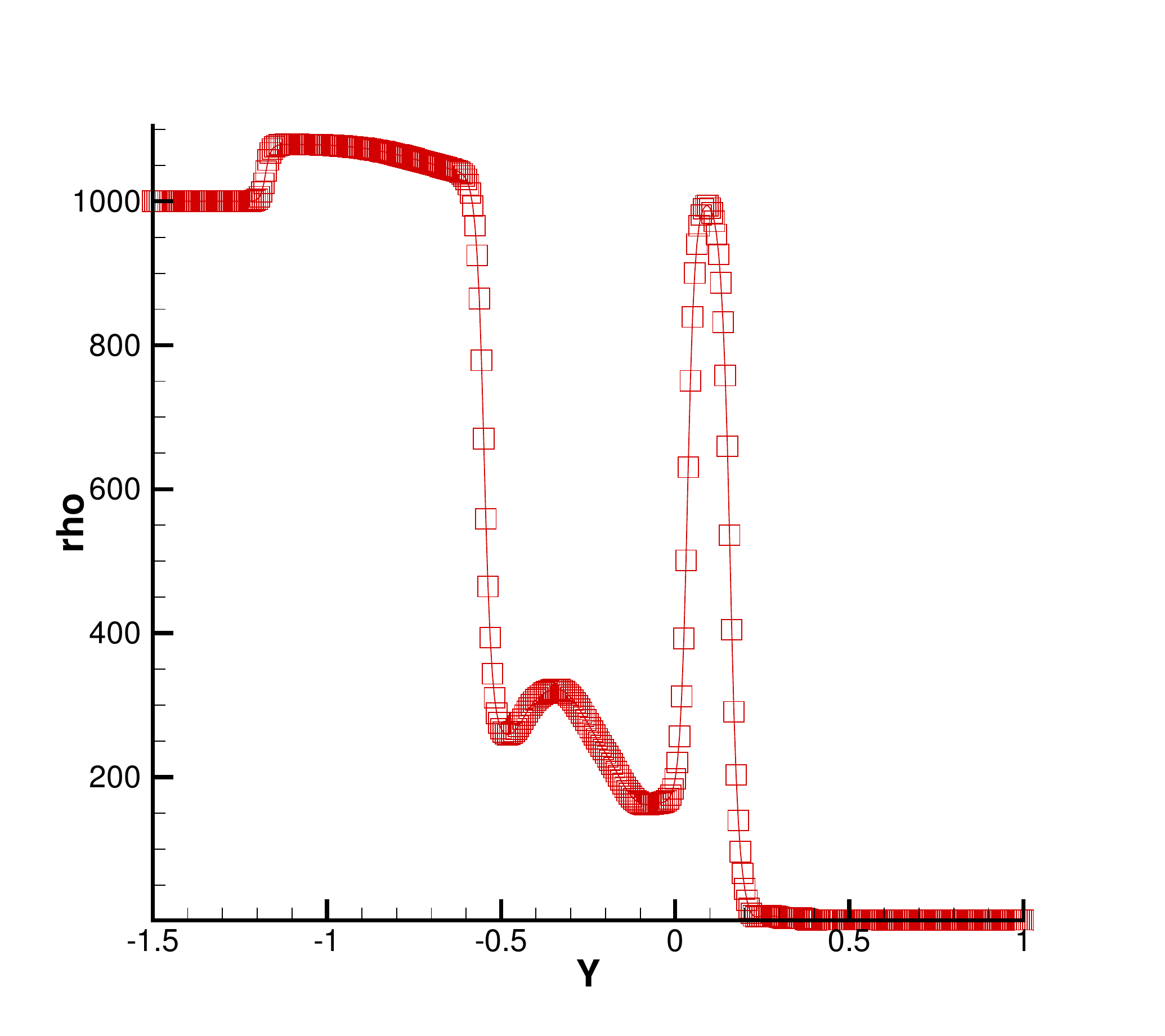}}\quad
   {\includegraphics[width=6cm]{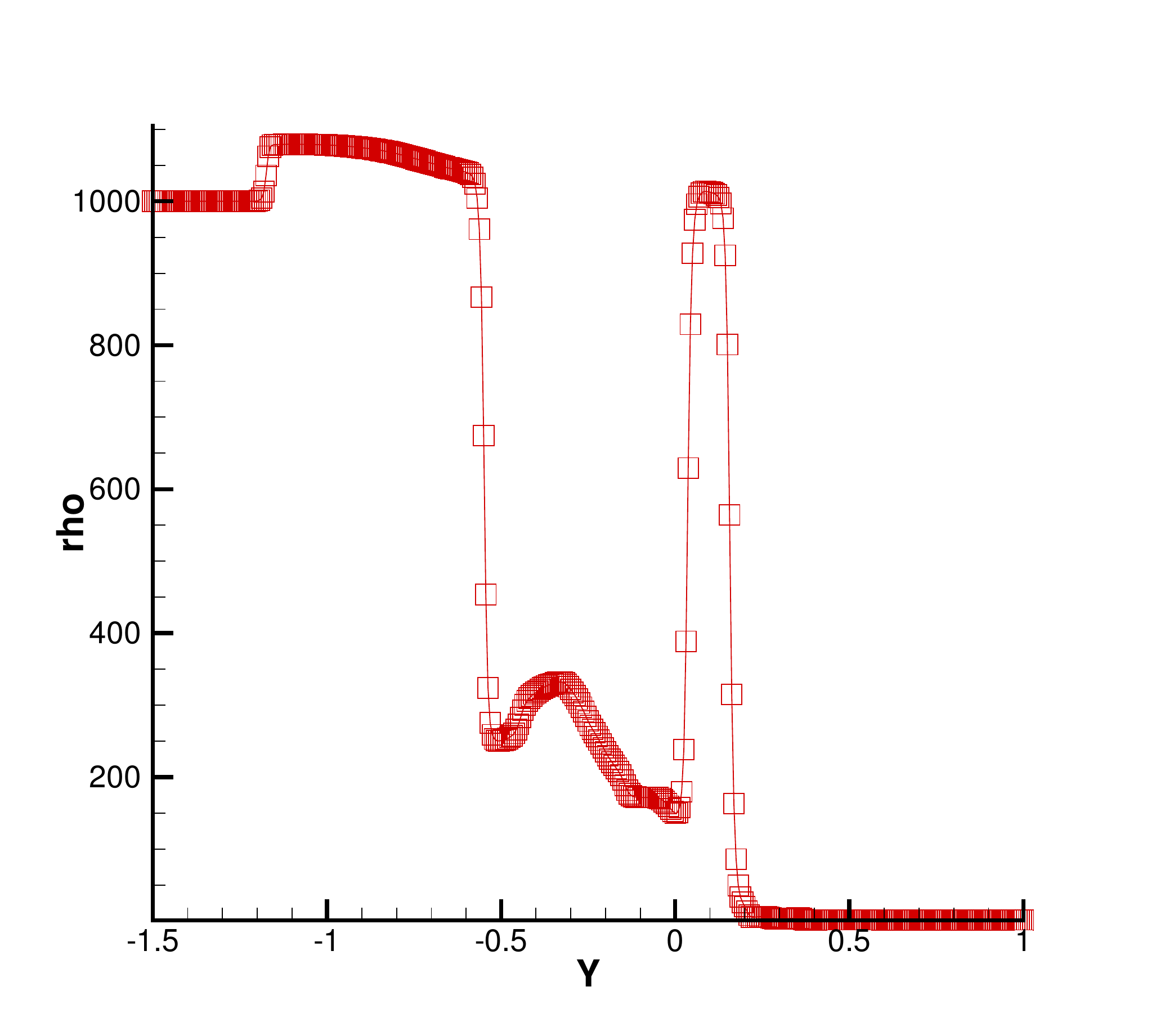}}
   }

   \mbox{

 {\includegraphics[width=6cm]{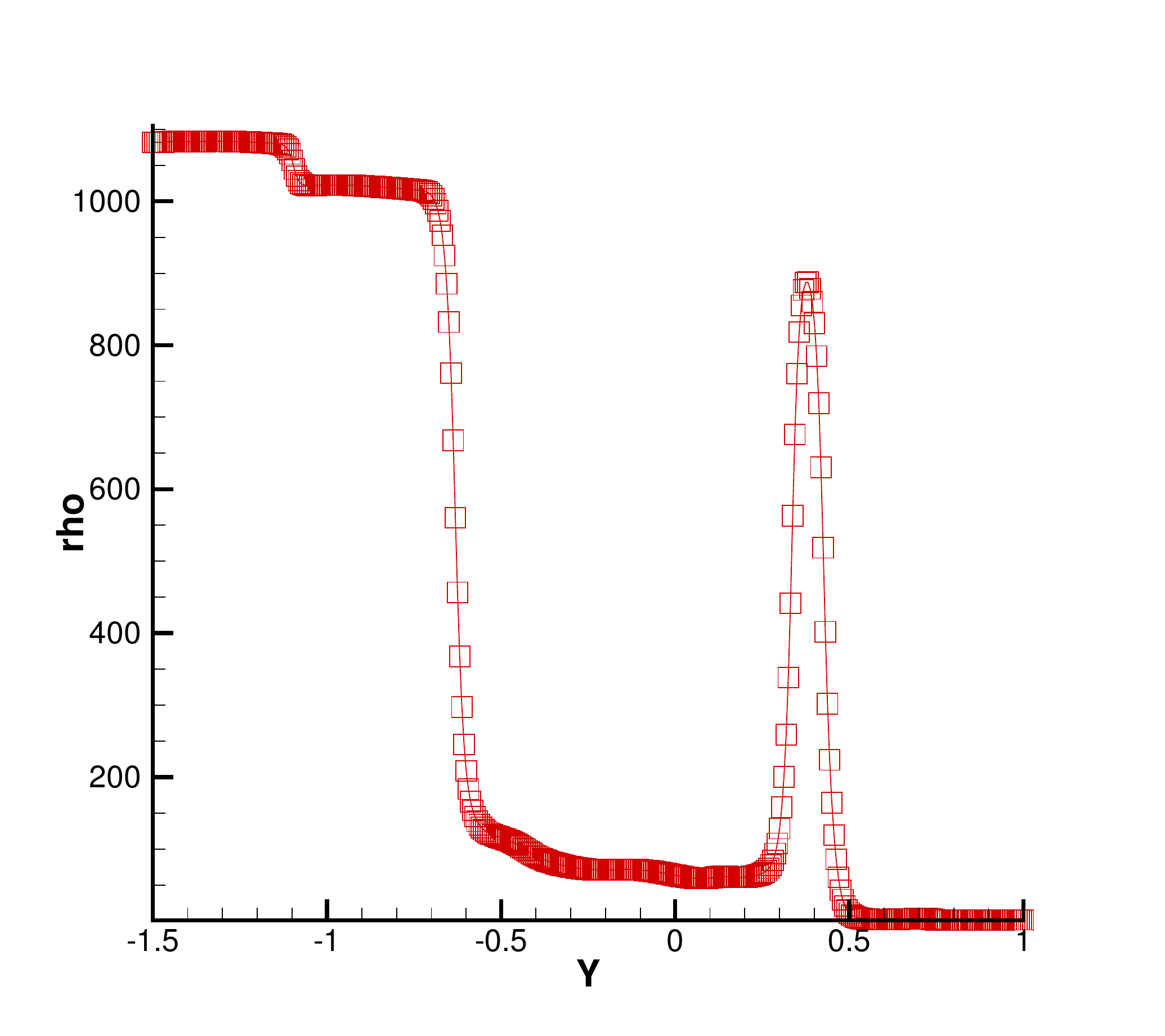}}\quad
   {\includegraphics[width=6cm]{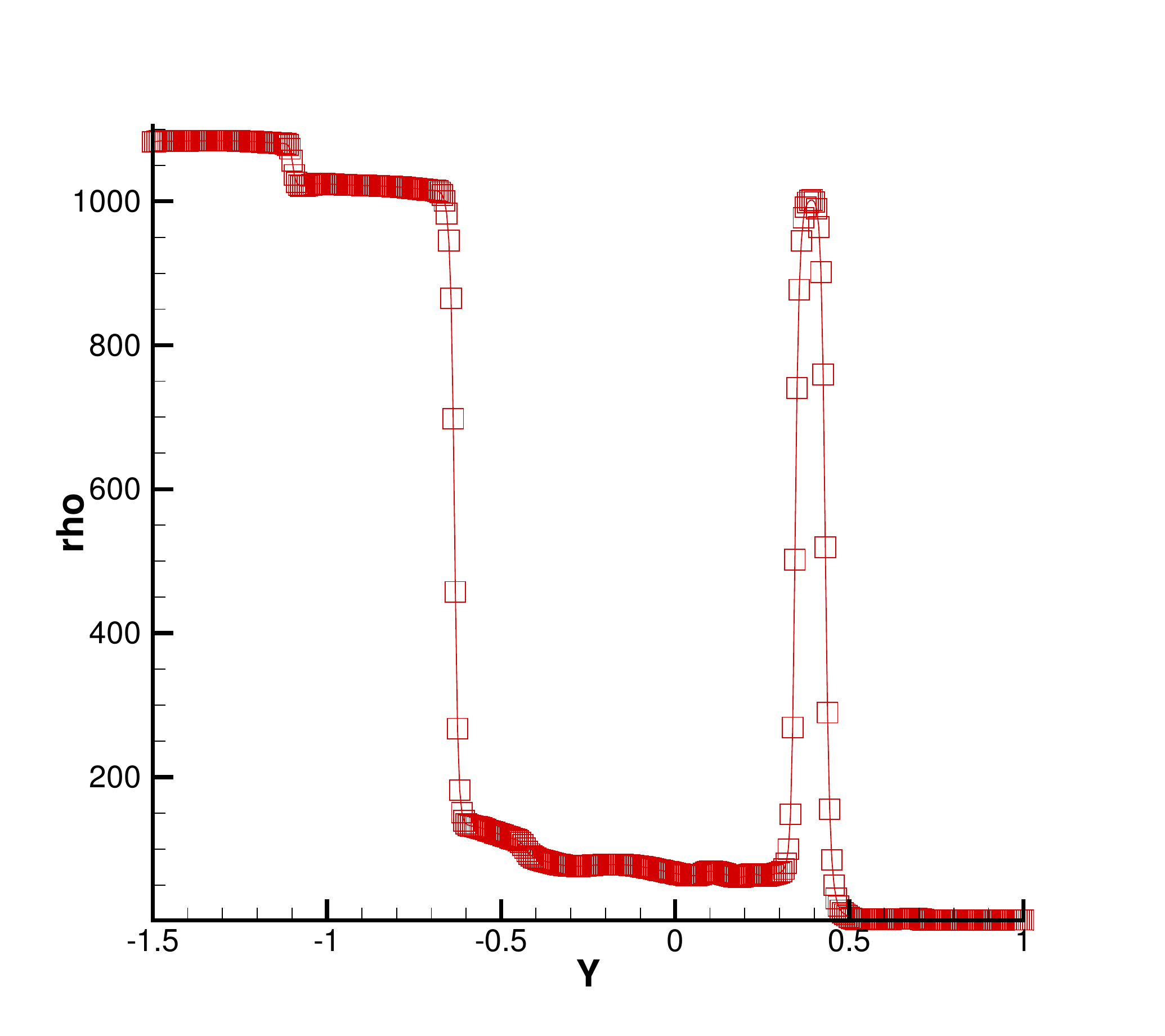}}
   }

   \mbox{
 {\includegraphics[width=6cm]{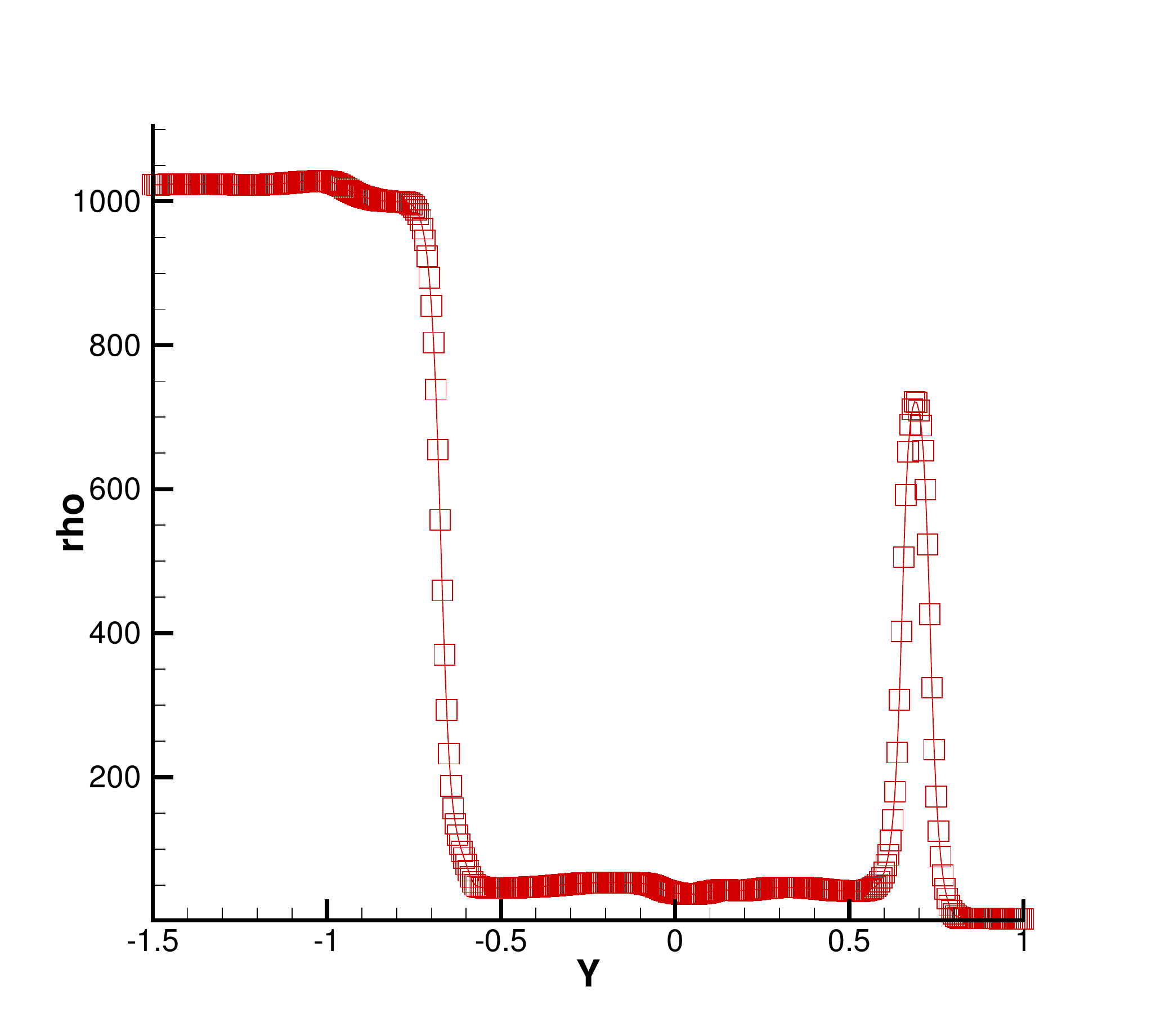}}\quad
   {\includegraphics[width=6cm]{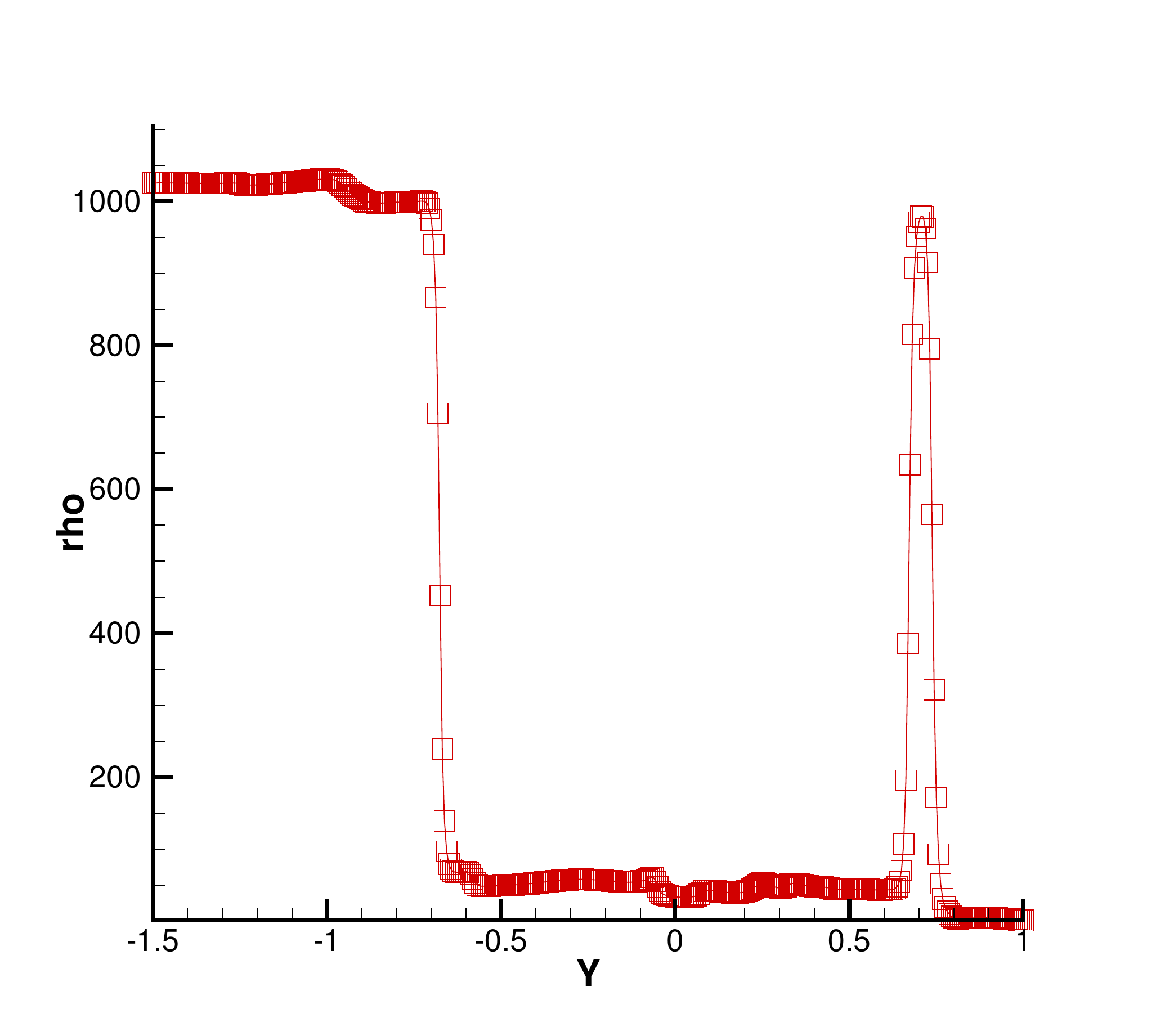}}
   }

   \caption{Cross-sectional plots of the results in Fig. \ref{figbwstiff1} along $x=0.$ From top to bottom: $T=$ 0.2, 0.4, 0.8, 1.2 ms. 
   Left: $P^1$ elements; Right: $P^2$ elements}
   \label{figbwstiff3}
   \end{center}
   \end{figure}

\begin{figure}[hbtp]
 \begin{center}
 \mbox{

 {\includegraphics[width=8cm]{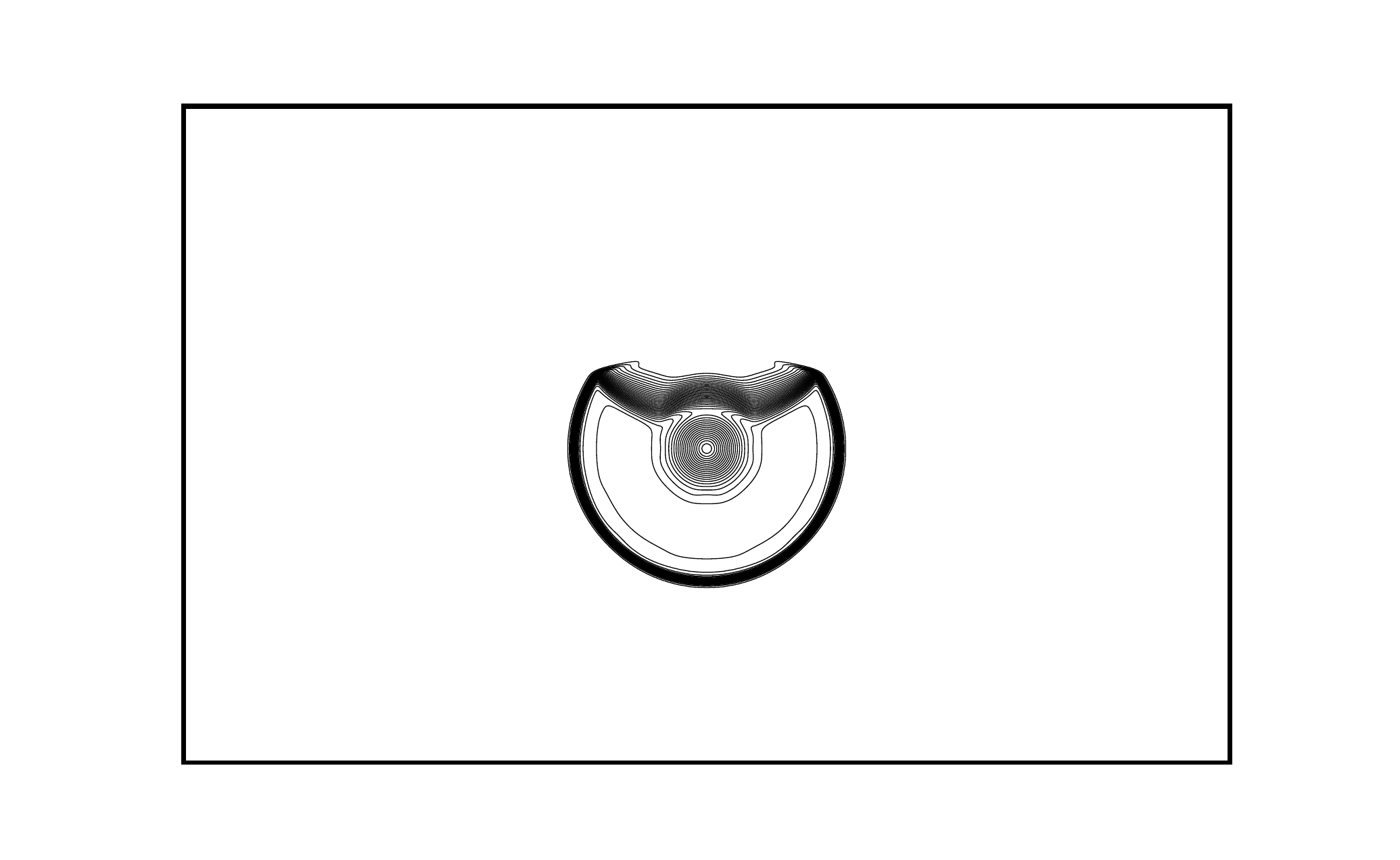}}\quad
   {\includegraphics[width=8cm]{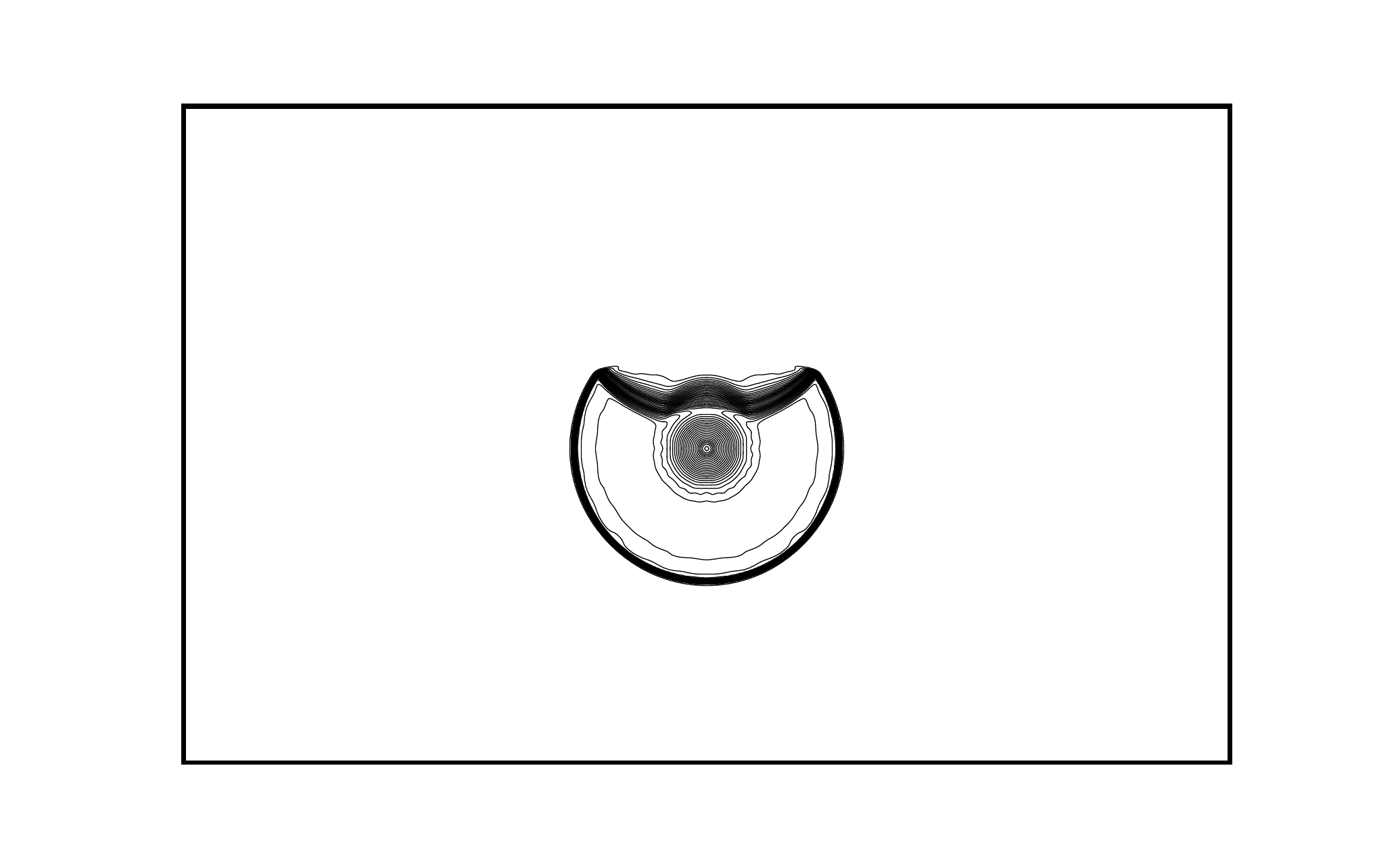}}
   }

   \mbox{

 {\includegraphics[width=8cm]{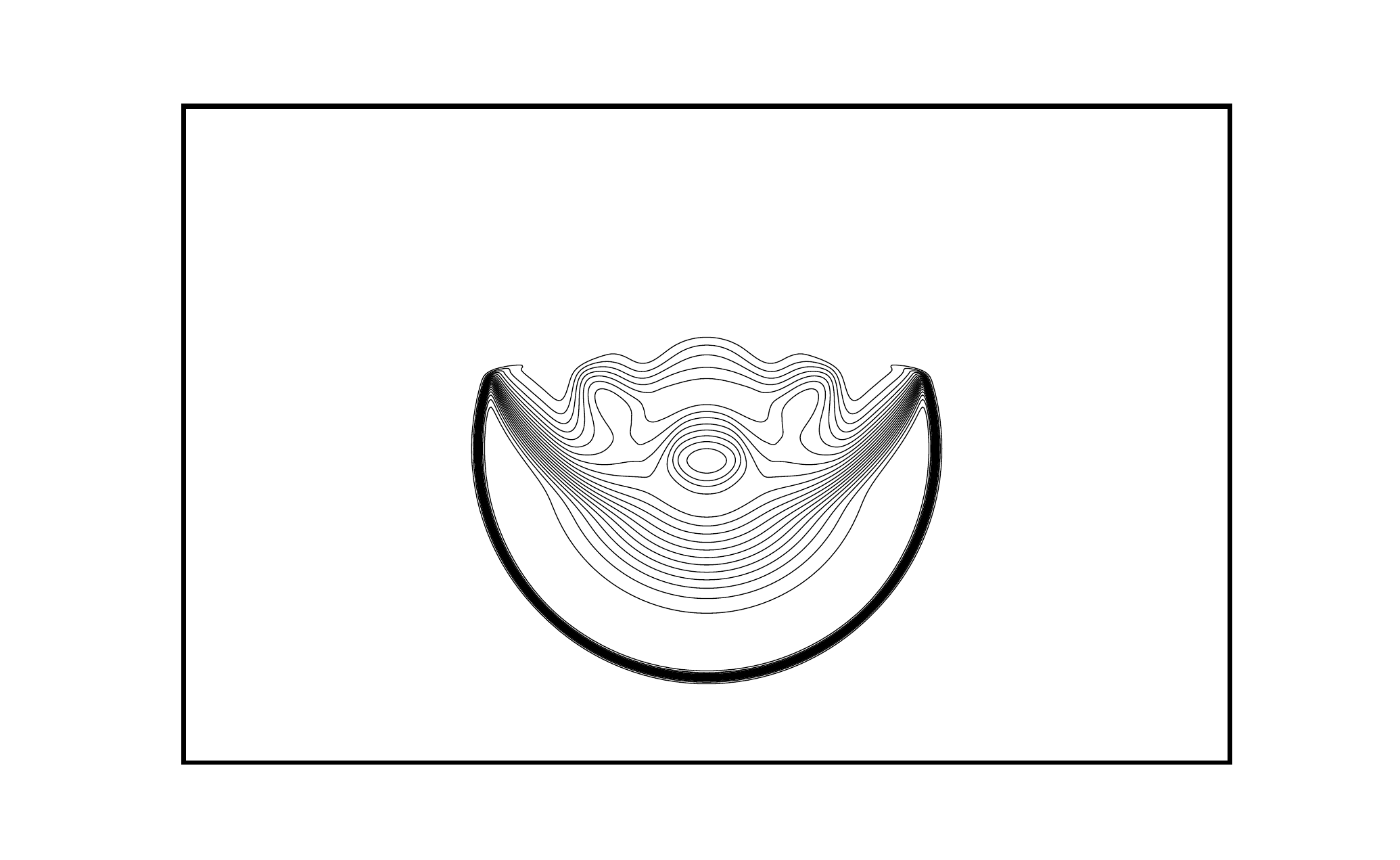}}\quad
   {\includegraphics[width=8cm]{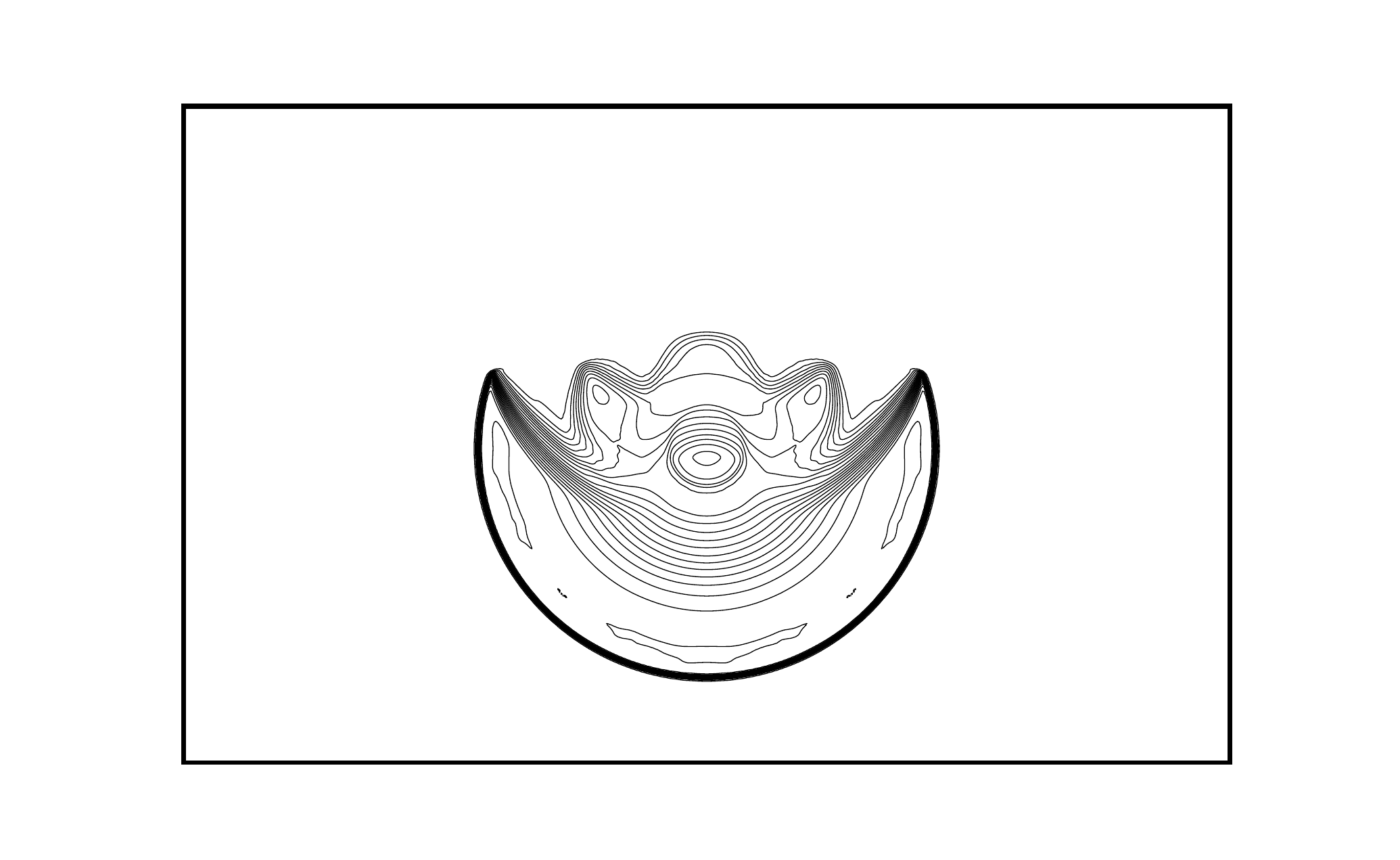}}
   }

   \mbox{

 {\includegraphics[width=8cm]{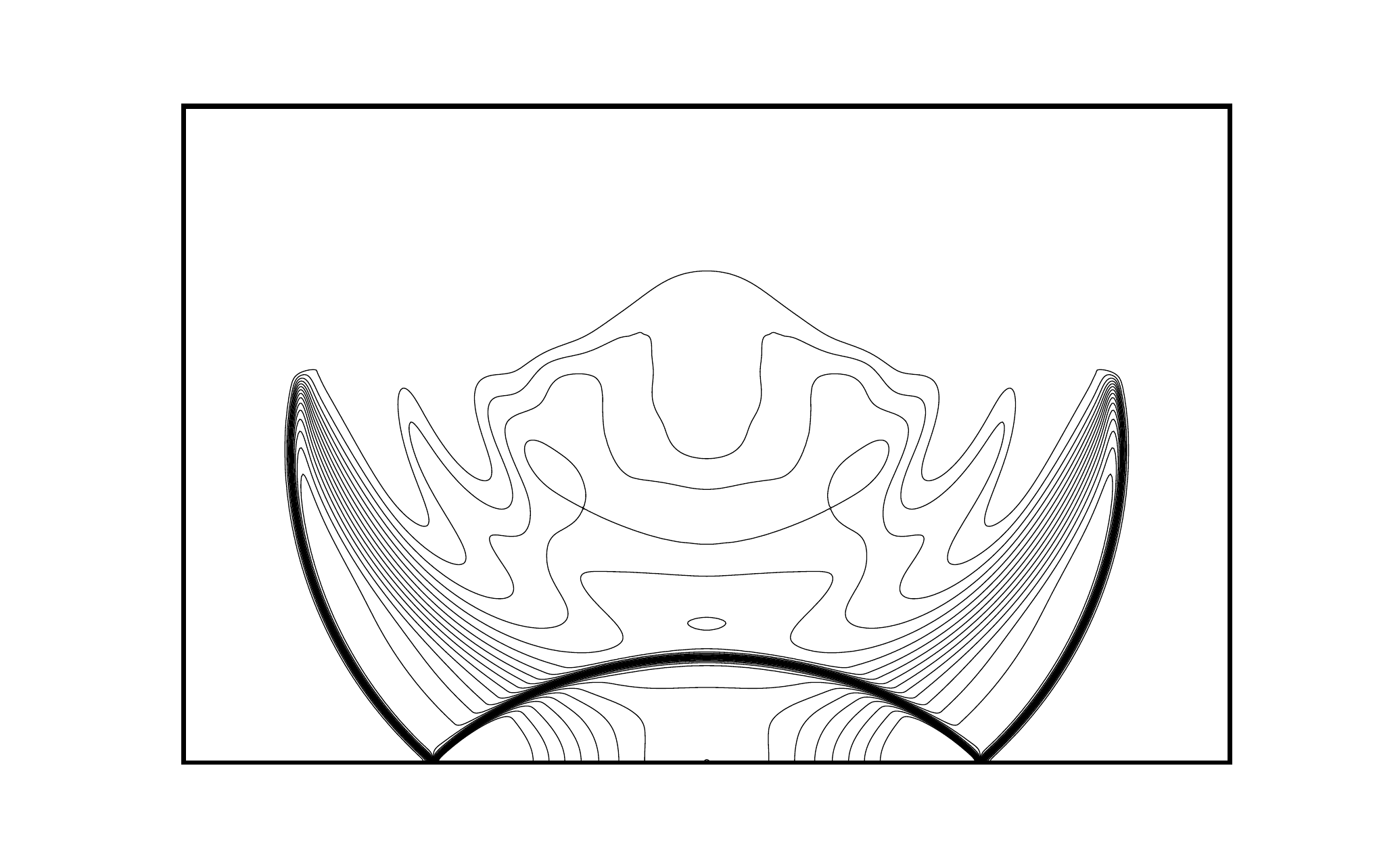}}\quad
   {\includegraphics[width=8cm]{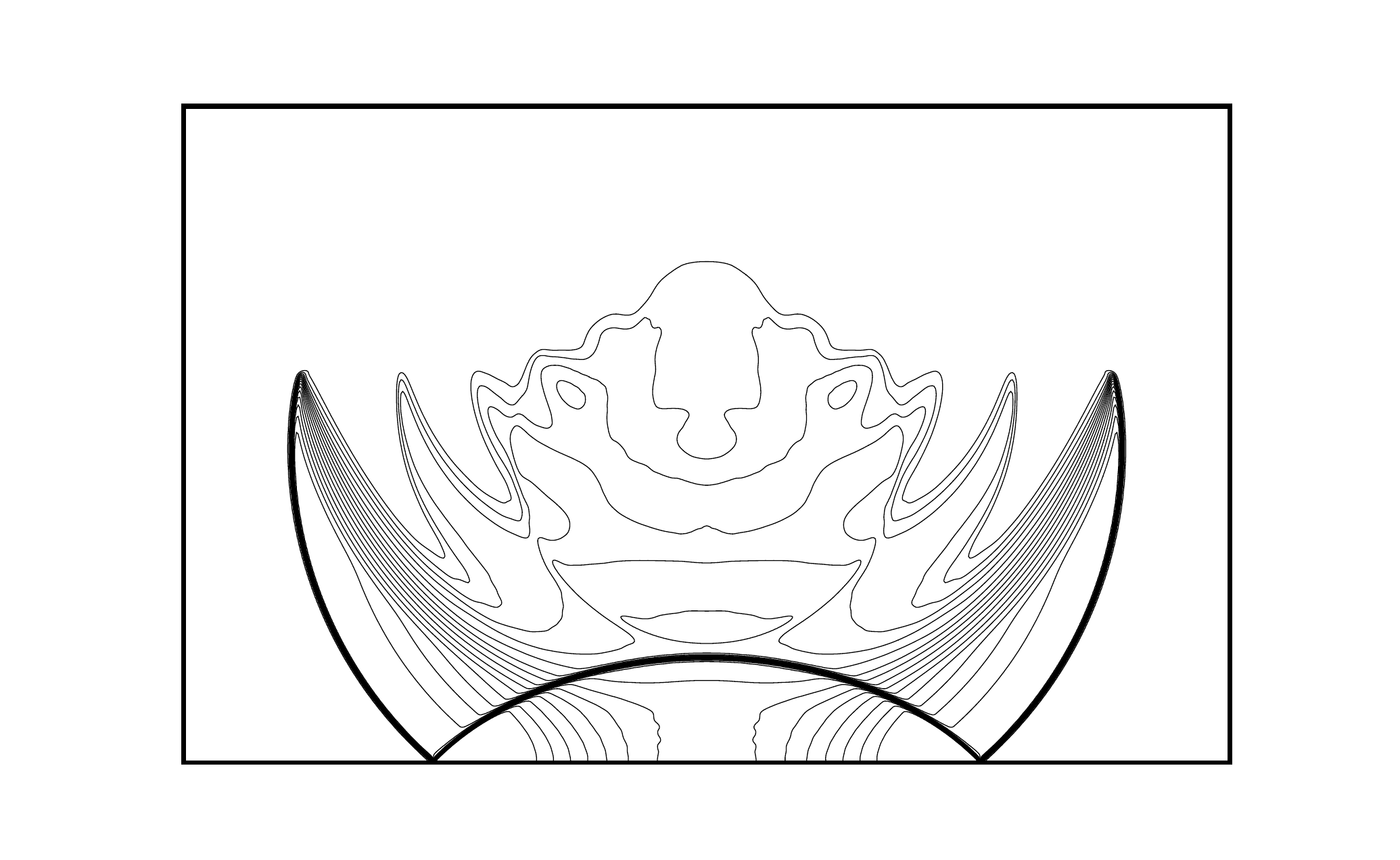}}
   }

   \mbox{

 {\includegraphics[width=8cm]{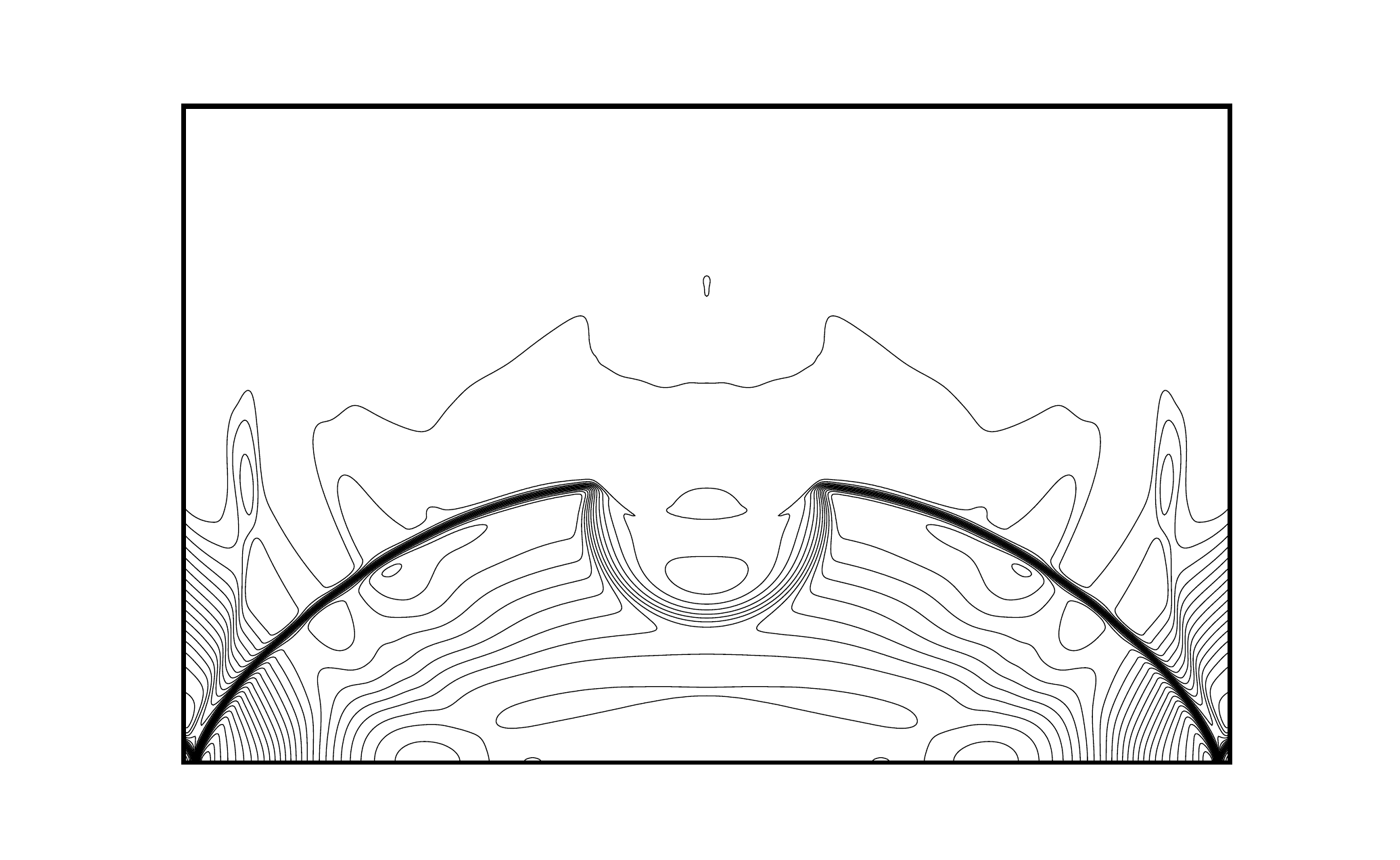}}\quad
   {\includegraphics[width=8cm]{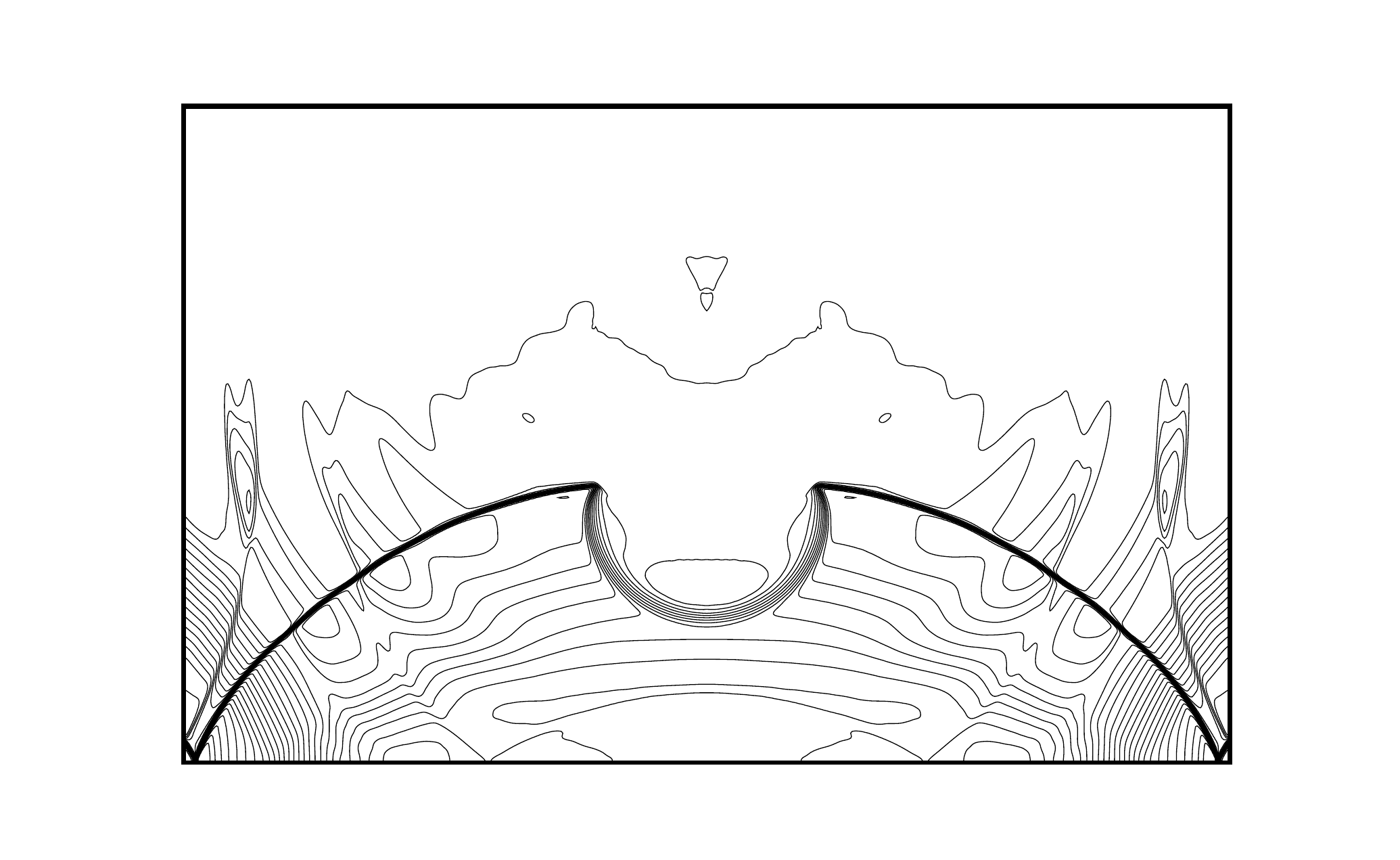}}
   }

   \caption{Example~\ref{exam4.10} The pressure contours. From top to bottom: $T=$ 0.2, 0.4, 0.8, 1.2 ms. Left: $P^1$ elements; Right: $P^2$ elements}
   \label{figbwstiff2}
   \end{center}
   \end{figure}

\begin{figure}[hbtp]
 \begin{center}
 \mbox{

 {\includegraphics[width=6cm]{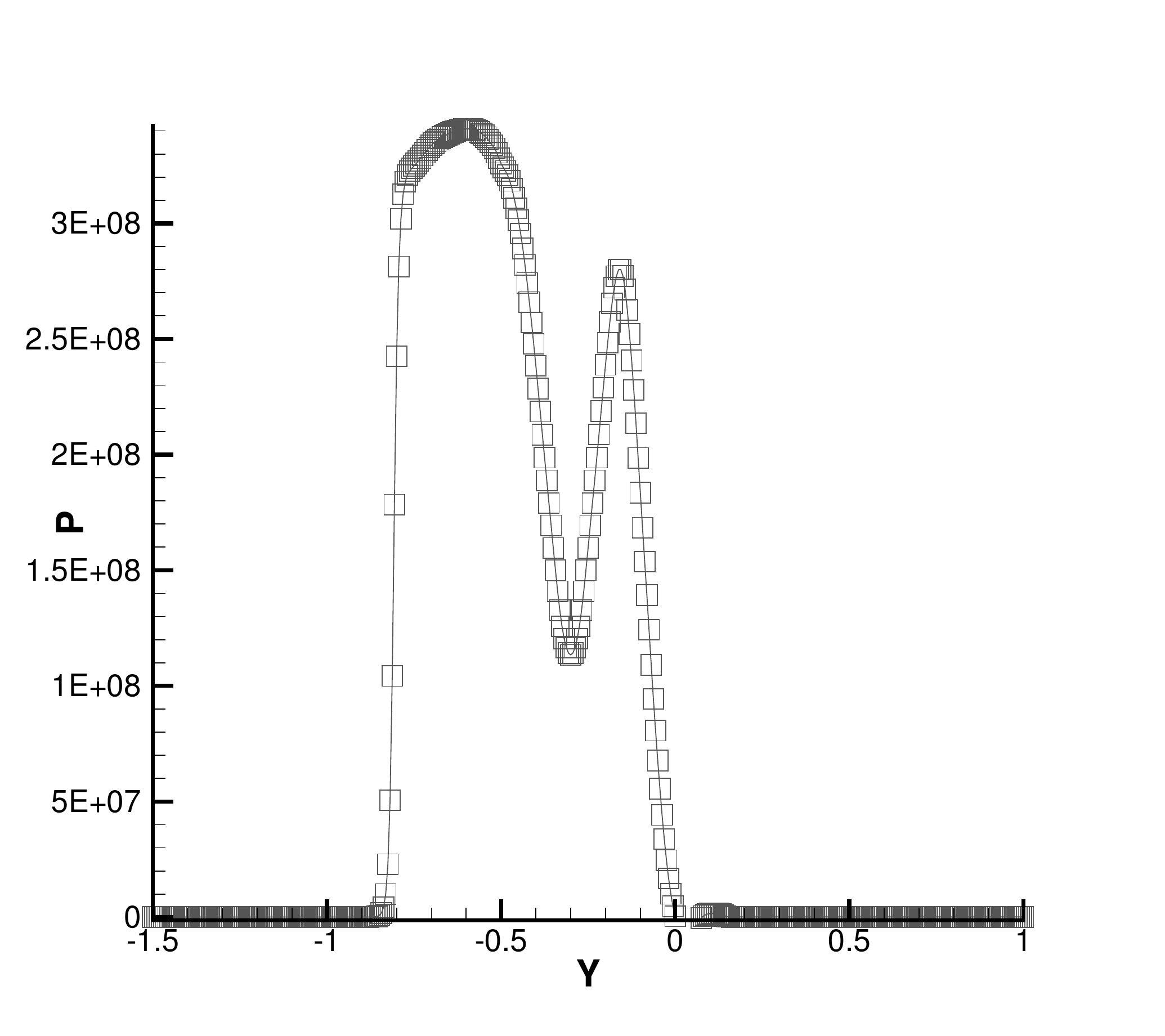}}\quad
   {\includegraphics[width=6cm]{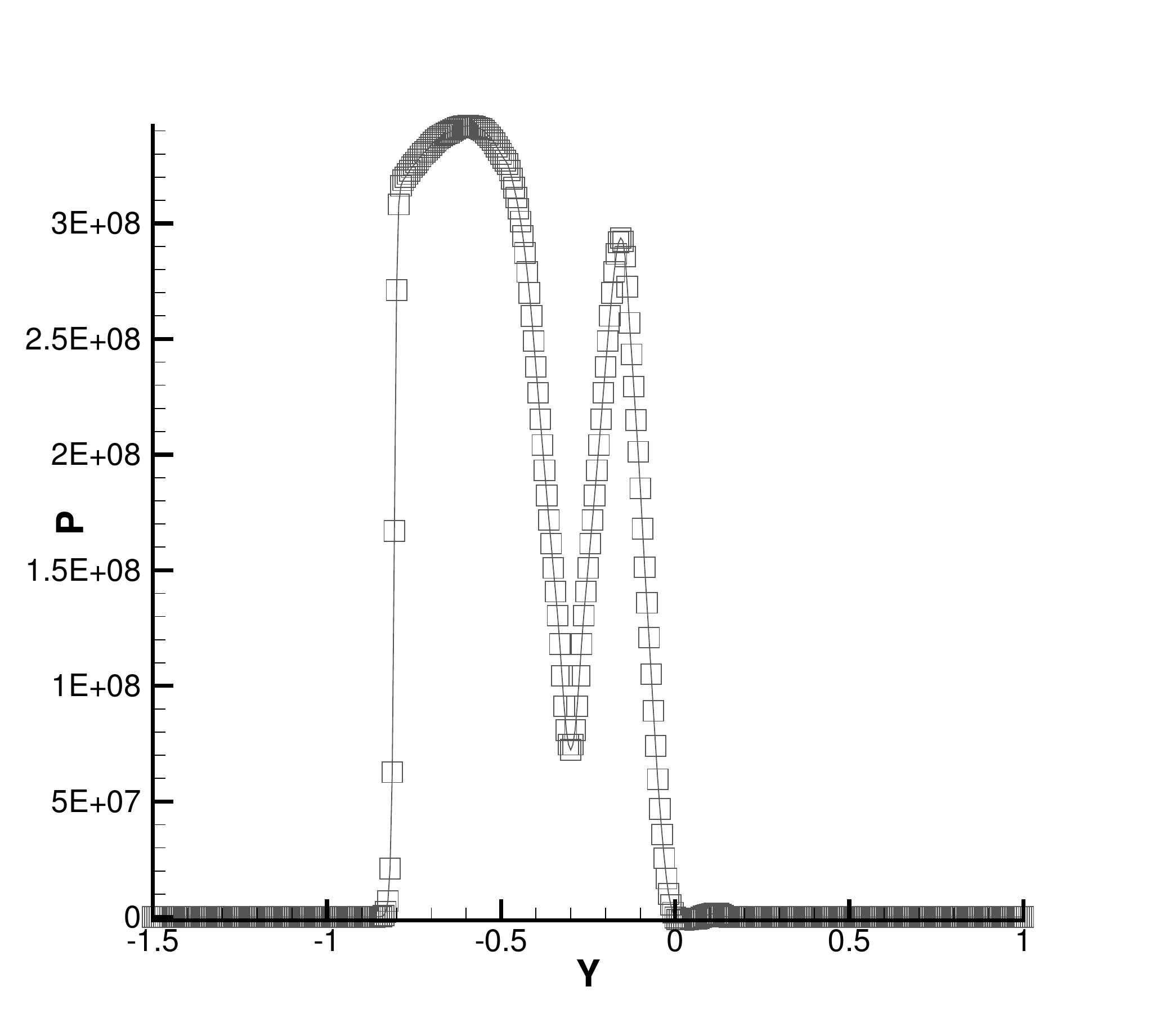}}
   }

   \mbox{

 {\includegraphics[width=6cm]{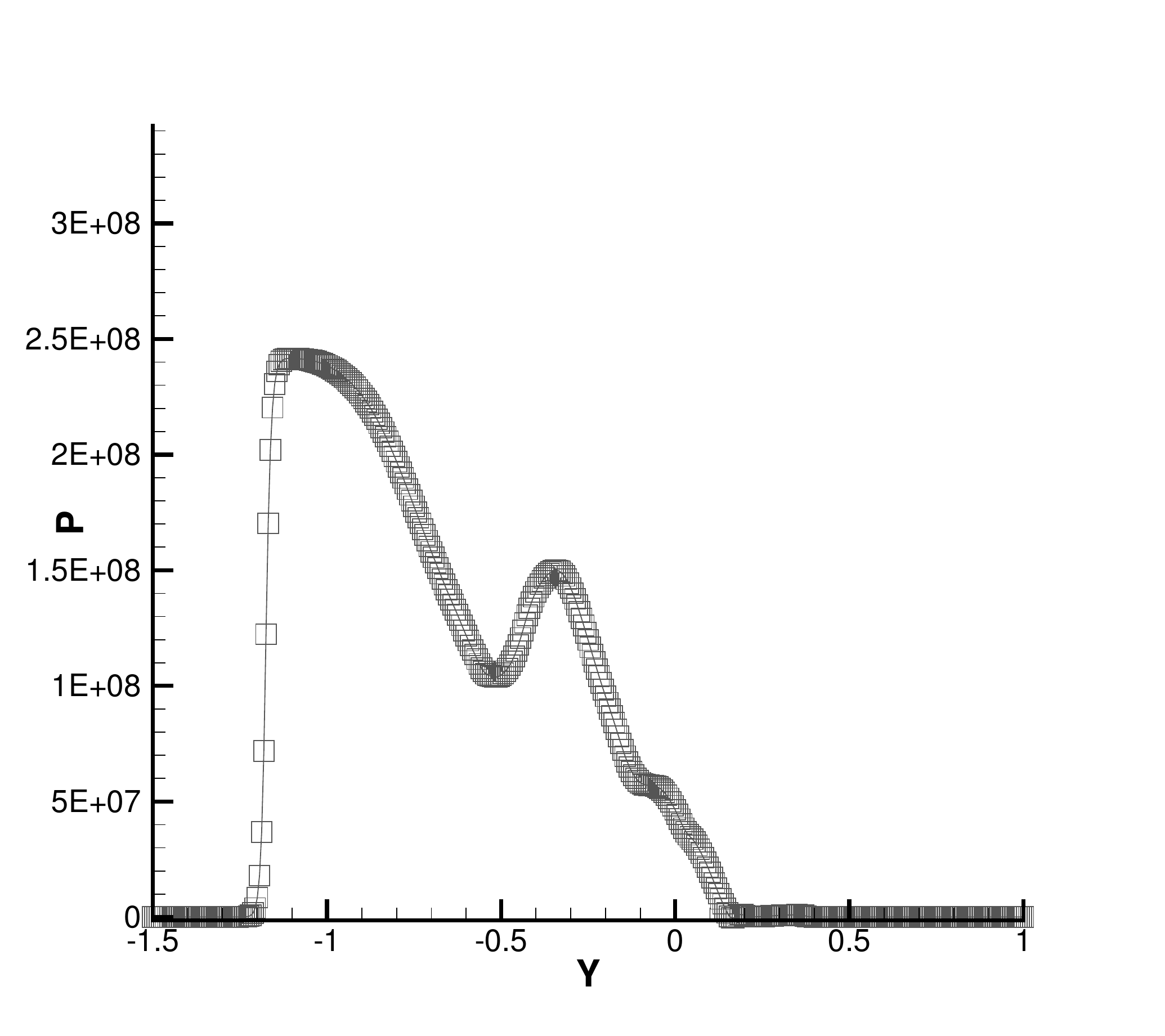}}\quad
   {\includegraphics[width=6cm]{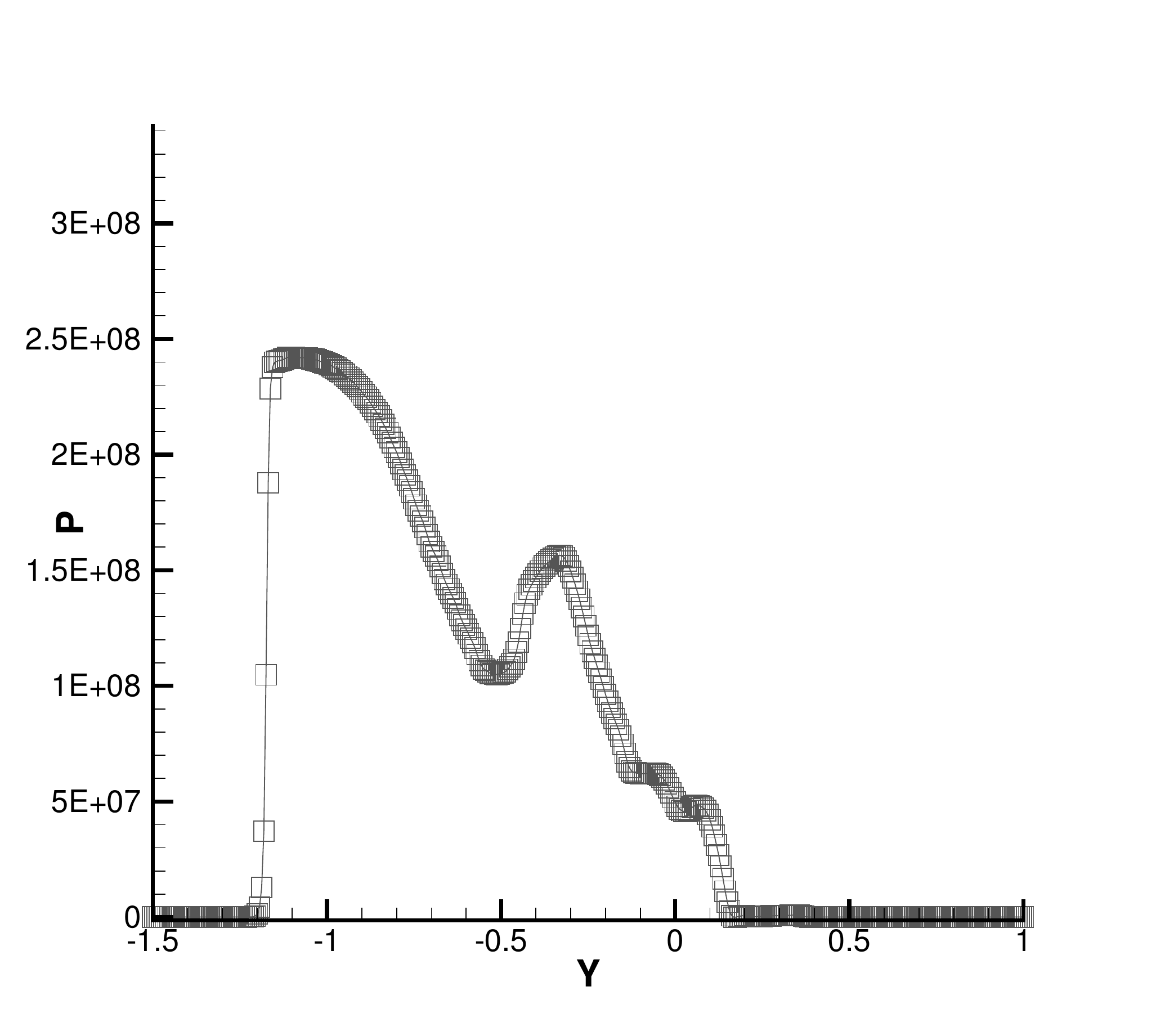}}
   }

   \mbox{

 {\includegraphics[width=6cm]{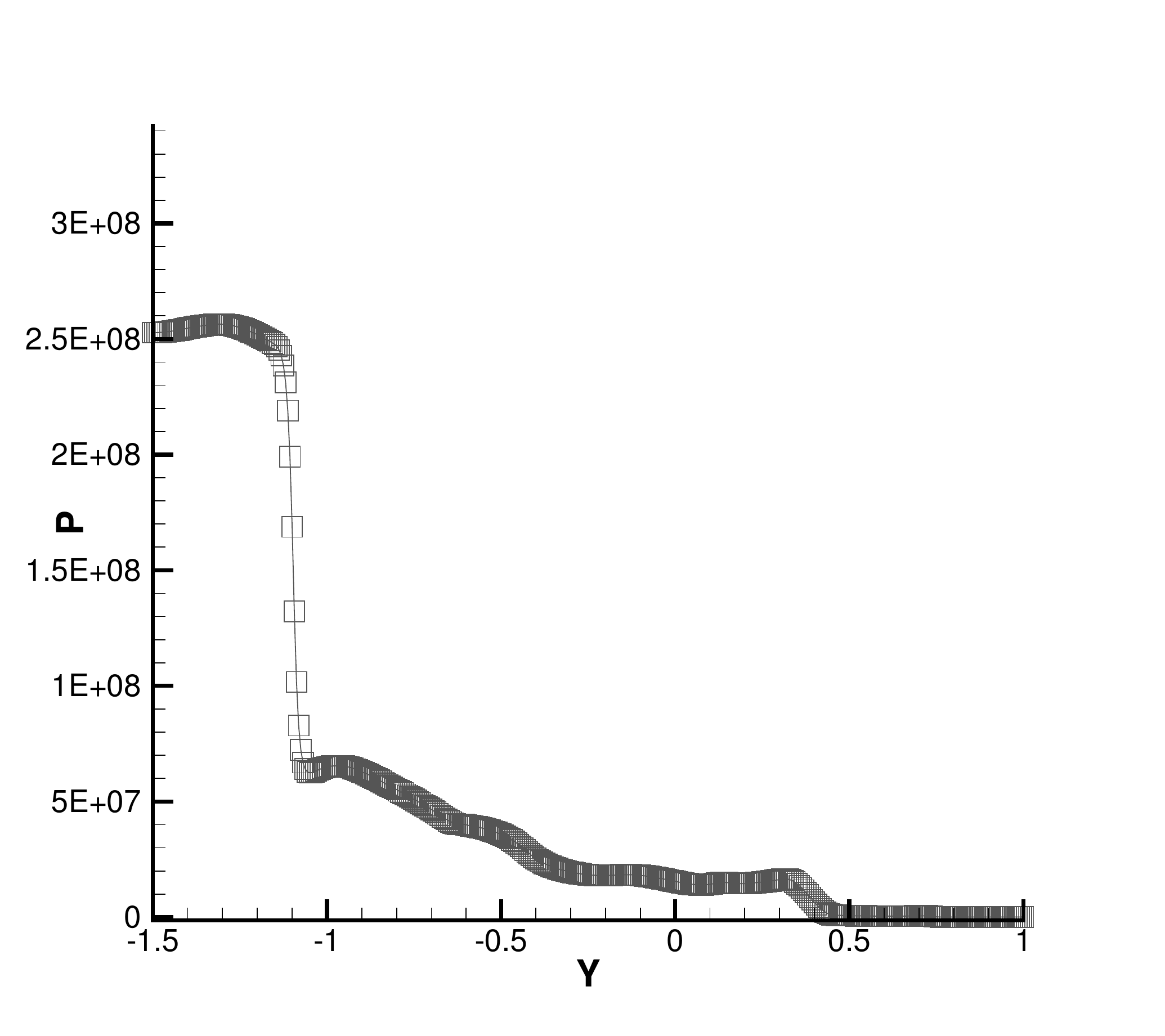}}\quad
   {\includegraphics[width=6cm]{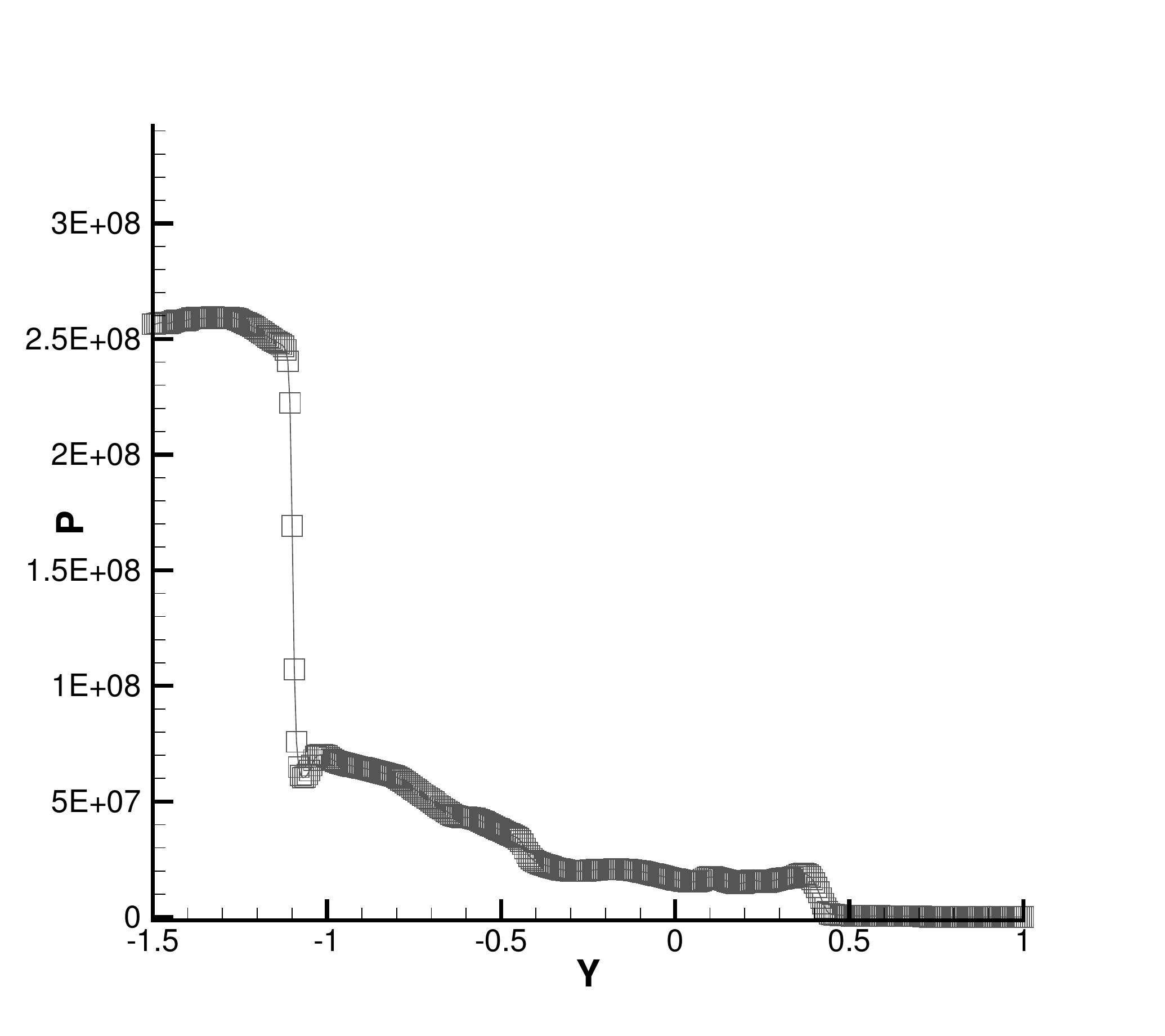}}
   }

   \mbox{

 {\includegraphics[width=6cm]{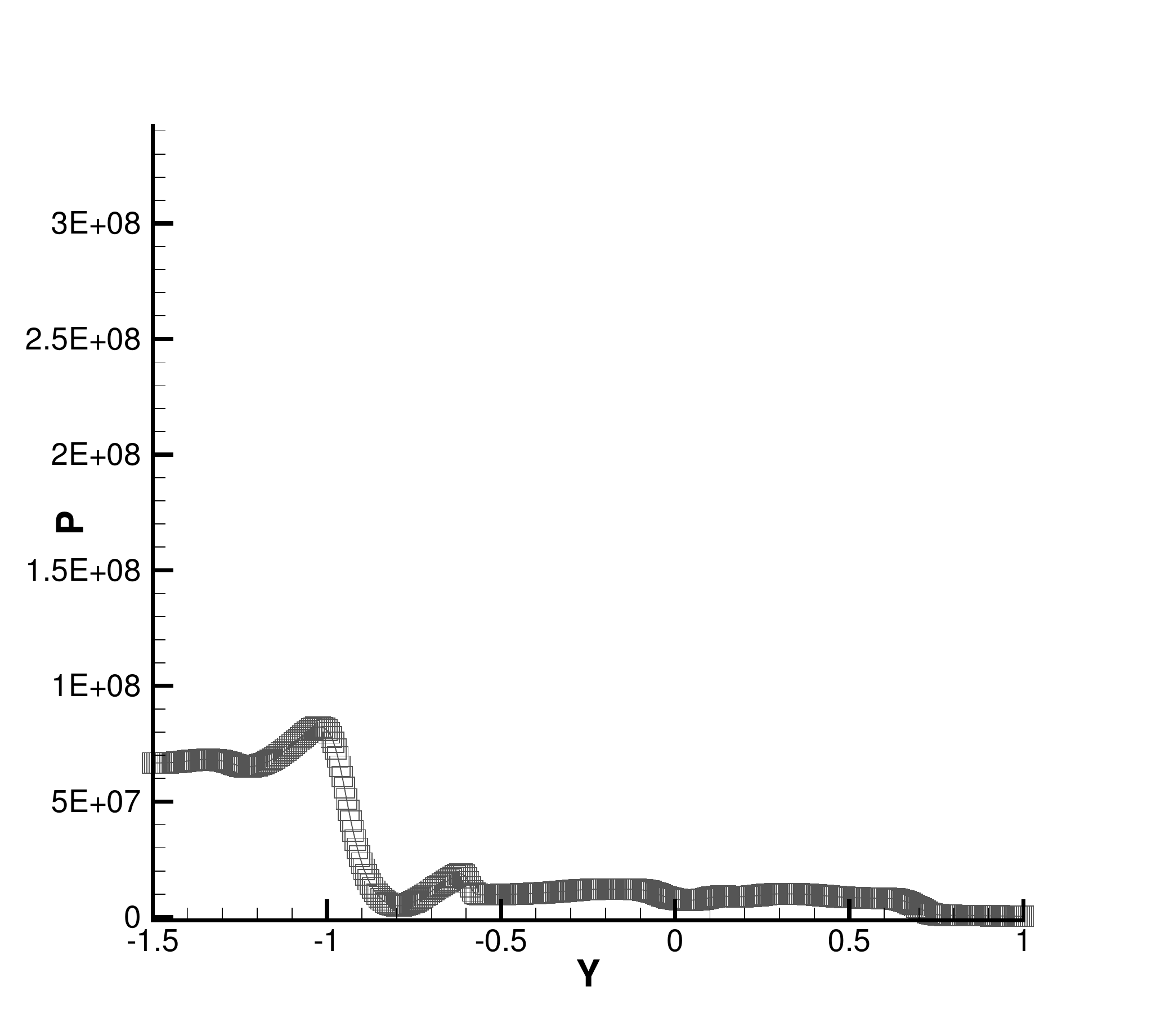}}\quad
   {\includegraphics[width=6cm]{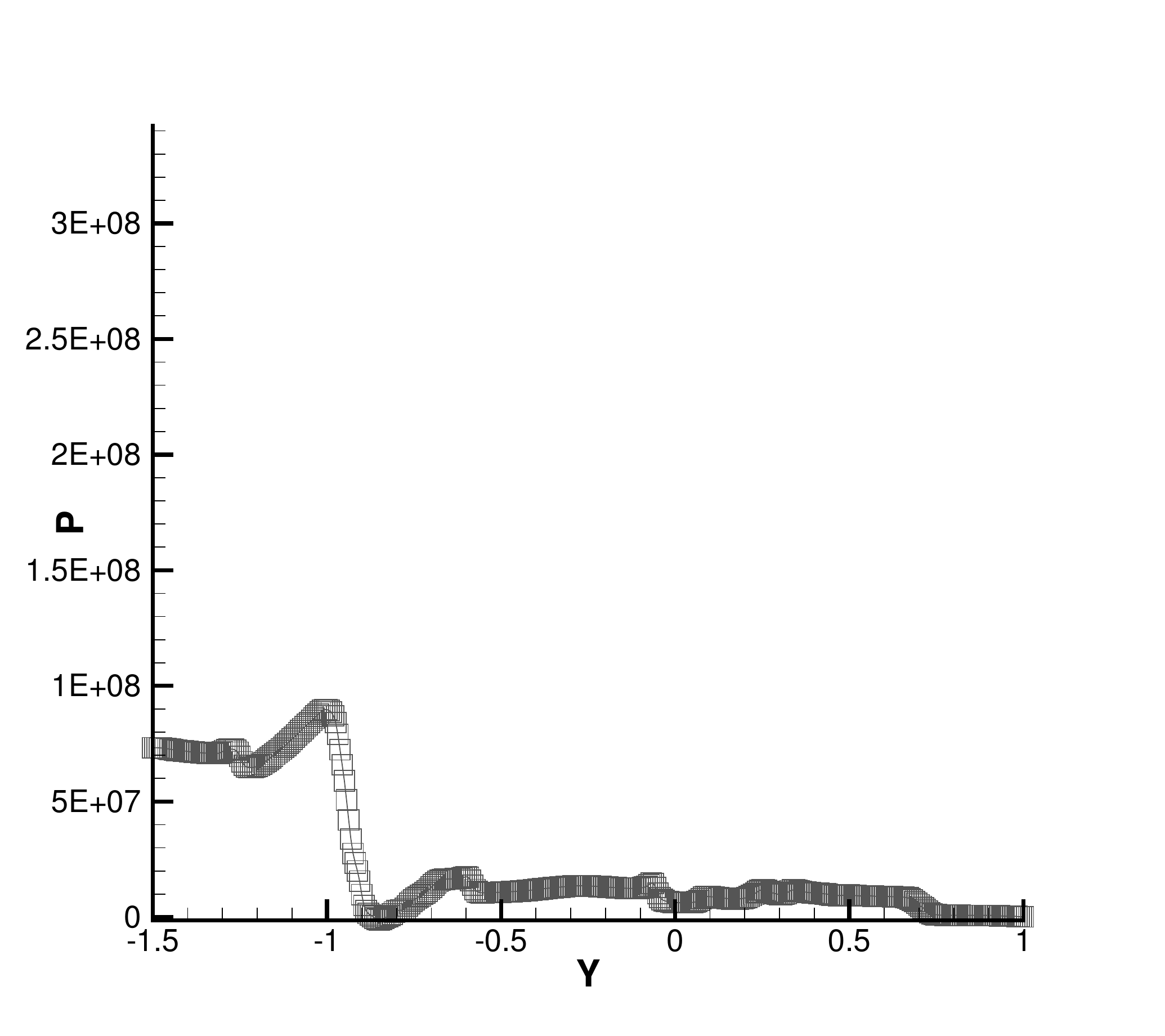}}
   }

   \caption{Cross-sectional plots of the results in Fig. \ref{figbwstiff2} along $x=0.$ From top to bottom: $T=$ 0.2, 0.4, 0.8, 1.2 ms. Left: $P^1$ elements; Right: $P^2$ elements}
   \label{figbwstiff4}
   \end{center}
   \end{figure}
}
\end{exam}

\begin{exam}{\em
\label{shock2d}
To show our method works with complex Mie-Gr\"uneisen EOS in two dimensions, we are concerned with interaction of a shock in molybdenum with a block of encapsulated mid-ocean ridge basalt (MORB) liquid \cite{miller1996,shyue2001, lee2009}. The computational domain is set to be an unit square. A Mach 1.163 rightward-moving shock is located at $x=0.3$ and to impact MORB contained in a rectangle of $[0.4,0.7]\times [0,0.5]$. Both materials are modeled by shock wave EOS \eqref{sw}. Reflecting boundary conditions are imposed on bottom and non-reflecting boundary conditions are used on the other three sides. For this problem, inside the region of the MORB liquid, we have the state variables
\[
(\rho,\mu,\nu,P)=(2260,0,0,0),
\]
and outside the MORB, the state variables in the preshock region are given by
\[
(\rho,\mu,\nu,P)=(9961,0,0,0),
\]
and the state variables in the postshock region are
\[
(\rho,\mu,\nu,P)=(11042,543,0,3\times 10^{10}).
\]

 The contours of the density and pressure at two different selected times $T=50$ and $T=100 \;\mu s$ with a uniform $400\times 400$ mesh are illustrated in Figs \ref{figshock2d} and \ref{figshock2dP}. In the density plot, we can see the incident shock in molybdenum and transmitted shock in MORB with the former moving faster than the latter at $T=50\; \mu s$. And the transmitted shock has not passed the MORB block completely at $T=100\; \mu s$. In addition, the structure of diffraction of the shock by MORB is well captured in the pressure graphs. From the displayed figures, one is easy to observe that the improved resolution of the numerical solution near the interface when $P^2$ elements is adopted in the test.

\begin{figure}[hbtp]
 \begin{center}
 \mbox{
{\includegraphics[width=8cm]{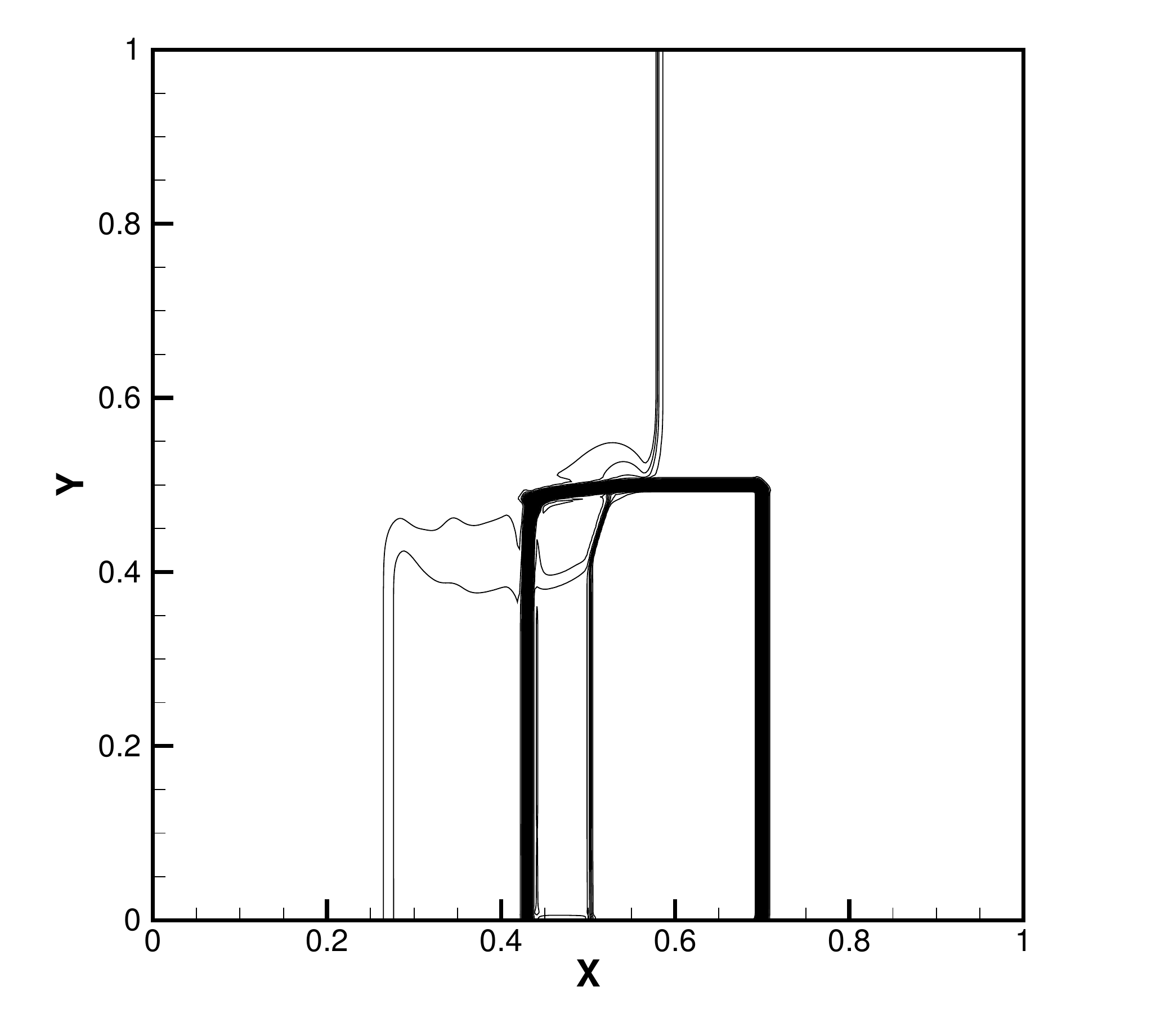}}\quad
{\includegraphics[width=8cm]{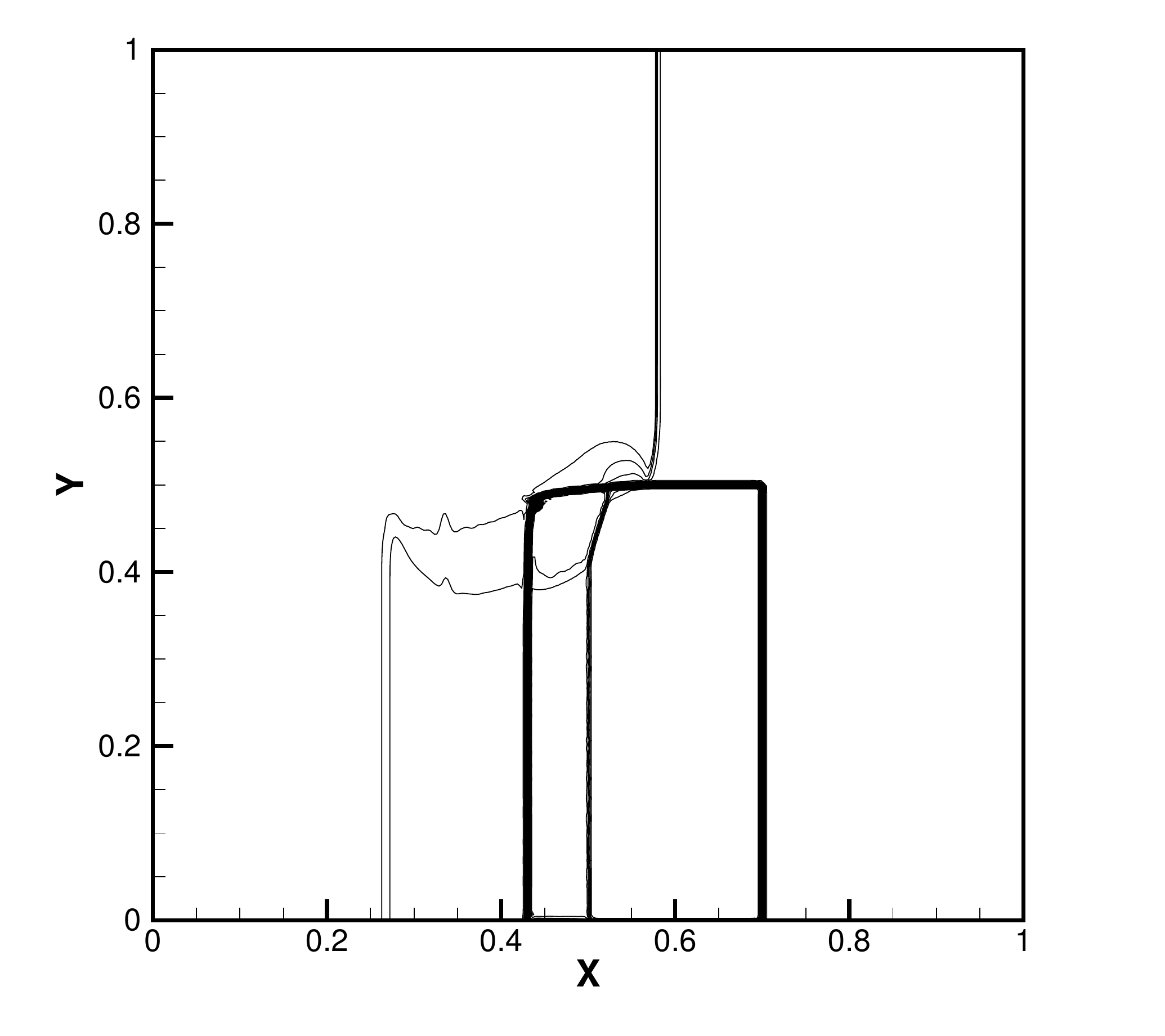}}

   }

   \mbox{
   {\includegraphics[width=8cm]{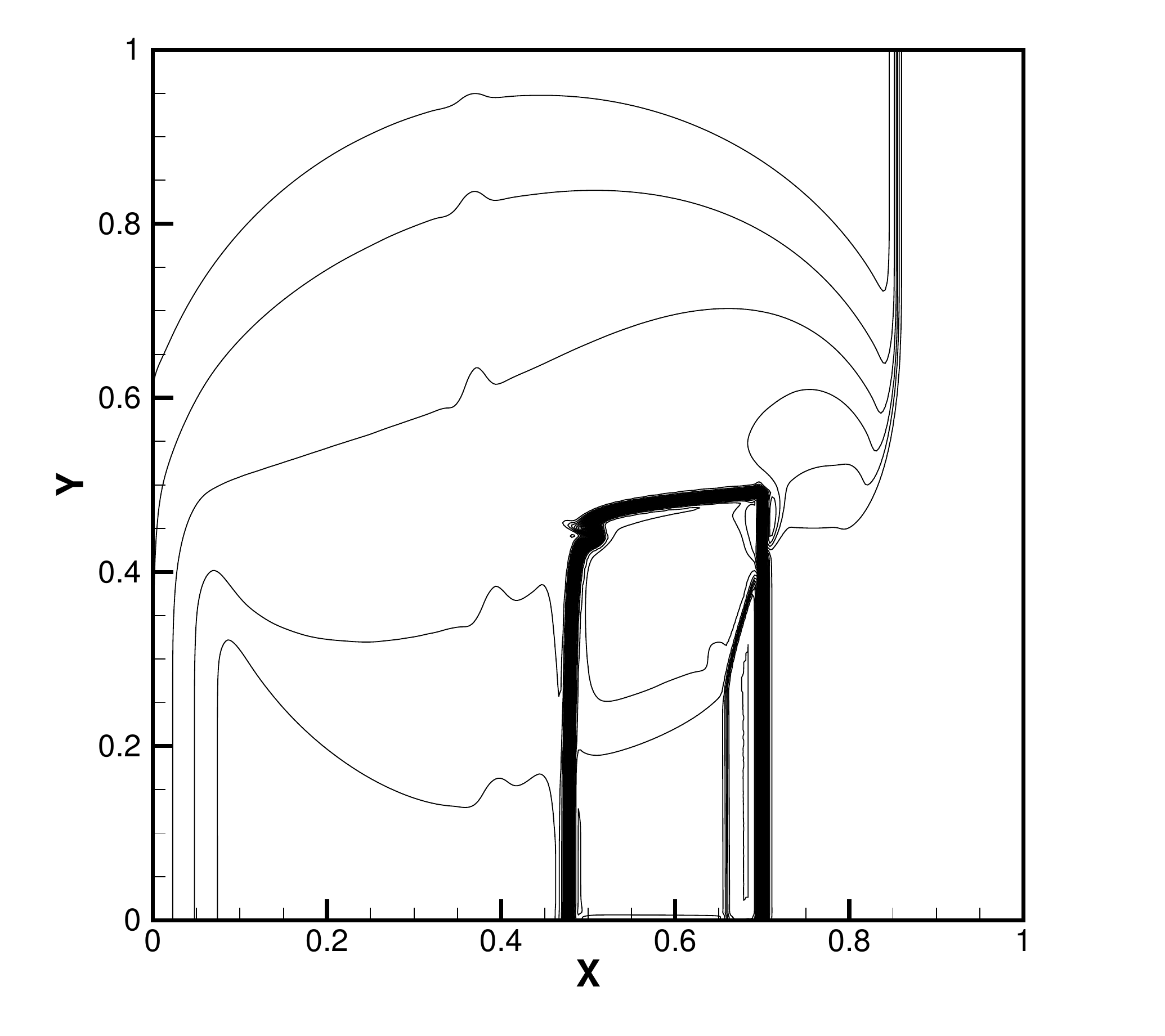}}\quad
      {\includegraphics[width=8cm]{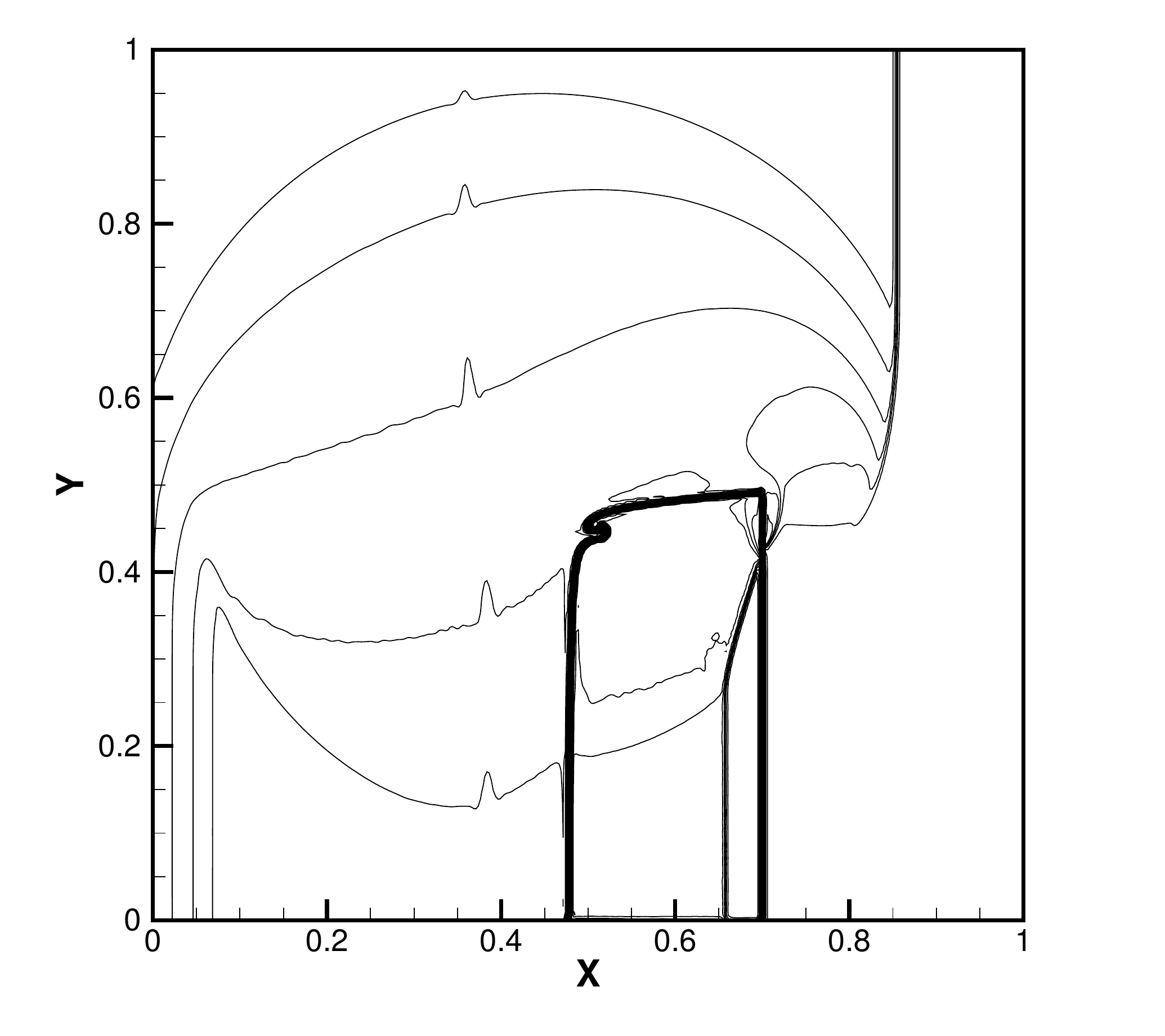}}

   }

   \caption{Example~\ref{shock2d} $N=400\times 400$, density contours. Top: $T=$ 50 $\mu$s; Bottom: $T=$ 100 $\mu$s. Left: $P^1$ elements; Right: $P^2$ elements}
   \label{figshock2d}
   \end{center}
   \end{figure}

\begin{figure}[hbtp]
 \begin{center}
 \mbox{
{\includegraphics[width=8cm]{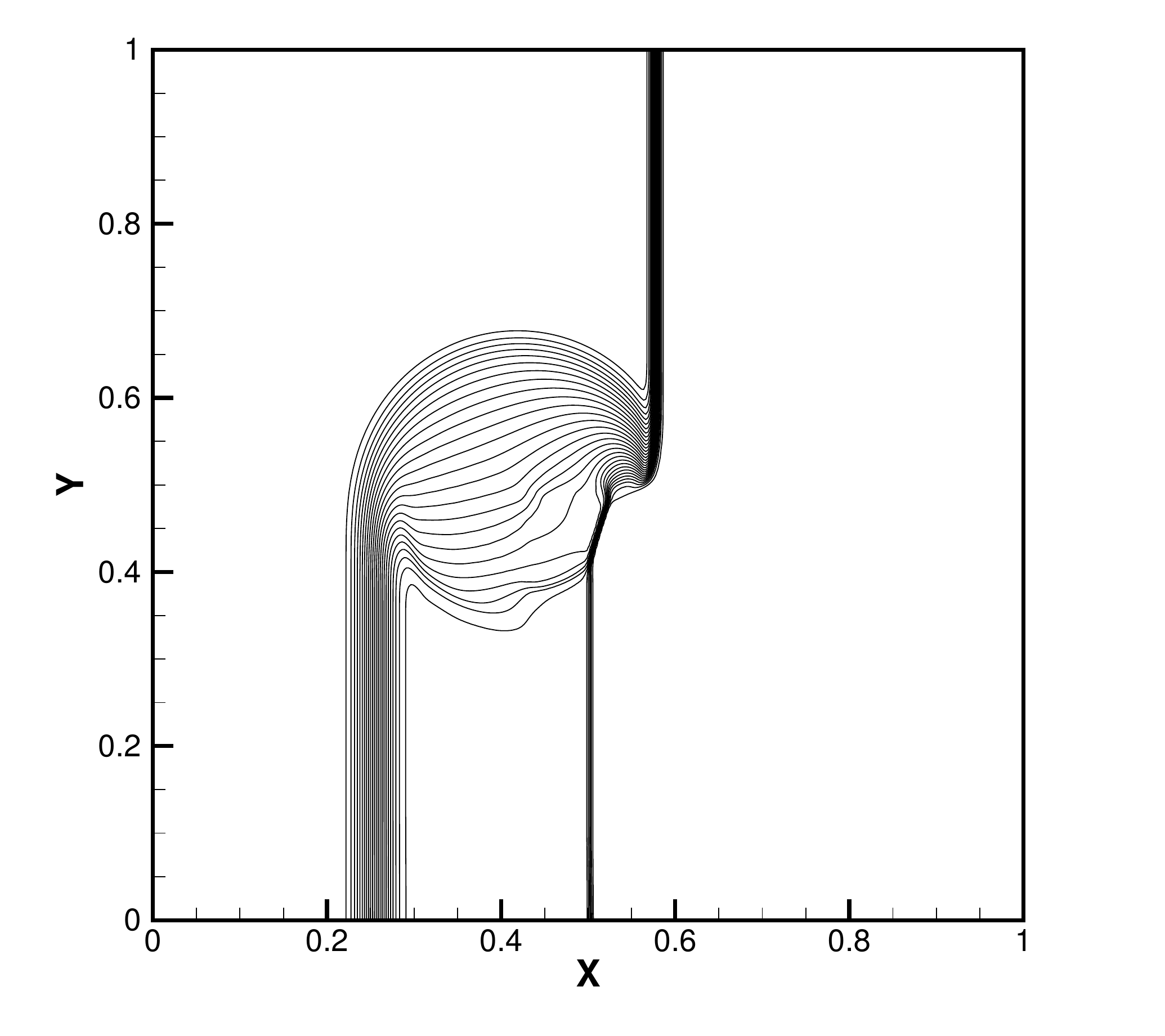}}\quad
{\includegraphics[width=8cm]{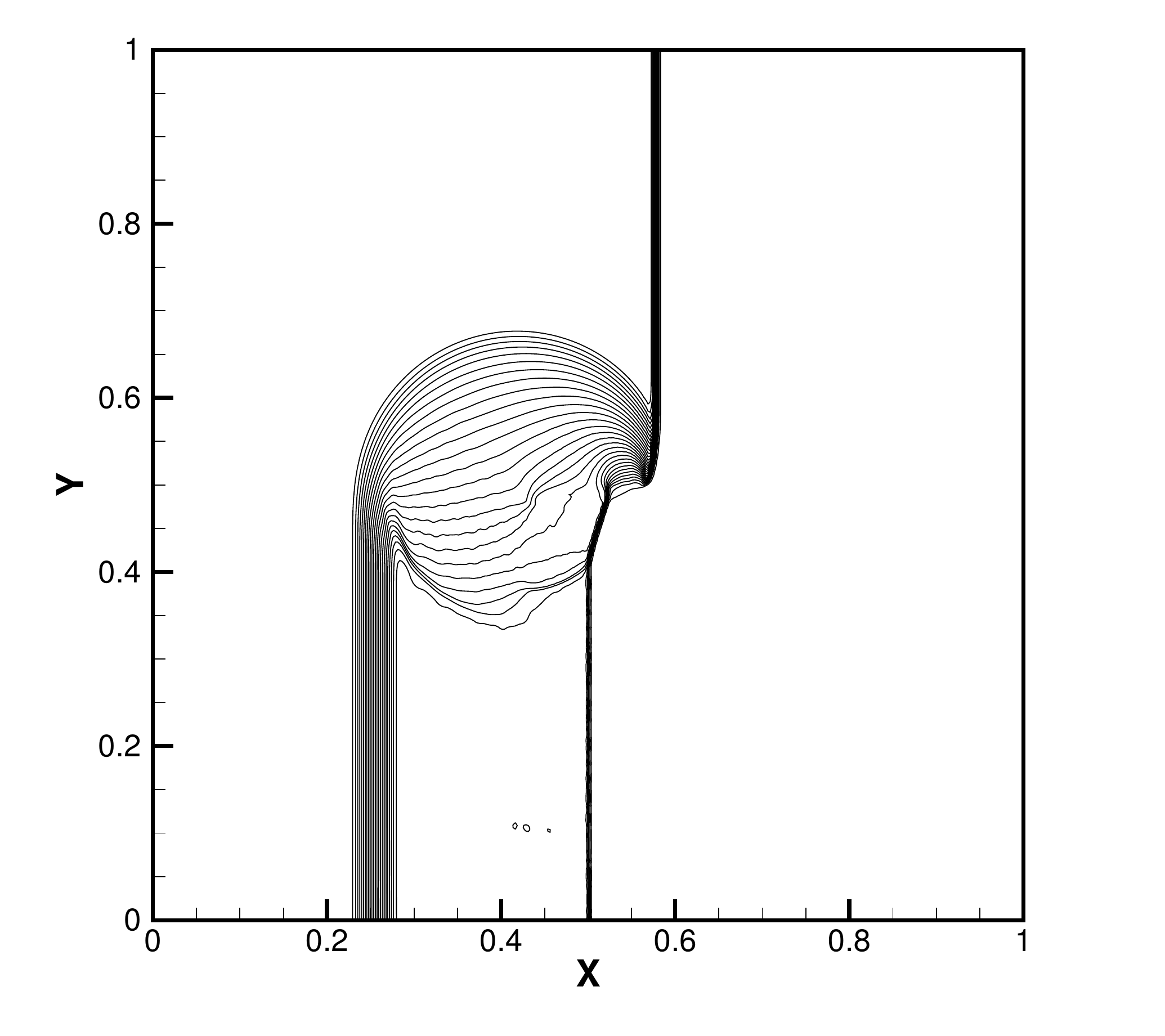}}

   }

   \mbox{
   {\includegraphics[width=8cm]{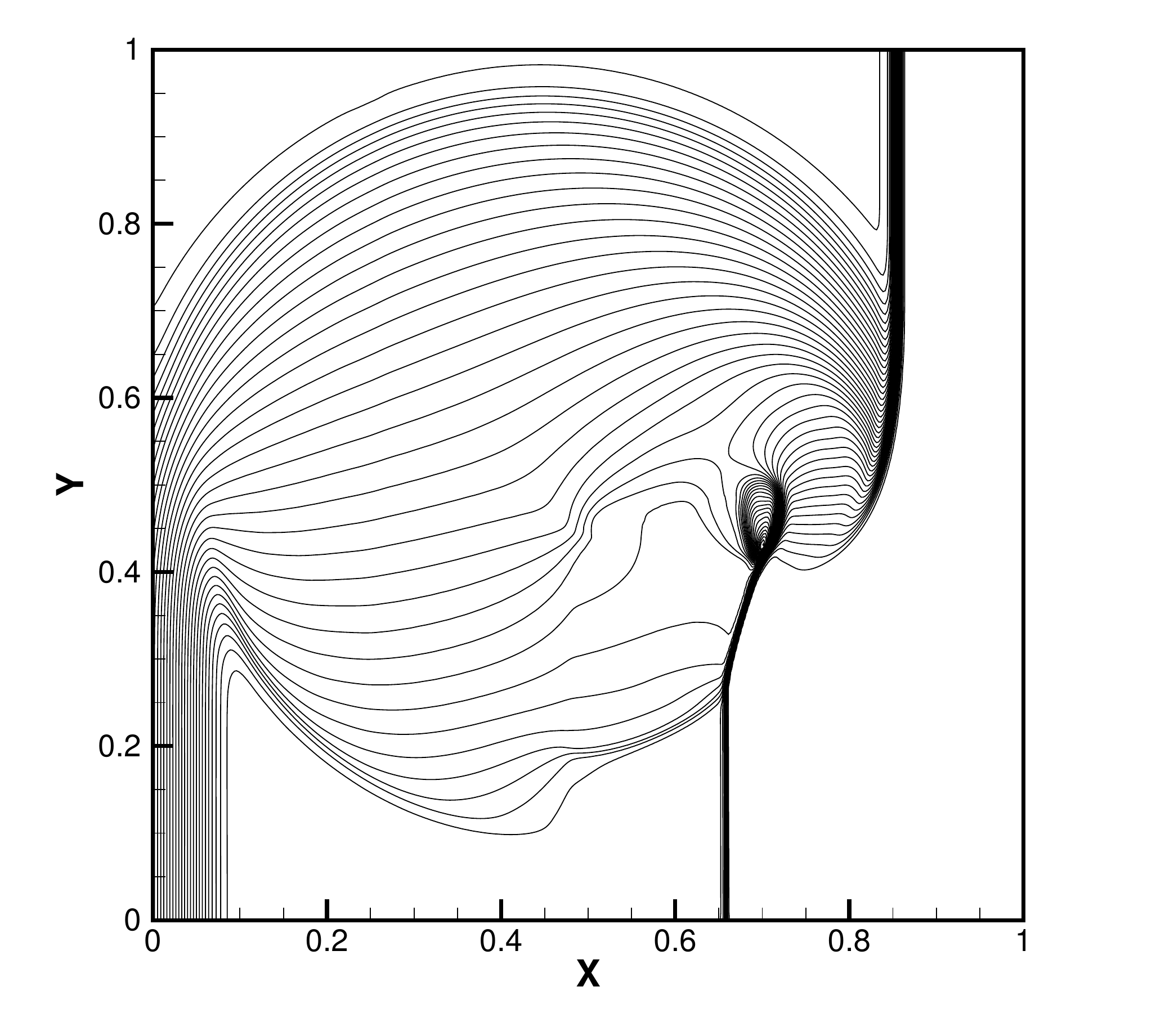}}\quad
      {\includegraphics[width=8cm]{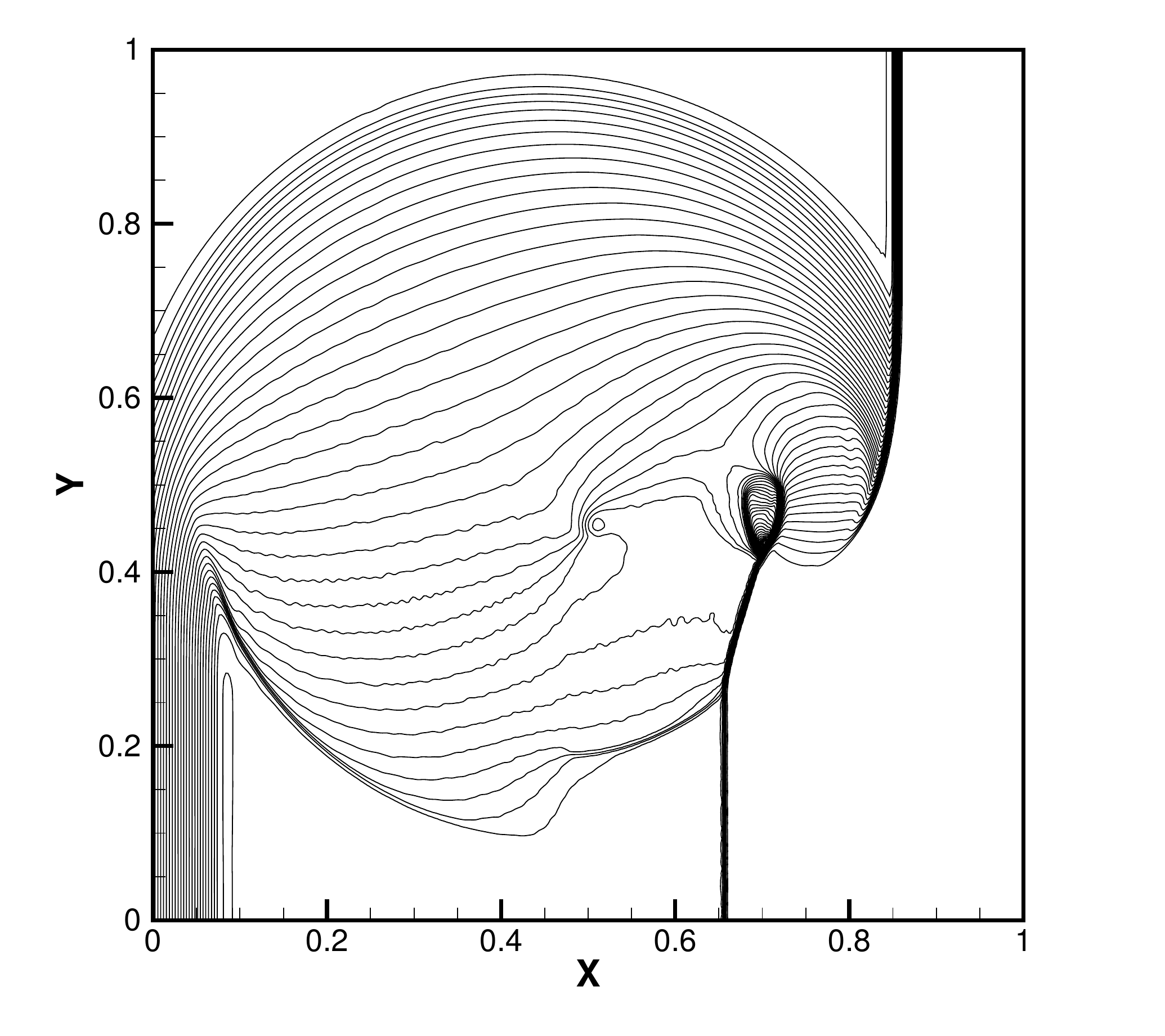}}

   }

   \caption{Example~\ref{shock2d} $N=400\times 400$, pressure contours. Top: $T=$ 50 $\mu$s; Bottom: $T=$ 100 $\mu$s. Left: $P^1$ elements; Right: $P^2$ elements}
   \label{figshock2dP}
   \end{center}
   \end{figure}

}
\end{exam}

\begin{exam}{\em
\label{threecoms}

Finally, we consider the three-component impact problem in two dimensions \cite{shyue2001}. The computation domain is taken as $(x,y)\in (0,1)\times (0,1)$. Initially, a leftward going copper plate traveling vertically in a shock tube with speed 1500 m/s from right to left in region $x\geqslant 0.6$, while in region $x<0.6$, we have a solid inert explosive on the top and a liquid water on the bottom separated by the interface at $y=0.5$. The solid inert explosive and liquid water are at rest and all three fluid components are in the usual atmospheric condition initially throughout the domain. The copper and explosive are modeled by the CC EOS \eqref{cc} while the water is modeled by JWL EOS \eqref{jwl}.

The numerical results are shown in Figs. \ref{copexpwatrho}-\ref{copexpwatpy}. Clearly, we observe that the shock speed in explosive is larger than the one in water from the figures since the acoustic impedance of explosive is greater than the one for the water. 

\begin{figure}[hbtp]
 \begin{center}
 \mbox{
{\includegraphics[width=8cm]{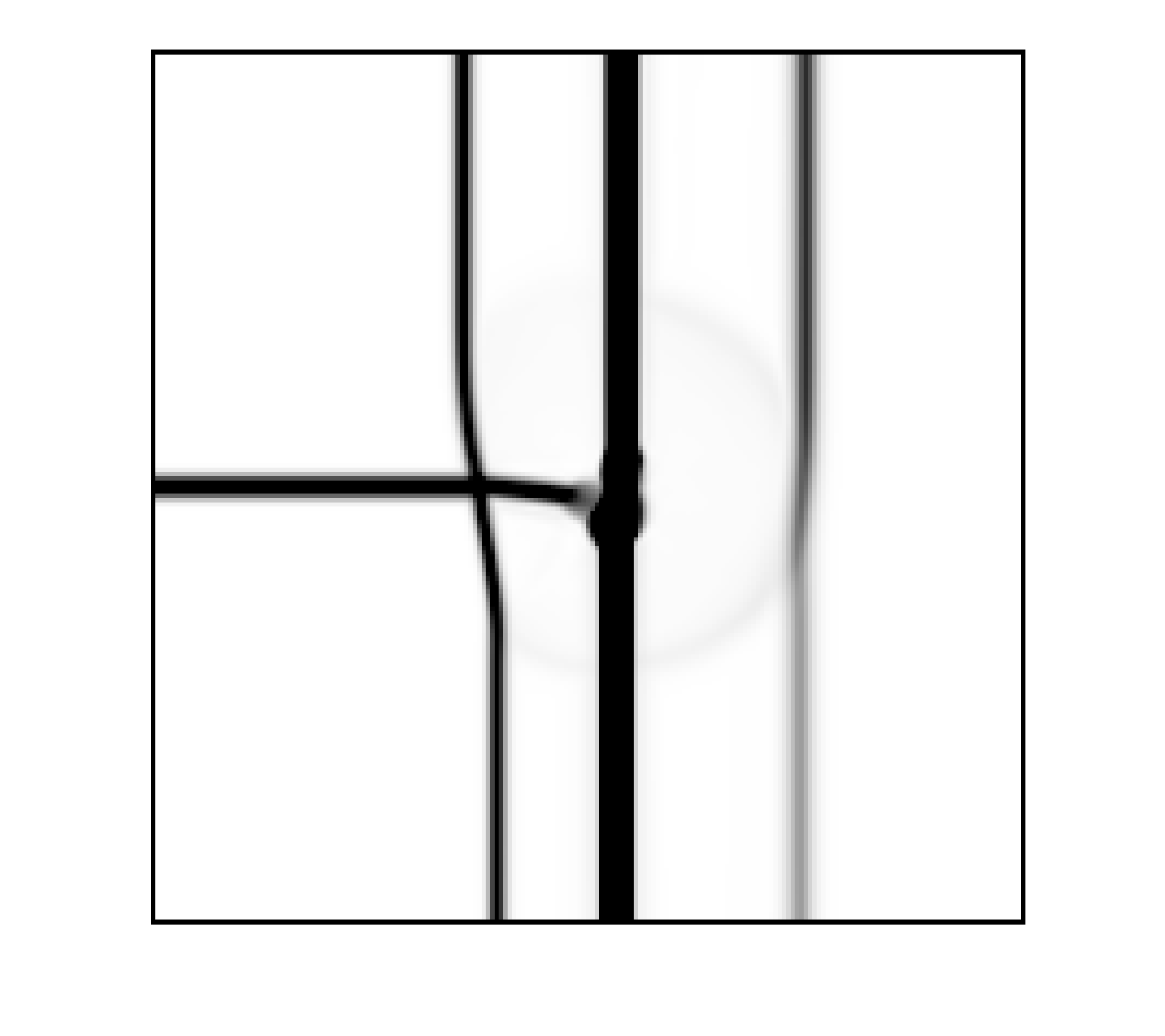}}\quad
{\includegraphics[width=8cm]{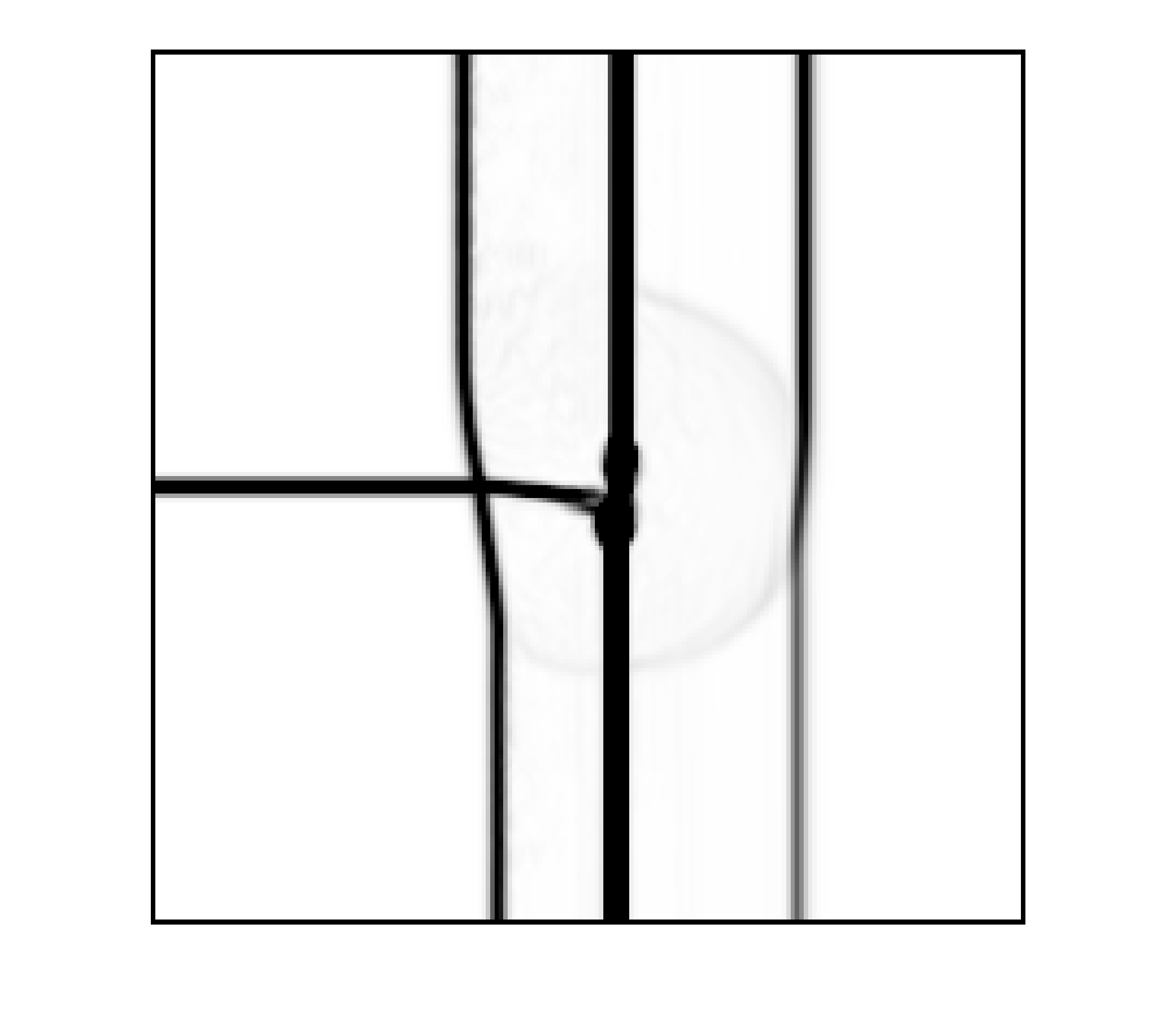}}

   }

   \mbox{
   {\includegraphics[width=8cm]{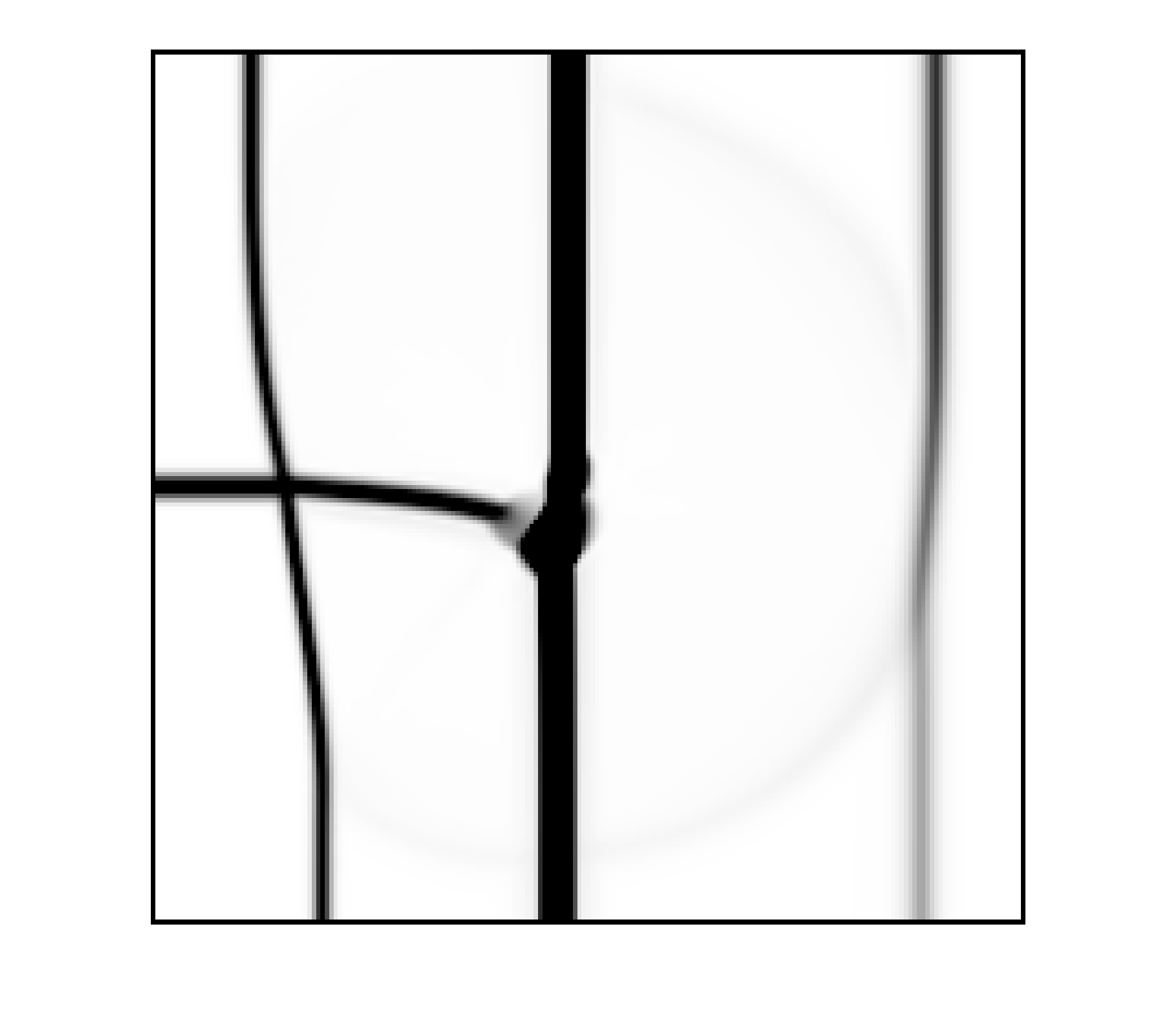}}\quad
      {\includegraphics[width=8cm]{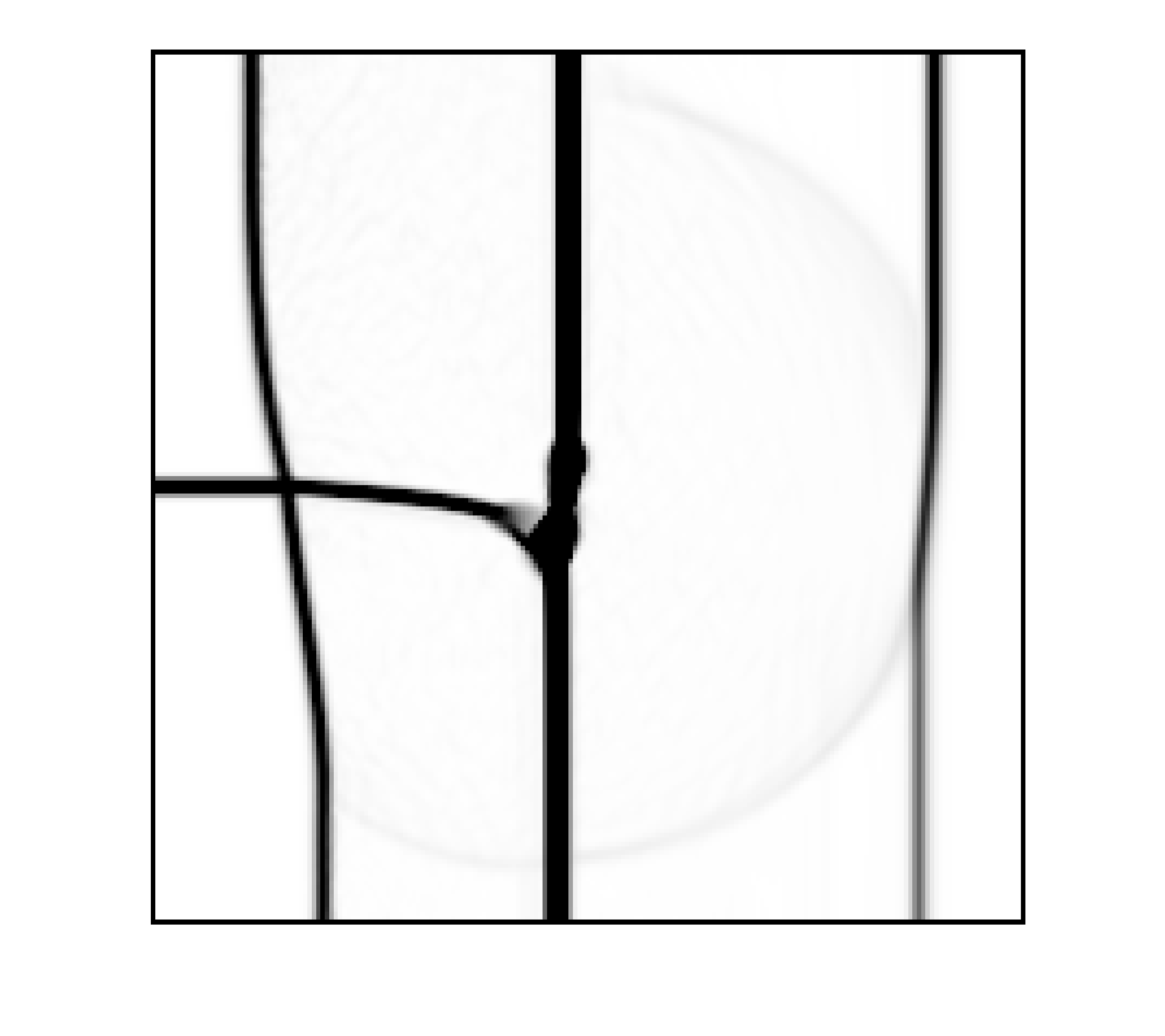}}

   }
  
   \caption{Example~\ref{threecoms} Schlieren-type images for the density using $N=200\times 200$. Top: $T=$ 50 $\mu$s; Bottom: $T=$ 100 $\mu$s. Left: $P^1$ elements; Right: $P^2$ elements}
   \label{copexpwatrho}
   \end{center}
   \end{figure}

\begin{figure}[hbtp]
 \begin{center}
 
 \mbox{
{\includegraphics[width=8cm]{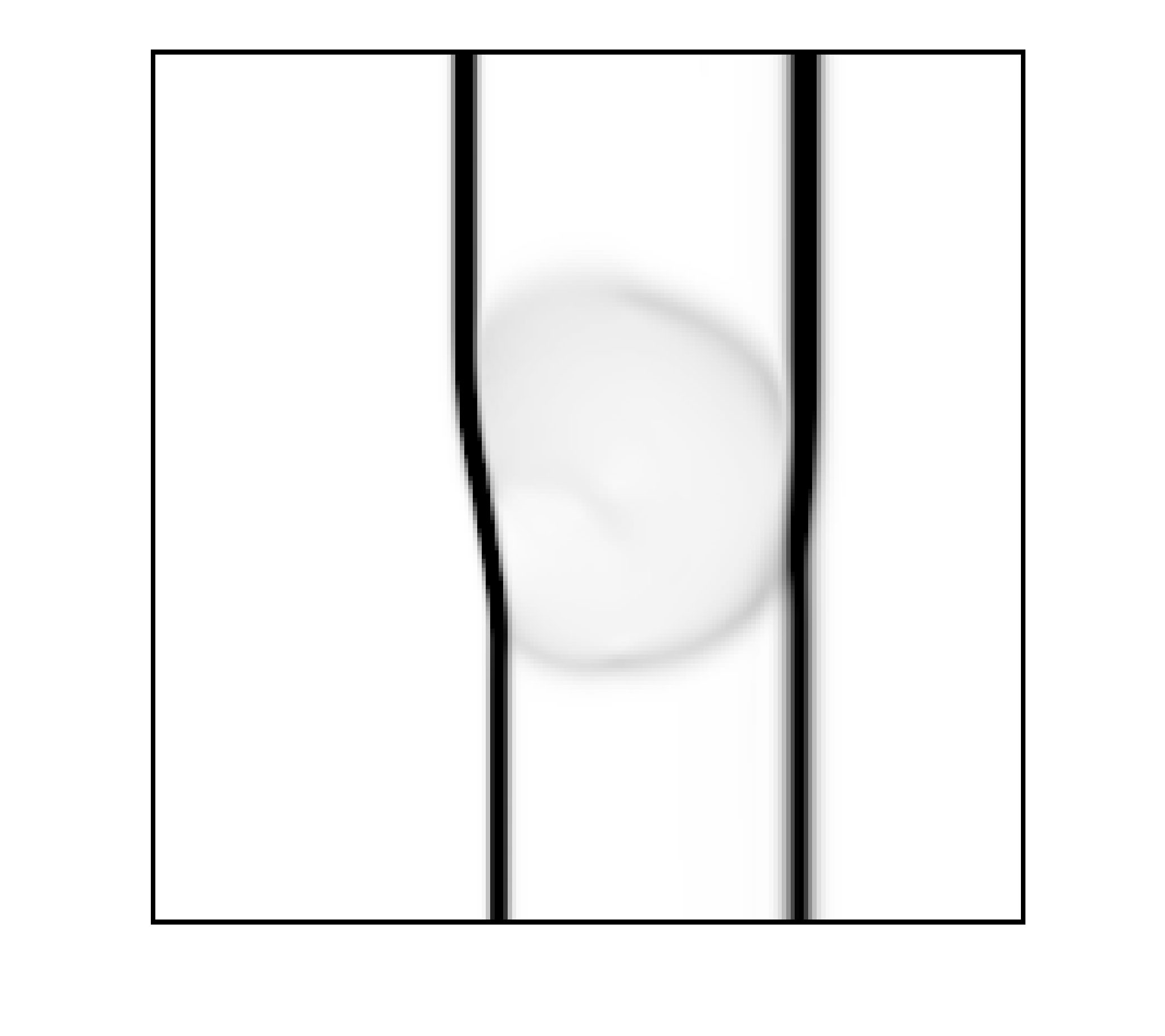}}\quad
{\includegraphics[width=8cm]{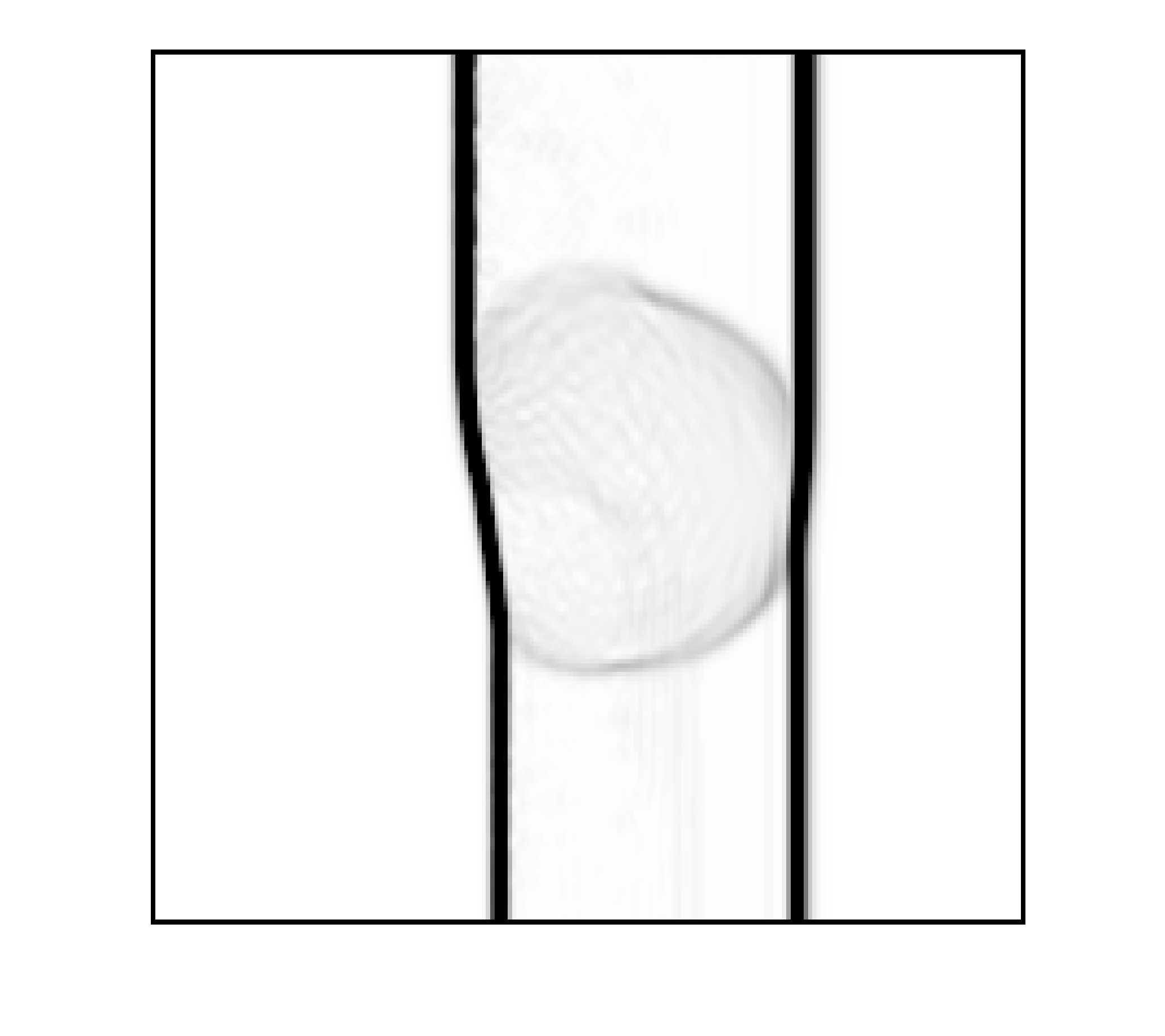}}

   }

   \mbox{
   {\includegraphics[width=8cm]{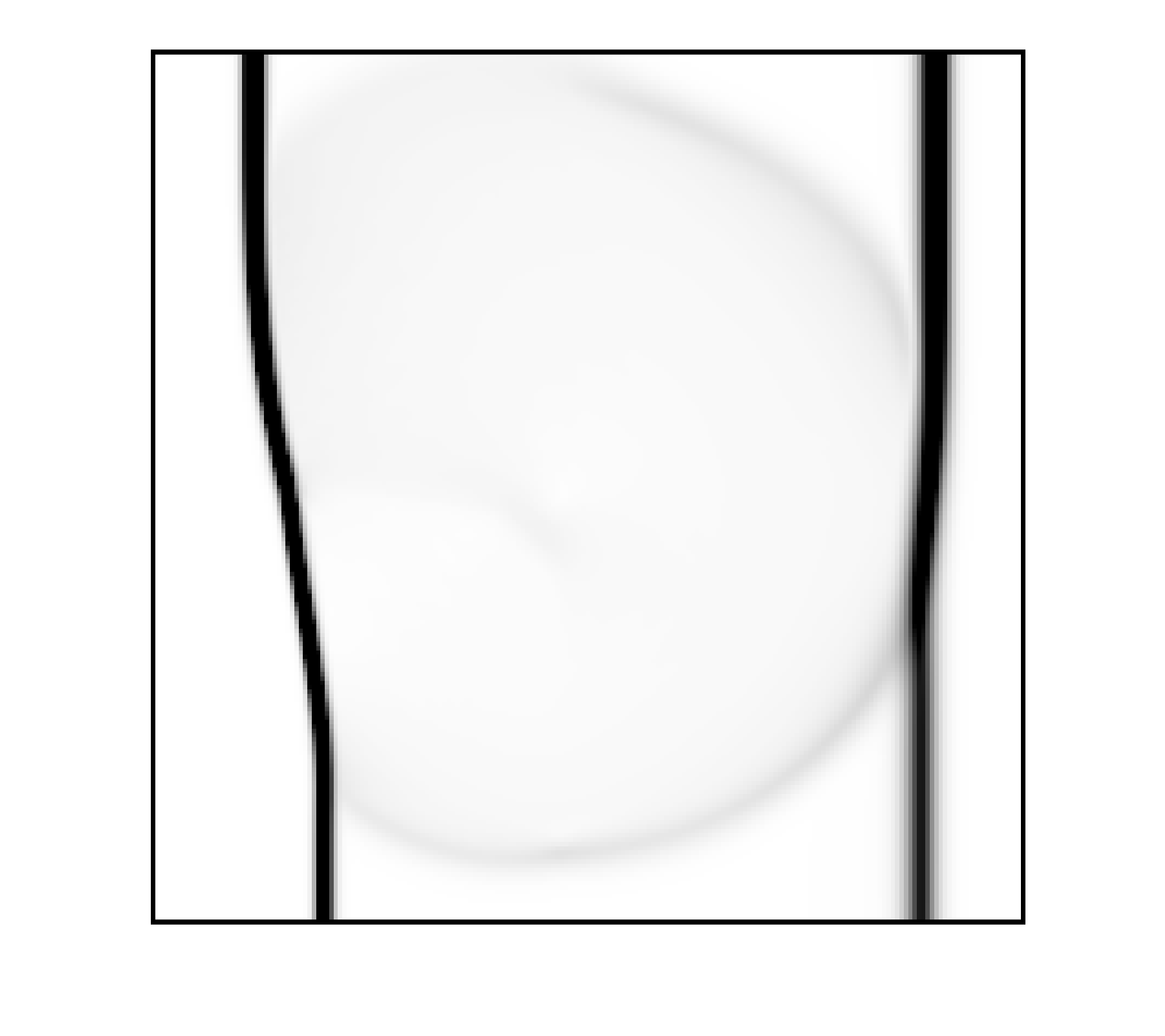}}\quad
      {\includegraphics[width=8cm]{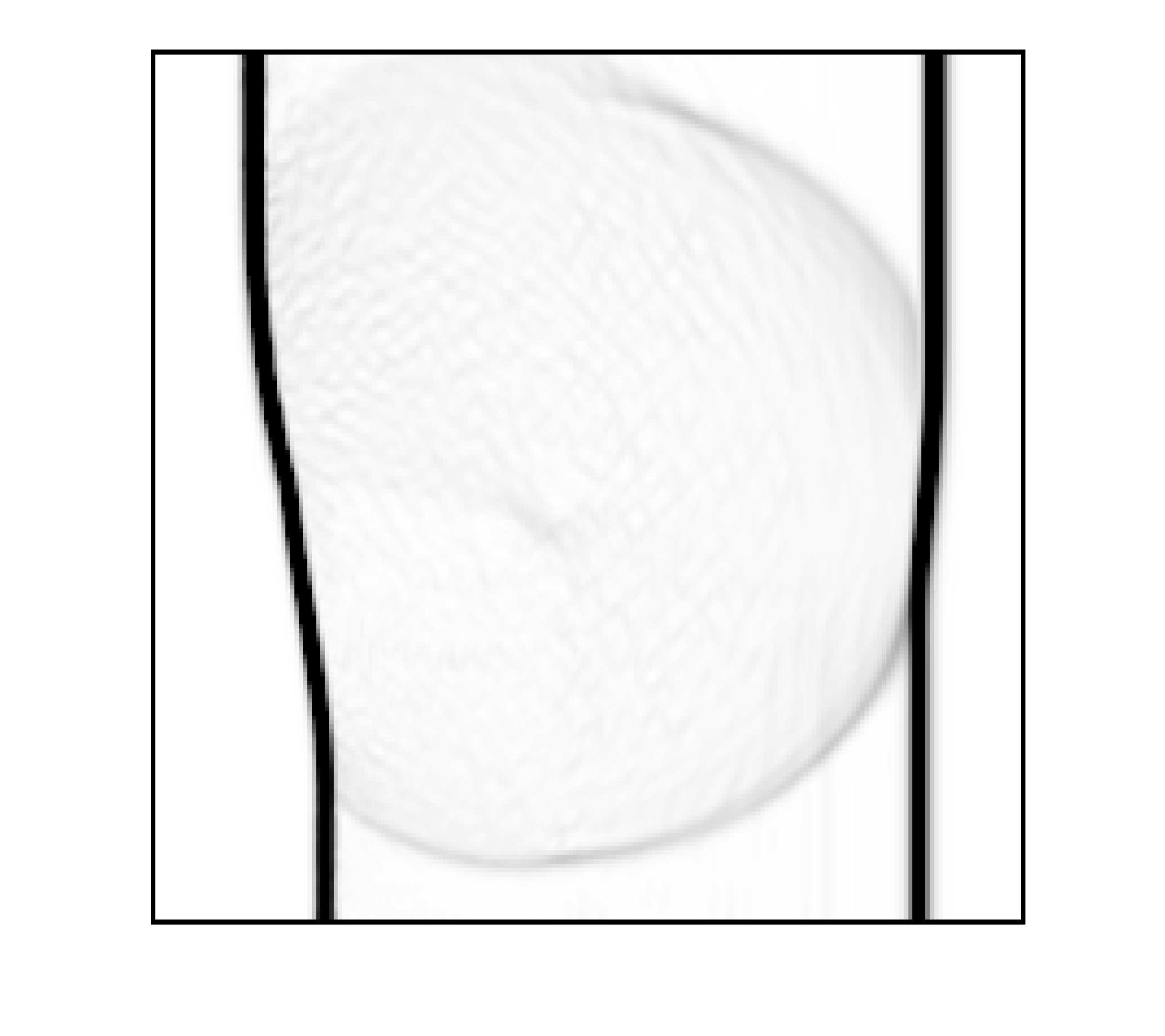}}

   }

   \caption{Example~\ref{threecoms} Schlieren-type images for the pressure using $N=200\times 200$. Top: $T=$ 50 $\mu$s; Bottom: $T=$ 100 $\mu$s. Left: $P^1$ elements; Right: $P^2$ elements}
   \label{copexpwatp}
   \end{center}
   \end{figure}

\begin{figure}[hbtp]
 \begin{center}
 \mbox{
{\includegraphics[width=8cm]{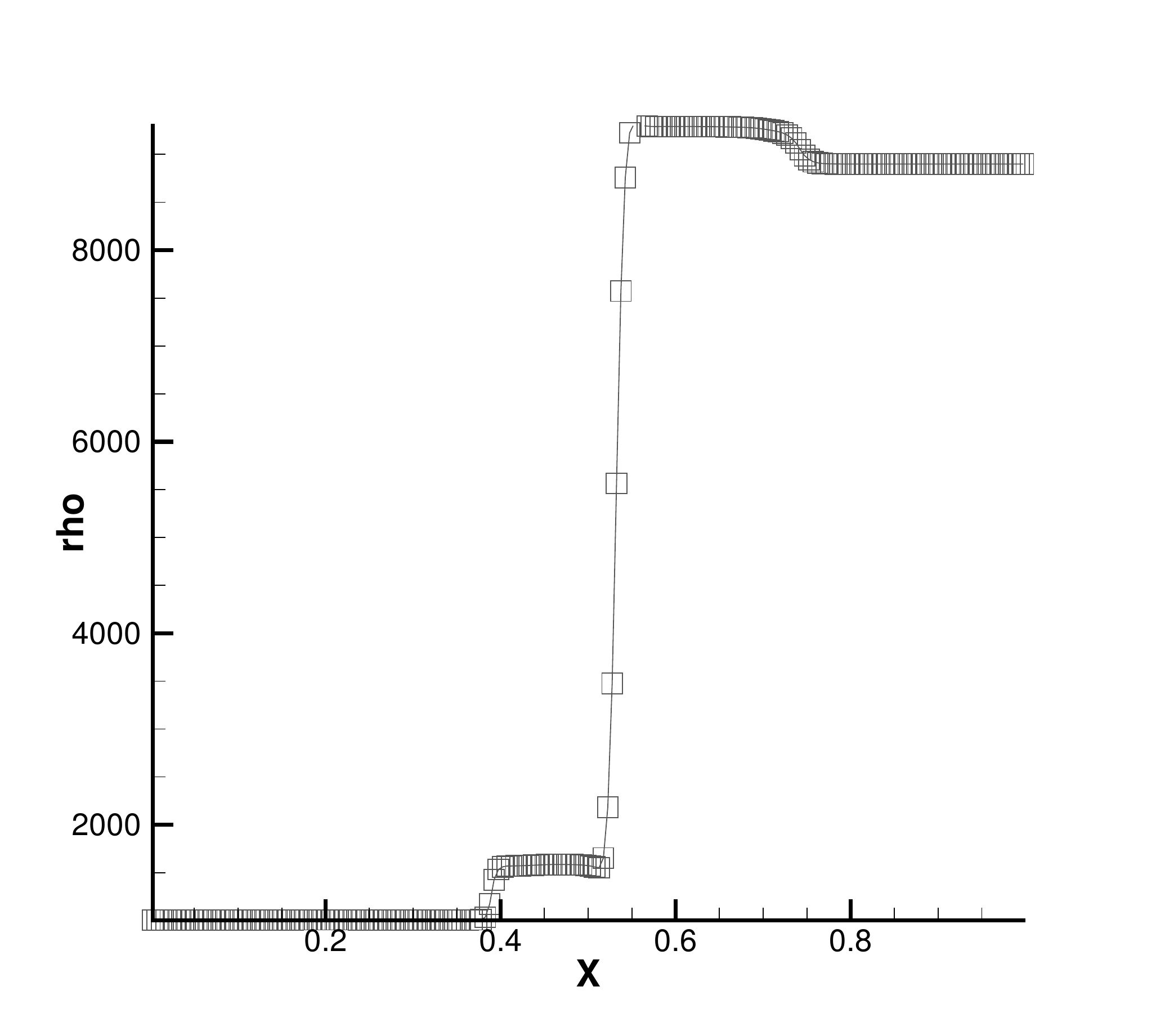}}\quad
{\includegraphics[width=8cm]{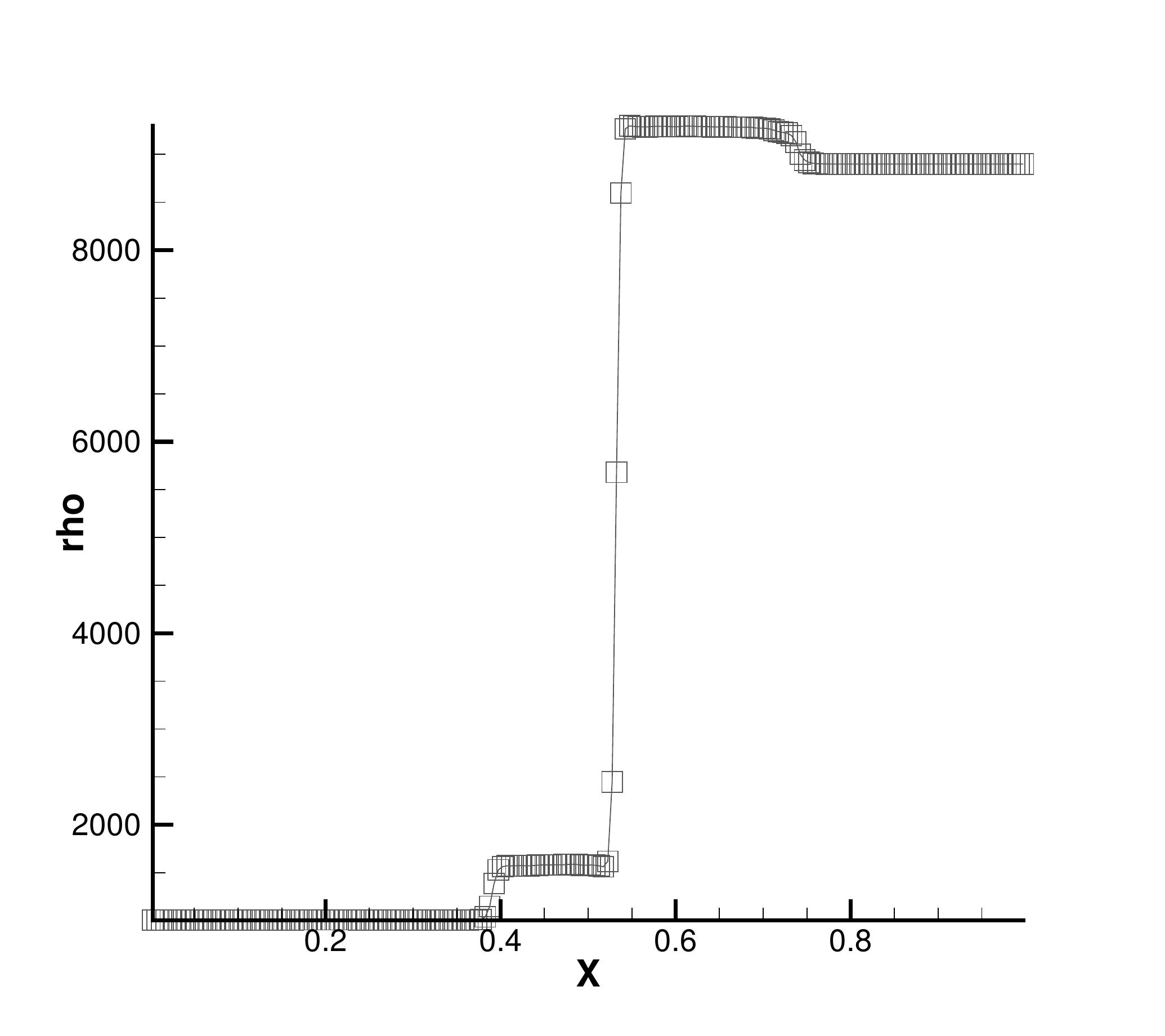}}

   }

   \mbox{
   {\includegraphics[width=8cm]{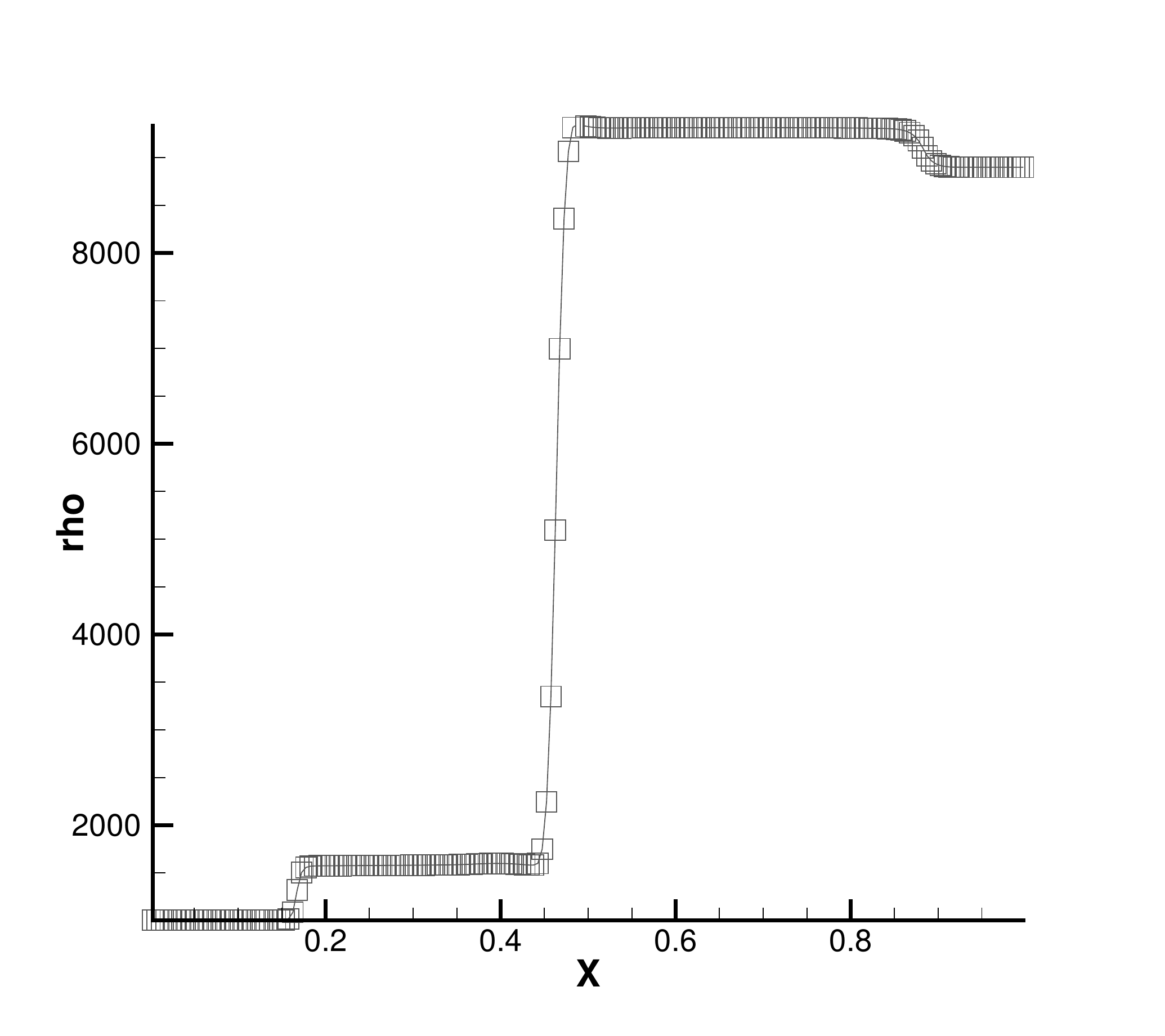}}\quad
      {\includegraphics[width=8cm]{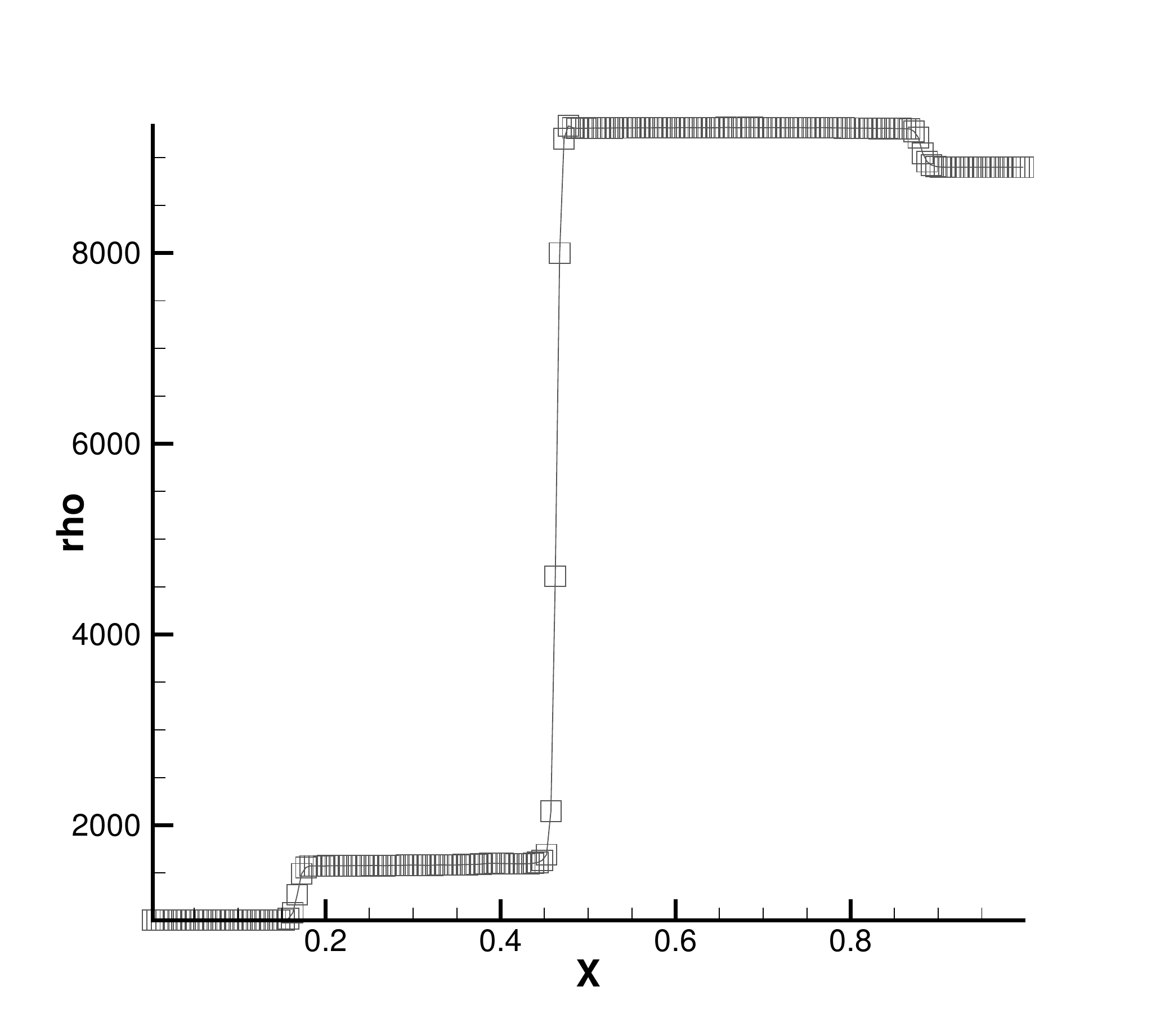}}

   }
  
   \caption{Example~\ref{threecoms} The cross-sectional plots of the results shown in Fig. \ref{copexpwatrho} along $y=0.4$. Top: $T=$ 50 $\mu$s; Bottom: $T=$ 100 $\mu$s. Left: $P^1$ elements; Right: $P^2$ elements}
   \label{copexpwatrhoy}
   \end{center}
   \end{figure}

\begin{figure}[hbtp]
 \begin{center}
 \mbox{
{\includegraphics[width=8cm]{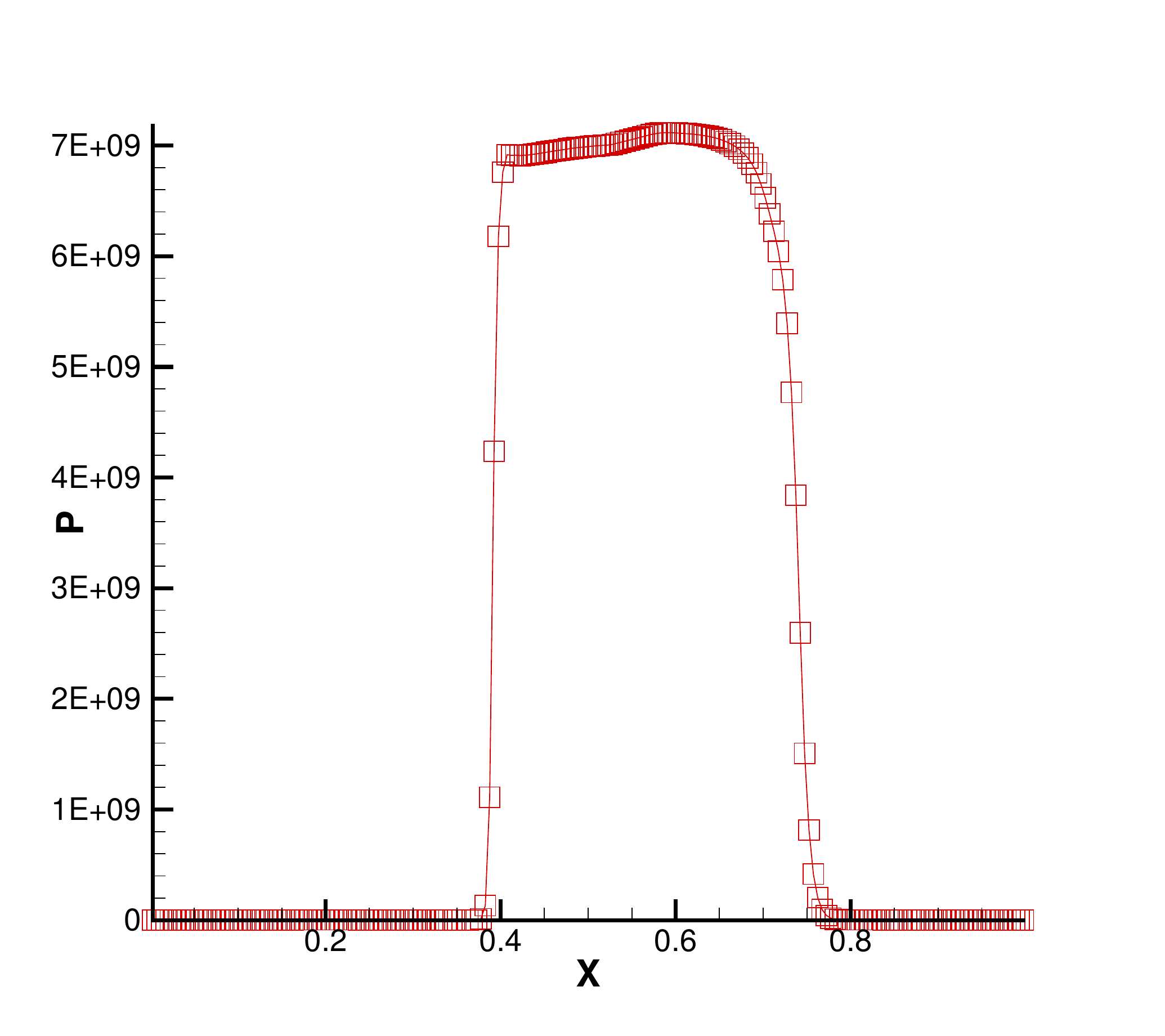}}\quad
{\includegraphics[width=8cm]{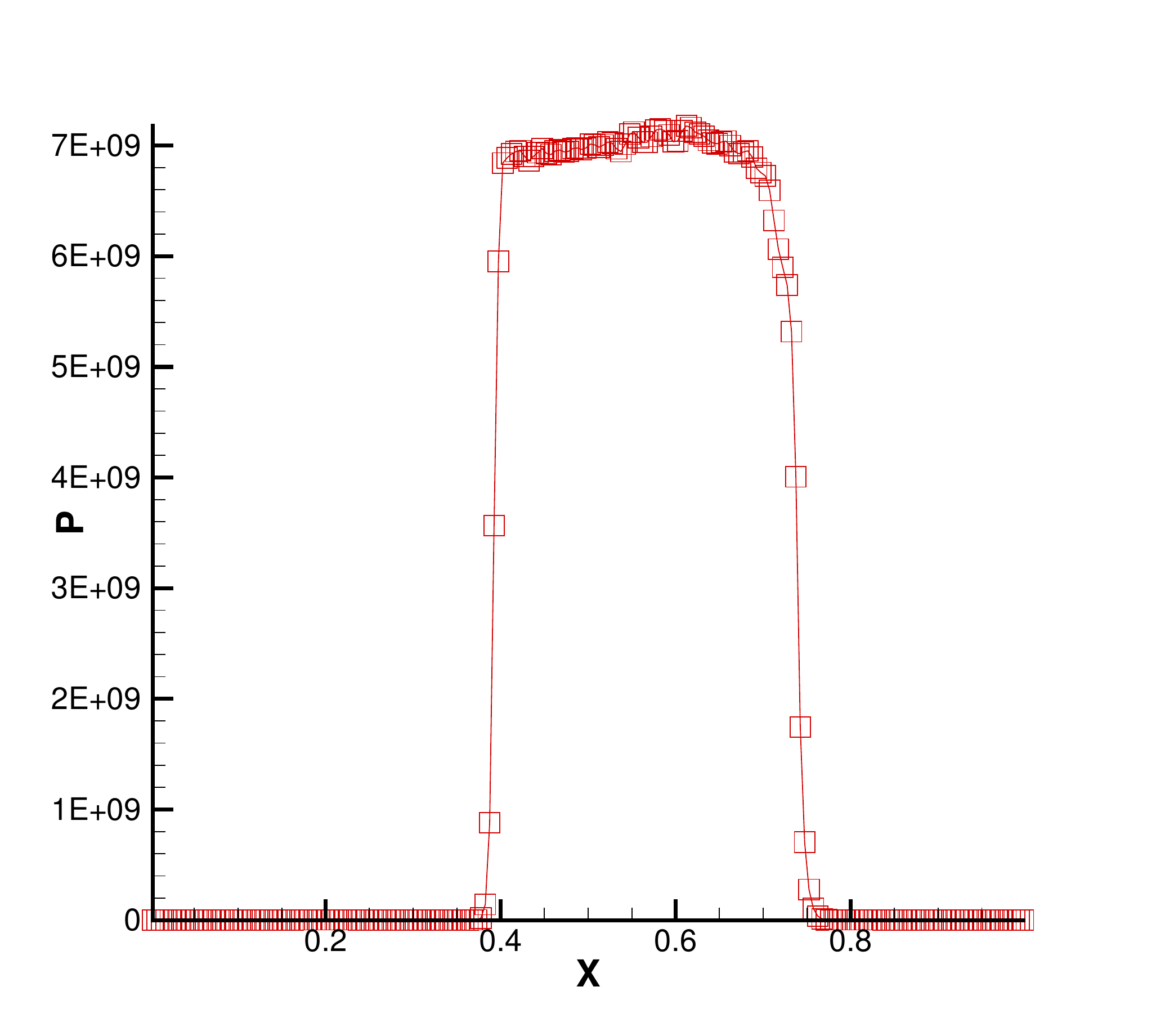}}

   }

   \mbox{
   {\includegraphics[width=8cm]{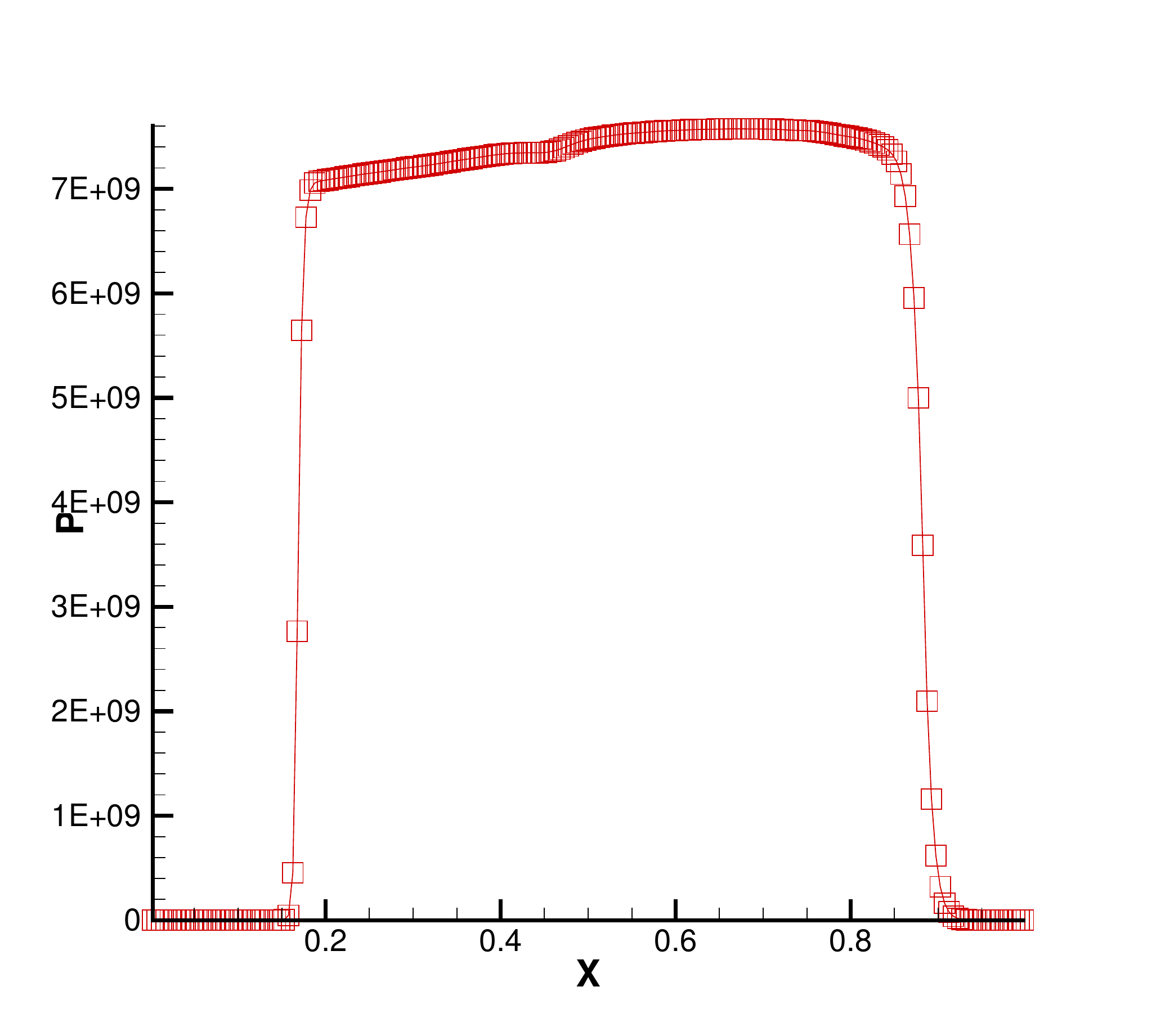}}\quad
      {\includegraphics[width=8cm]{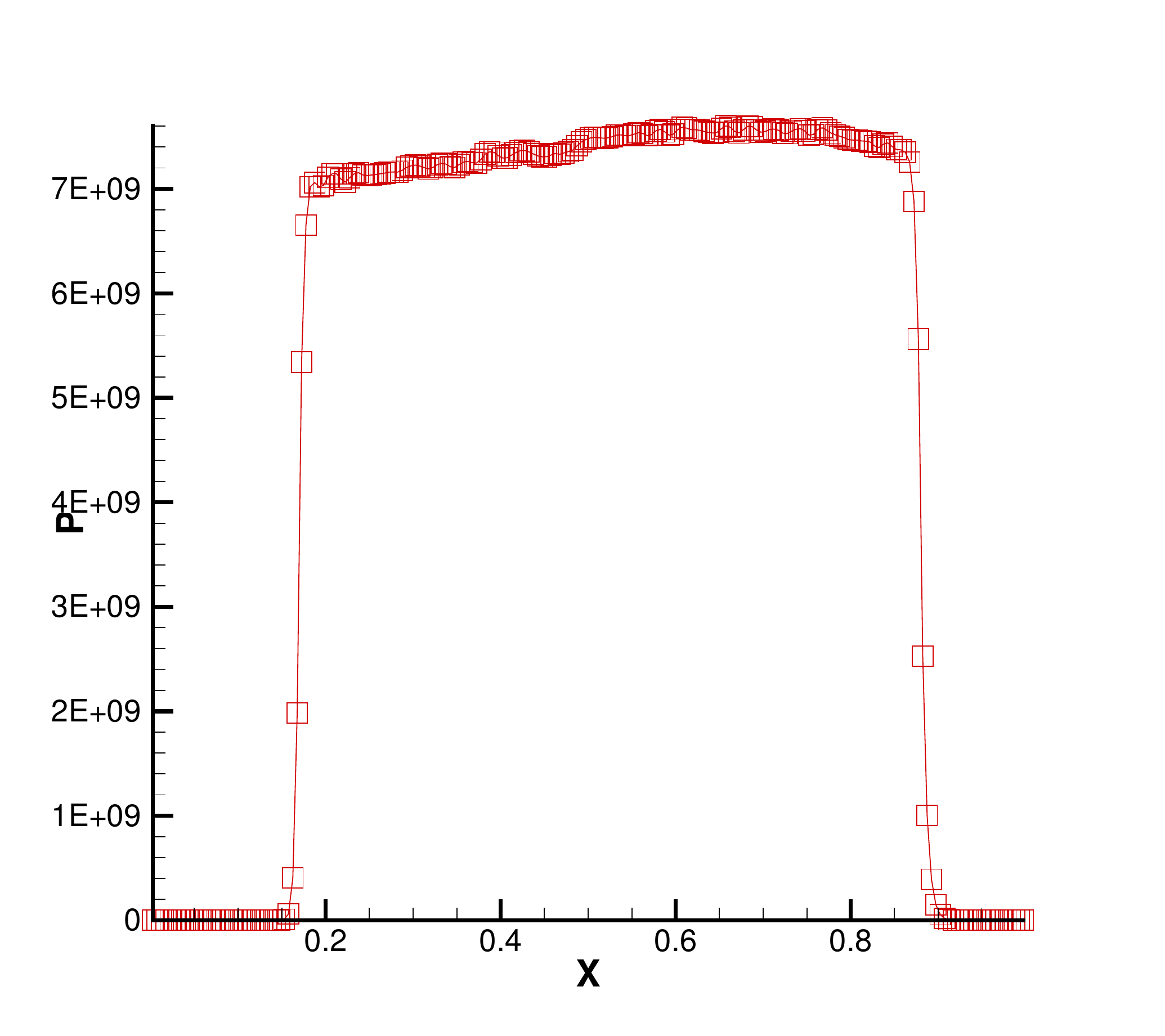}}

   }
  
   \caption{Example~\ref{threecoms} The cross-sectional plots of the results shown in Fig. \ref{copexpwatp} along $y=0.4$. Top: $T=$ 50 $\mu$s; Bottom: $T=$ 100 $\mu$s. Left: $P^1$ elements; Right: $P^2$ elements}
   \label{copexpwatpy}
   \end{center}
   \end{figure}

}
\end{exam}

\section{Conclusions}
\label{sec4}
\setcounter{equation}{0}
\setcounter{figure}{0}
\setcounter{table}{0}

We have presented a high order quasi-conservative DG method for compressible multi-component flows with Mie-Gr\"uneisen equation of state based on the 5-equation model in the previous sections. In this paper the NOK flux is used to compute the numerical flux, which is free from constructing Riemann solver. Then, a DG scheme is defined for the volume fraction equations according to the procedure of the quasi-conservative method, which can keep the velocity and pressure oscillation-free at the interface. In addition, a maximum-pricinple-satisfying limiter is employed to ensure that the volume fraction does not go out of the range. Numerical results in one and two dimensions shown in the paper demonstrate the ability of the method to capture shocks and material interfaces and be high order in smooth regions. In the future, we plan to further extend the method to the unstructured mesh. In order to reduce the numerical diffusion further, we will extend the quasi-Lagrangian moving DG method \cite{luo2019} to the 5-equation model of multi-component flows.

\section*{Acknowledgements}
{
This work is supported by National Natural Science Foundation of China (Grant Nos. 11671050), Science Challenge Project (Grant No. TZ2016002), and National Natural Science Foundation of China (Grant Nos. U1630247).
}

\end{document}